\documentclass[twocolumn]{aastex63}
\usepackage{appendix}
\usepackage{natbib}
\usepackage{subfigure}
\usepackage{booktabs}
\usepackage{xspace}
\usepackage[encapsulated]{CJK}

\usepackage{amsmath} % or simply amstext
\newcommand{\angstrom}{\text{\,\normalfont\AA}\xspace}

\newcommand{\oiii}{[\ion{O}{3}]\xspace}
\newcommand{\nii}{[\ion{N}{2}]\xspace}

\newcommand{\feii}{\ion{Fe}{2}\xspace}
\newcommand{\nev}{[\ion{Ne}{5}]\xspace}
\newcommand{\msun}{$M_{\odot}$}
\newcommand{\sersic}{S\'ersic}

\newcommand{\mbh}{$M_{\rm BH}$}

\newcommand{\jwst}{\ensuremath{JWST}}

\newcommand{\source}{A2744$-$45924\xspace}
\newcommand{\hide}[1]{}

\def\halpha{\ensuremath{\mathrm{H}\alpha}\xspace}
\def\hbeta{\ensuremath{\mathrm{H}\beta}\xspace}

\def\MgII{Mg\,{\sc ii}\xspace}

\def\FeII{Fe\,{\sc ii}\xspace}

\def\HeII{He\,{\sc ii}\xspace}

\newcommand{\jw}{\emph{JWST}}
    
\usepackage[encapsulated]{CJK}

\begin{document}
   
\title{An unambiguous AGN and a Balmer break in an Ultraluminous Little Red Dot at z=4.47 from Ultradeep UNCOVER and All the Little Things Spectroscopy}

\author[0000-0002-2057-5376]{Ivo Labbe}
\affiliation{Centre for Astrophysics and Supercomputing, Swinburne University of Technology, Melbourne, VIC 3122, Australia}

\author[0000-0002-5612-3427]{Jenny E. Greene}
\affiliation{Department of Astrophysical Sciences, Princeton University, 4 Ivy Lane, Princeton, NJ 08544, USA}

\author[0000-0003-2871-127X]{Jorryt Matthee}
\affiliation{Institute of Science and Technology Austria (ISTA), Am Campus 1, Klosterneuburg, Austria}

\author[0000-0003-0660-9776]{Helena Treiber}
\affiliation{Department of Astrophysical Sciences, Princeton University, 4 Ivy Lane, Princeton, NJ 08544, USA}

\author[0000-0002-5588-9156]{Vasily Kokorev}
\affiliation{Department of Astronomy, The University of Texas at Austin, Austin, TX 78712, USA}

\author[0000-0001-8367-6265]{Tim B. Miller}
\affiliation{Center for Interdisciplinary Exploration and Research in Astrophysics (CIERA), Northwestern University, IL 60201, USA}

\author[0000-0001-5346-6048]{Ivan Kramarenko}
\affiliation{Institute of Science and Technology Austria (ISTA), Am Campus 1, Klosterneuburg, Austria}

\author[0000-0003-4075-7393]{David J. Setton}
\thanks{Brinson Prize Fellow}
\affiliation{Department of Astrophysical Sciences, Princeton University, 4 Ivy Lane, Princeton, NJ 08544, USA}

\author[0000-0002-0463-9528]{Yilun Ma (\begin{CJK*}{UTF8}{gbsn}马逸伦\ignorespacesafterend\end{CJK*})}
\affiliation{Department of Astrophysical Sciences, Princeton University, 4 Ivy Lane, Princeton, NJ 08544, USA}

\author[0000-0003-4700-663X]{Andy D. Goulding}
\affiliation{Department of Astrophysical Sciences, Princeton University, 4 Ivy Lane, Princeton, NJ 08544, USA}

\author[0000-0001-5063-8254]{Rachel Bezanson}
\affiliation{Department of Physics and Astronomy and PITT PACC, University of Pittsburgh, Pittsburgh, PA 15260, USA}

\author[0000-0003-3997-5705]{Rohan~P.~Naidu}
\altaffiliation{NASA Hubble Fellow}
\affiliation{MIT Kavli Institute for Astrophysics and Space Research, 77 Massachusetts Ave., Cambridge, MA 02139, USA}

\author[0000-0003-2919-7495]{Christina C.\ Williams}
\affiliation{NSF National Optical-Infrared Astronomy Research Laboratory, 950 North Cherry Avenue, Tucson, AZ 85719, USA}

\author[0000-0002-7570-0824]{Hakim Atek}
\affiliation{Institut d'Astrophysique de Paris, CNRS, Sorbonne Universit\'e, 98bis Boulevard Arago, 75014, Paris, France}

\author[0000-0003-2680-005X]{Gabriel Brammer} \affiliation{Cosmic Dawn Center (DAWN), Niels Bohr Institute, University of Copenhagen, Jagtvej 128, K{\o}benhavn N, DK-2200, Denmark}

\author[0000-0002-7031-2865]{Sam E. Cutler}\affiliation{Department of Astronomy, University of Massachusetts, Amherst, MA 01003, USA}

\author[0009-0009-9795-6167]{Iryna Chemerynska}
\affiliation{Institut d'Astrophysique de Paris, CNRS, Sorbonne Universit\'e, 98bis Boulevard Arago, 75014, Paris, France}

\author[0000-0001-9978-2601]{Aidan P. Cloonan}
\affiliation{Department of Astronomy, University of Massachusetts, Amherst, MA 01003, USA}

\author[0000-0001-8460-1564]{Pratika Dayal}
\affiliation{Kapteyn Astronomical Institute, University of Groningen, P.O. Box 800, 9700 AV Groningen, The Netherlands}

\author[0000-0002-2380-9801]{Anna de Graaff}
\affiliation{Max-Planck-Institut f\"ur Astronomie, K\"onigstuhl 17, D-69117, Heidelberg, Germany}

\author[0000-0001-7440-8832]{Yoshinobu Fudamoto} 
\affiliation{Center for Frontier Science, Chiba University, 1-33 Yayoi-cho, Inage-ku, Chiba 263-8522, Japan}

\author[0000-0001-7201-5066]{Seiji Fujimoto}\altaffiliation{Hubble Fellow}
\affiliation{Department of Astronomy, The University of Texas at Austin, Austin, TX 78712, USA}

\author[0000-0001-6278-032X]{Lukas J. Furtak}
\affiliation{Department of Physics, Ben-Gurion University of the Negev, P.O. Box 653, Be'er-Sheva 84105, Israel}

\author[0000-0002-3254-9044]{Karl Glazebrook}\affiliation{Centre for Astrophysics and Supercomputing, Swinburne University of Technology, PO Box 218, Hawthorn, VIC 3122, Australia}

\author[0000-0002-9389-7413]{Kasper~E.~Heintz} % -- from ALT
\affiliation{Cosmic Dawn Center (DAWN), Denmark}
\affiliation{Niels Bohr Institute, University of Copenhagen, Jagtvej 128, 2200 Copenhagen N, Denmark}
\affiliation{Department of Astronomy, University of Geneva, Chemin Pegasi 51, 1290 Versoix, Switzerland}

\author[0000-0001-6755-1315]{Joel Leja}
\affiliation{Department of Astronomy \& Astrophysics, The Pennsylvania State University, University Park, PA 16802, USA}
\affiliation{Institute for Computational \& Data Sciences, The Pennsylvania State University, University Park, PA 16802, USA}
\affiliation{Institute for Gravitation and the Cosmos, The Pennsylvania State University, University Park, PA 16802, USA}

\author[0000-0001-9002-3502]{Danilo Marchesini} \affiliation{Department of Physics \& Astronomy, Tufts University, MA 02155, USA}

\author[0000-0003-2804-0648 ]{Themiya Nanayakkara}
\affiliation{Centre for Astrophysics and Supercomputing, Swinburne University of Technology, PO Box 218, Hawthorn, VIC 3122, Australia}

\author[0000-0002-7524-374X]{Erica J. Nelson}
\affiliation{Department for Astrophysical and Planetary Science, University of Colorado, Boulder, CO 80309, USA}

\author[0000-0001-5851-6649]{Pascal A. Oesch}
\affiliation{Department of Astronomy, University of Geneva, Chemin Pegasi 51, 1290 Versoix, Switzerland}
\affiliation{Cosmic Dawn Center (DAWN), Niels Bohr Institute, University of Copenhagen, Jagtvej 128, K{\o}benhavn N, DK-2200, Denmark}

\author[0000-0002-9651-5716]{Richard Pan}\affiliation{Department of Physics and Astronomy, Tufts University, 574 Boston Ave., Medford, MA 02155, USA}

\author[0000-0002-0108-4176]{Sedona H. Price}\affiliation{Department of Physics and Astronomy and PITT PACC, University of Pittsburgh, Pittsburgh, PA 15260, USA}

\author[0000-0003-4702-7561]{Irene Shivaei}
\affiliation{Centro de Astrobiolog\'ia (CAB), CSIC–INTA, Cra. de Ajalvir Km.~4, 28850- Torrej\'on de Ardoz, Madrid, Spain}

\author[0000-0001-8823-4845]{David Sobral}
\affiliation{Departamento de F\'isica, Faculdade de Ci\^encias, Universidade de Lisboa, Edif\'icio C8, Campo Grande, PT1749-016 Lisbon, Portugal}
\affiliation{BNP Paribas Corporate \& Institutional Banking, Torre Ocidente Rua Galileu Galilei, 1500-392 Lisbon, Portugal}

\author[0000-0002-1714-1905]{Katherine A. Suess}
\altaffiliation{NHFP Hubble Fellow}
\affiliation{Kavli Institute for Particle Astrophysics and Cosmology and Department of Physics, Stanford University, Stanford, CA 94305, USA}

\author[0000-0002-8282-9888]{Pieter van Dokkum}
\affiliation{Astronomy Department, Yale University, 52 Hillhouse Ave,
New Haven, CT 06511, USA}

\author[0000-0001-9269-5046]{Bingjie Wang (\begin{CJK*}{UTF8}{gbsn}王冰洁\ignorespacesafterend\end{CJK*})}
\affiliation{Department of Astronomy \& Astrophysics, The Pennsylvania State University, University Park, PA 16802, USA}
\affiliation{Institute for Computational \& Data Sciences, The Pennsylvania State University, University Park, PA 16802, USA}
\affiliation{Institute for Gravitation and the Cosmos, The Pennsylvania State University, University Park, PA 16802, USA}

\author[0000-0003-1614-196X]{John R. Weaver}
\affiliation{Department of Astronomy, University of Massachusetts, Amherst, MA 01003, USA}

\author[0000-0001-7160-3632]{Katherine E. Whitaker} \affiliation{Department of Astronomy, University of Massachusetts, Amherst, MA 01003, USA}\affiliation{Cosmic Dawn Center (DAWN), Denmark}

\author[0000-0002-0350-4488]{Adi Zitrin}
\affiliation{Department of Physics, Ben-Gurion University of the Negev, P.O. Box 653, Be'er-Sheva 84105, Israel}

\date{November 2024}

\begin{abstract}
We present a detailed exploration of the most optically-luminous Little Red Dot ($L_{\mathrm{H}\alpha} = 10^{44}$~erg/s, $L_V=10^{45}$~erg/s, $\mathrm{F444W} = 22$ AB) discovered by JWST to date. Located in the {\em UNCOVER} Abell~2744 field, source \source\ was observed by NIRSpec/PRISM with ultradeep spectroscopy reaching $\langle \text{SNR} \rangle \sim 100 \, \text{pix}^{-1}$, high-resolution $3-4$~\micron\ NIRCam/Grism spectroscopy from the {\em All the Little Things (ALT)} program, and multi-band NIRCam Medium Band imaging from {\em Mega Science}. The NIRCam spectra reveal high rest-frame equivalent width $W_{\mathrm{H}\alpha,0,\mathrm{broad}} > 800$~\angstrom, broad  H$\alpha$ emission ($\mathrm{FWHM} \sim 4500$~km/s), on top of narrow, complex H$\alpha$ absorption. NIRSpec data show exceptionally strong rest-frame ultraviolet (UV) to near-infrared (NIR) \ion{Fe}{2} emission ($W_{\mathrm{FeII-UV},0} \sim 340$~\AA), UV \ion{N}{4}]~$\lambda \lambda 1483, 1486$ and \ion{N}{3}]~$\lambda 1750$, and broad NIR \ion{O}{1}$\lambda8446$ emission. The line spectra unambiguously demonstrate a broad-line region (BLR) associated with a possible $M_{\mathrm{BH}} \sim 10^9 M_\odot$ supermassive black hole embedded in dense gas, which might explain the non-detection in ultradeep Chandra X-ray data ($\gtrsim10\times$ underluminous relative to broad $L_{\halpha}$). Strong UV Nitrogen lines suggest supersolar N/O ratios due to rapid star formation or intense radiation fields near the AGN. The observed continuum shows a clear Balmer break at rest-frame $3650$~\AA, which cannot be accounted for by power-law AGN alone. A stellar population model produces an excellent fit with a reddened Balmer break and would imply a massive ($M_* \approx 8 \times 10^{10}~M_{\odot}$), old $\sim 500$ Myr, compact stellar core, among the densest stellar systems known ($\rho \approx 3 \times 10^6~M_{\odot}$/pc$^2$ for $R_{e,\mathrm{optical}} = 70 \pm 10$ pc), and AGN emission lines with extreme intrinsic equivalent width $W_{\mathrm{H}\alpha,0} \gg 1000$~\angstrom. However, although high $M_*$ and $M_{\mathrm{BH}}$ are supported by evidence of a galaxy overdensity containing 40 galaxies at $z=4.41-4.51$, deep high-resolution spectroscopy is required to confirm stellar absorption and rule out that the Balmer break is instead caused by dense gas surrounding the AGN.
\end{abstract}

\keywords{Active galactic nuclei (16), High-redshift galaxies (734), Intermediate-mass black holes (816), Early universe (435)}

\section{Introduction}
\label{sec:intro}

One of the most exciting findings of \jw has been the abundance of high redshift ($z>4$) active galactic nuclei \citep[AGN; e.g.,][]{Harikane:2022,Furtak:2022,Ubler:2023,Larson:2023,Kocevski:2023}. Rest-UV selections from the ground have characterized the most massive and rare quasars \citep[e.g.,][]{Fan:2023}. \jw\ is enabling the discovery of new populations of accreting black holes in up to $10\%$ of the galaxy population at $4<z<7$ \citep{Maiolino:2023,Scholtz:2024} extending to $z = 10$ 
\citep{Larson:2023,Kokorev:2023,Goulding:2023}. More and more, these high-redshift AGN populations are straining models \citep{Habouzit:2022}, both to grow the black holes so early and simultaneously to explain the apparently high ratios of black hole to galaxy mass \citep[e.g.,][]{Furtak:2023nature,Kokorev:2023,Pacucci:2023,Dayal:2024}.
 
We focus here on one particular subset of the new broad-line selected AGN. The combination of red sensitivity and high angular resolution of \jw\ has revealed a population of dust-reddened broad-line AGN \citep[e.g.,][]{Kocevski:2023}, with high abundances that comprise a few percent of the galaxy population \citep[e.g.,][]{Matthee:2023,Kokorev:2024}. In photometry, they are distinguished by a combination of compact morphology and a characteristic ``v-shaped'' spectral energy distribution (SED), featuring a very red continuum in the rest-frame optical and faint blue light in the rest UV \citep[][]{Barro:2023,Labbe:2023uncover,Killi:2024,Kocevski:2024}. In spectroscopy, they are characterized by prominent broad Balmer lines suggesting dense gas orbiting a supermassive black hole \citep[][]{Matthee:2023,Greene:2024,Wang:2024brd,Wang:2024UB}.  Due to their red color and compact size, they have been dubbed ``little red dots'' \citep{Matthee:2023}. Photometric selections based on red colors in the rest-frame optical along with compact sizes have been deployed to select large photometric samples of little red dots \citep[e.g.,][]{Labbe:2023uncover,Kokorev:2024,Kocevski:2024,Akins:2024}. We find a remarkably high incidence of broad Balmer lines in photometrically selected samples \citep{Labbe:2023uncover}. Spectroscopic follow-up reveals that $>80\%$ of the reddest objects in \jw\ (F277W-F444W$>1.7$ AB mag) also harbor broad lines \citep{Greene:2024,Wang:2024UB}. 

There is not yet a complete picture of the phenomenology of the little red dots. Perhaps we are finding reddened versions of UV-selected quasars \citep{Glikman:2012,Banerji:2015}, or very compact dusty starbursts \citep{PerezGonzalez:2024}, or maybe we are probing a new astrophysical phenomenon. An important next step is to characterize the properties of the host galaxies of the little red dots. It is quite challenging to do this robustly, because the sources are so compact and because it is challenging to decompose the spectra into the contributions from galaxy versus AGN light \citep{Wang:2024brd,Ma:2024lens}. Fortunately, one of the most luminous and intriguing little red dots, \source, happens to be behind the A2744 field. Thanks to a range of \jwst\ programs, including UNCOVER \citep{Bezanson:2022}, Medium Bands, MegaScience \citep{Suess:2024}, and ALT \citep{Naidu24}, we have deep NIRCam imaging for all 20 broad and medium bands, along with F356W grism data at the position of \source.

In this paper, we highlight one of the brightest (F444W~$=22$~mag~AB) and most H$\alpha$-luminous little red dots from \citet{Labbe:2023uncover} found so far. \source\ is in the Abell 2744 field (R.A.=3.584758, Dec=$-$30.343630). We will show that this extreme object may have something to teach us about the nature of little red dots more generally. We present the data in \S \ref{sec:data}, the compact galaxy morphology in \S \ref{sec:morphology}, the fits of the spectral energy distribution in \S \ref{sec:sed}, the fits of the emission lines in \S \ref{sec:lines}, the quantification of the environment in \S \ref{sec:environ}, and the discussion in \S \ref{sec:discussion}.

Throughout this work, we assume a flat $\Lambda$CDM cosmology \citep[e.g.][]{planck20} with $\Omega_{\mathrm{m},0}=0.3$, $\Omega_{\mathrm{\Lambda},0}=0.7$ and H$_0=70$ km s$^{-1}$ Mpc$^{-1}$, and a \citet{Chabrier:2003} initial mass function (IMF) between $0.1-100$ $M_{\odot}$. All magnitudes are expressed in the AB system \citep{oke1983}.

\section{Data}
\label{sec:data}

\subsection{UNCOVER Photometry, Spectroscopy, and Lensing Calculations}

For spectroscopic targeting, we relied on seven-band NIRCam photometry and NIRSpec/PRISM spectroscopy taken as part of the Cycle 1 Treasury program Ultradeep NIRSpec and NIRCam Observations before the Epoch of Reionization \citep[UNCOVER;][]{Bezanson:2022}. The UNCOVER data have already yielded  exciting new discoveries such as a triply imaged $z=7.04$ little red dot \citep{Furtak:2023,Furtak:2023nature}, $z>12$ galaxies \citep{Wang:2023highz}, faint galaxies in the epoch of reionization \citep{Atek:2023eor}, and their mass-metallicity relation \citep{Chemerynska:2024}, and overdensities at $z>7$ \citep{Fujimoto:2023overdense}. We have identified some ALMA bright and \emph{HST}-dark objects \citep{Fujimoto:2023,Price:2023}. Follow-up of objects identified in \citet{Labbe:2023uncover} resulted in a dramatic yield of broad lines in little red dots \citep[][]{Greene:2024,Furtak:2023nature}, one at $z=8.5$ broad-line AGN \citep{Kokorev:2023}, as well as deep spectra of three brown dwarfs \citep{Langeroodi:2023,Burgasser:2023}.

\begin{figure*}
\hspace{10mm}
\centering
\includegraphics[width=0.79\textwidth]{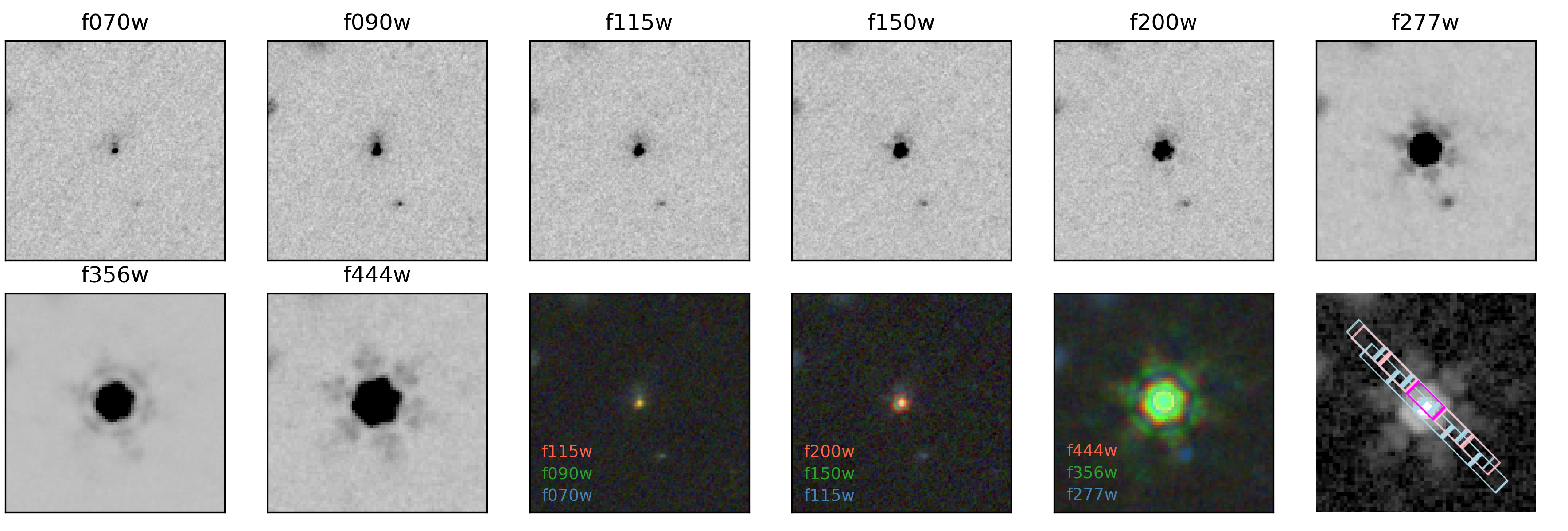}
\centering
\includegraphics[width=0.99\textwidth]{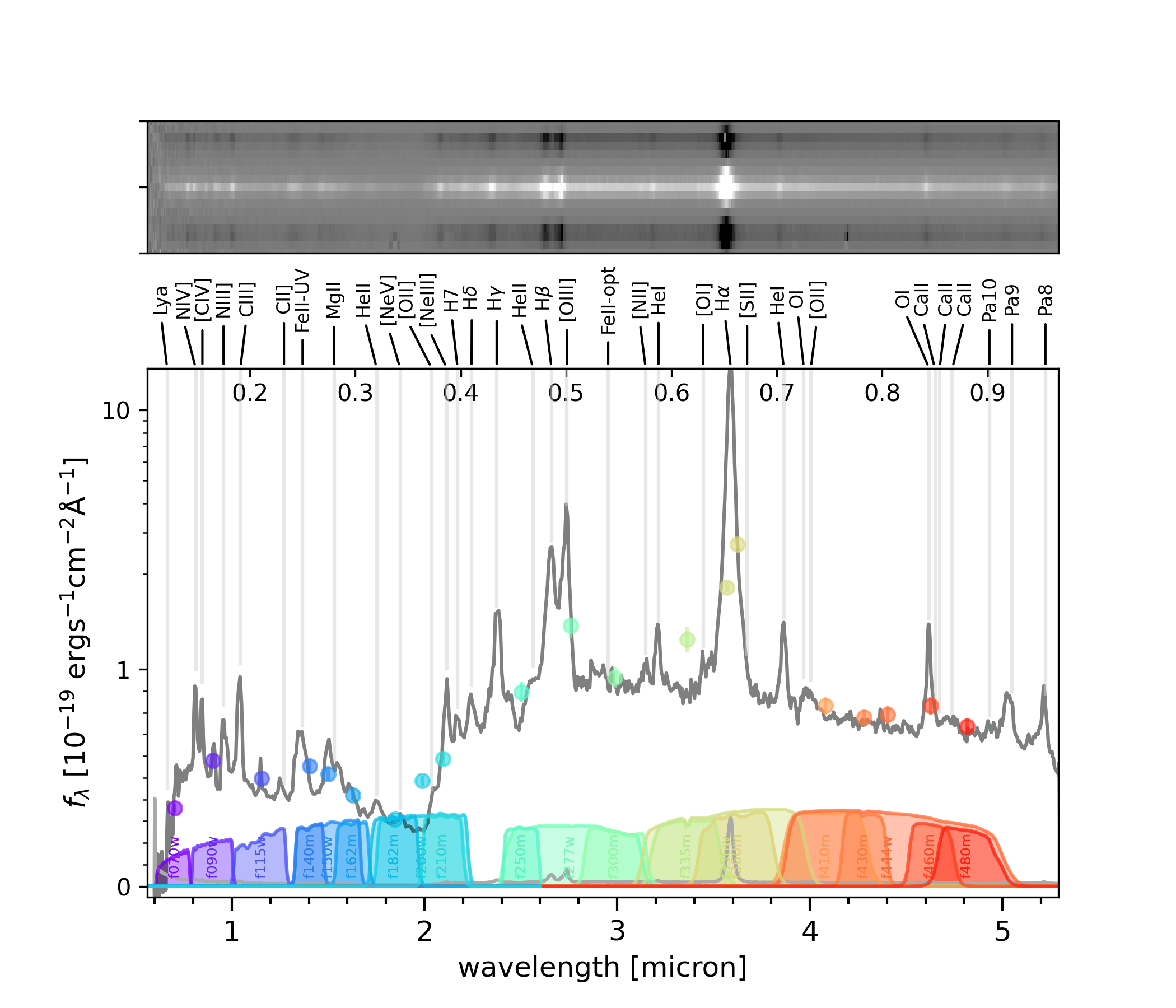}
\caption{Top: broadband NIRCam images of \source\ in the short wavelength channel filters F070W, F090W, F115W, F150W, F200W at 20 mas pixel size and long wavelength channel filters F277W, F356W, F444W at 40 mas pixel size. The images are $3\arcsec$ on a side. North is up, East is left. Also shown are color composite images and an overlay of the micro-shutter array locations of the NIRSpec PRISM observations. Middle: the 2D NIRSpec/PRISM spectrum used for analysis. Bottom: observed NIRSpec/PRISM spectrum (black) and error array (grey). Positions of key spectroscopic features are indicated. Top axis is rest-frame wavelength in micron.}
\label{fig:thedata}
\end{figure*}

UNCOVER imaging was performed over 45 arcmin$^2$ in the Hubble Frontier Field \citep{Lotz:2017} cluster Abell 2744, chosen for the large high-magnification area. Original photometry was performed in seven NIRCam filters (F115W, F150W, F200W, F277W, F356W, F410M, F444W) with 4-6 hour integrations leading to depths of $\sim 30$~mag (31.5 mag accounting for magnification). \citet{Weaver:2023} presents the photometric catalogs, also including available \emph{HST} data, and the lens model is presented by \citet{Furtak:2022b_SLmodel}. \source\ was selected from the DR1 images \citep{Labbe:2023uncover}. Photometric redshifts are presented by \citet{Wang:2023}. The bluest NIRCam filters (F070W, F090W) were observed by both the Medium Bands, Mega Science \citep{Suess:2024} and ALT \citep{Naidu24} surveys, for a total integration time of $\sim$6-8~hr and a depth of $\sim 30$~mag. Medium Bands, Mega Science also observed all twelve of NIRCam's medium-band filters for $\sim$2.3~hr each, leading to depths of $\sim \, 27.4 - 29$~mag \citep[see Table 1 of][]{Suess:2024}. Updated photometric redshift catalogs following the same methods as \citet{Weaver:2023} and \citet{Wang:2023} but including new Cycle 2 imaging data are presented in \citet{Suess:2024} and publicly available on the UNCOVER website\footnote{\url{https://jwst-uncover.github.io/}}.

\source\ was observed as part of the NIRSpec/PRISM Micro-Shutter Array (MSA) component of UNCOVER \citep{{Price:2024}}. Three of the seven MSA configurations included \source. The MSA observations employed a 2-POINT-WITH-NIRCam-SIZE2 dither pattern, and 3 shutter slitlet nod pattern at an aperture angle of $\sim 44$ degrees. The source was observed in configurations MSA 4, 5, 6, and 7 for a total integration time of 16.3 hours. 
The PRISM data for each MSA are reduced with \texttt{msaexp} \citep[v0.6.10,][]{Brammer:22}. We begin with the level 2 products provided by MAST\footnote{Available from: \url{http://dx.doi.org/10.17909/8k5c-xr27}} and then correct for 1/f noise, mask snowballs, and remove the bias on a per-frame basis. WCS, slit-identification and flat-fielding are applied through the \jwst\ pipeline, along with photometric correction. Two-dimensional stacked spectra are made by drizzling each MSA observation onto a common grid, aligning, and then stacking them.

The spectra are extracted optimally \citep[e.g.,][]{Horne:1986}, and flux calibration is performed by convolving the spectrum with the full suite of 20 broad and medium passbands from the combination of UNCOVER \citep{Weaver:2023} and Medium Bands, MegaScience \citep{Suess:2024} and solving for a fifth-order multiplicative polynomial to reproduce the photometry in $0.\arcsec32$ diameter apertures of MegaScience catalog v5.2.0 \citep{Price:2024}.
 
The resulting four spectra are inspected for calibration issues and artifacts. The source was well-centered in MSA 4, but less so in MSA \{5, 6, 7\}, leading to larger wavelength-dependent photometric corrections. In addition, spectra 5, 6, 7 show spurious discontinuities or undulations of the continuum at the $\sim10\%$ level. These artifacts would not impact the main conclusions of this paper but can affect individual lines and features. Finally, there are small wavelength shifts between the spectra ($\sim$50\AA\ at $3\mu$m; 0.2\%). These $\sim 1$~pixel offsets between spectra are similar to those noted in \citet{Degraaff:2024RUBIES} to arise for compact sources depending on location within the slit \citep[see also][]{Ferruit:2022,DEugenio:2024}.
To minimize the impact of calibration errors and other systematics on the analysis, we therefore focus on the 4.4 hour spectrum observed on MSA 4 for the remainder of this work. This spectrum has highest S/N of the four, S/N$=50-200$ per 1D-extracted spectral pixel, sufficient for our purposes, is well-centered in the shutter, well-calibrated, and does not show any obvious artifacts. 

Throughout, when we model the NIRSpec/PRISM data, we must forward-model the spectral resolution. In brief, synthetic observations are produced by convolving the model spectra with the NIRSpec/PRISM resolution using published values from JDOX \citep{Jakobsen:2022} (appropriate for uniformly illuminated slits) increased by a factor 1.3 because our target is a point source. Point sources have considerably better resolution, by up to a factor of $\approx 2$ \citep{DeGraaff:2024LSF}, but some gains are lost by the large pixel size and processing in the reduction process and coadding of data. The model is then resampled onto the observed wavelength grid with flux conservation.

The magnification of \textcolor{blue}{$\mu = 1.7 \pm 0.2$} for \source\ is calculated based on its position and spectroscopic redshift using the \texttt{v2.0} UNCOVER strong-lensing model of A2744 from \citet{Furtak:2022b_SLmodel}, which has been updated with UNCOVER NIRSpec and ALT NIRCam grism spectroscopic redshifts \citep{Price:2024}. This version of the lens model is also publicly available on the UNCOVER website at \url{https://jwst-uncover.github.io/DR4.html#LensingMaps}.

\subsection{ALT Survey}
 
The JWST Cycle 2 program All the Little Things (ALT; PID 3516 PIs Matthee \& Naidu, \citealt{Naidu24}) observed a mosaic around the Abell 2744 field with NIRCam imaging (F070W, F090W, F356W) and, in particular, grism spectroscopy (GrismR in the F356W filter). The mosaic consists of two sets of four visits, where the two sets are rotated by 5 degrees with respect to each other. These two angles facilitate the removal of contamination and source confusion that is typical for grism spectroscopy. The total on-source integration times range from 6.9-27.2 hours (typically 13.6 hours). The total field of view overlaps with a large part of the UNCOVER mosaic and yields spectral coverage at $\lambda=3.05-3.95 \mu$m with a nominal resolution of R$\sim1600$. Importantly for this paper, this resolution is sufficient to disentangle H$\alpha$ from [\ion{N}{2}] emission and measure line width. These data yield spectroscopic redshifts for $\sim1600$ galaxies at $z=0.3-8.5$ (see \citealt{Naidu24} for details on the survey design and analysis).

\source\ is covered by the ALT data with a total on source integration time of 6.9 hours. The grism spectrum in F356W covers rest-frame wavelengths 0.57-0.72 micron at $z=4.47$. The reduction of the grism data and extraction of the two-dimensional spectrum follows the method outlined in \cite{Kashino:2023}. The continuum level is removed from the spectrum as follows: at a certain wavelength, the continuum level is estimated by calculating the running median over a boxcar filter with a hole around the central wavelength. Similar to \cite{Matthee:2023}, we optimize the method to prevent over-subtraction of the broad wings of the lines. We increase the width of the filter and the hole (to 160 and 9 nm, respectively) and explicitly mask $\pm 5000$ km s$^{-1}$ around the H$\alpha$ line. The grism spectrum of \source\ is totally uncontaminated in the data from one of the two rotation angles. This allows us to verify the H$\alpha$ line profile was not modified by the continuum removal procedure. The one-dimensional spectrum is optimally extracted \citep[e.g.,][]{Horne:1986} using a Gaussian fit to the spatial extent of the spectrum collapsed over the H$\alpha$ wavelength. The typical rms noise level of the spectrum is $6\times10^{-20}$ erg s$^{-1}$ cm$^{-2}$ \AA$^{-1}$.

\section{Morphology}
\label{sec:morphology}

Initial visual inspection of the images in Figure \ref{fig:thedata} shows two distinct components: a main compact source and blue diffuse emission extending to the North.  To quantify the morphology of the source we perform \sersic\ profile fitting to the images using the package \texttt{pysersic}~\citep{Pasha2023}. We focus on three bands: F070W representing the rest-far-UV continuum (although note that the full band extends blueward of Ly$\alpha$), F200W near the 4000\AA\ break, and F300M for the line-free optical continuum. We use cutouts from the mosaics, weight maps and Point Spread Functions presented in \citet{Suess:2024} using methods presented in \citet{Weaver:2023}. We use the native pixel scale of 0.02 arcsec/pixel for the F070W and F200W images, and the 0.04 arcsec/ pixel scale for F300M. In order to not bias the measurements of the main bright compact source we perform a two component fit, where both are \sersic\ profiles, with the second component representing the nearby low-surface brightness emission seen in the SW filters. We employ a double \sersic\ model in order to determine whether the core is resolved or unresolved. For each band we sample the posterior distributions for the parameters using a No U-Turn sampler \citep{Hoffman2014,Phan2019}, with 2 chains for 1000 warm-up steps and 2000 sampling steps each. To ensure robust sampling we re-parameterize variables using a multivariate normal distribution fit to the posterior, following \citet{Hoffman2019}. We test to make sure the results converge by ensuring the $\hat{r}$ metric and effective sample size are $<1.02$ and $>400$ respectively for the parameters of interest \citep{Vehtari2021}.

\begin{figure*}
\hspace{-1mm}
\includegraphics[width=0.5\textwidth]{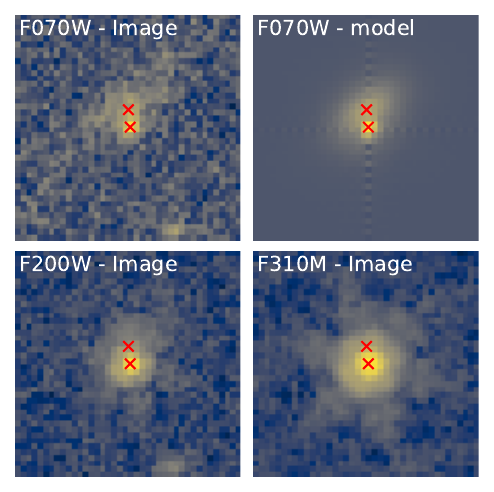}
\includegraphics[width=0.5\textwidth]{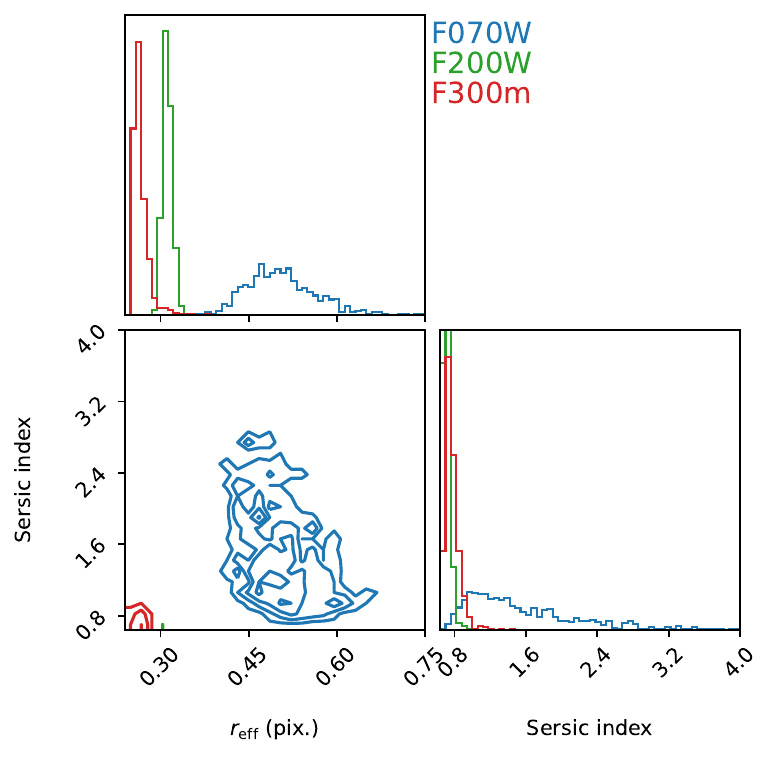}
\caption{Results of fitting two-component models to the F070W, F200W, and F300M images of \source\ using \texttt{pysersic}, orientation has North up as above in Figure \ref{fig:thedata}. We show zoomed in cutouts of the three bands modeling alongside the joint posterior distribution of the Sersic index and effective radius for the bright compact component. We find that \source\ is marginally resolved in F070W and F200W, but point-like at longer wavelengths. We adopt the F200W size, as it is most likely to be dominated by evolved stars \S \ref{sec:contfit}.}
\label{fig:imfit}
\end{figure*}

From these two component fits, the light is dominated at all wavelengths by the central compact component (Figure \ref{fig:imfit}), which is centered in the MSA slit (Figure \ref{fig:thedata}). This compact component appears to be marginally resolved in the F070W and F200W filters, but is completely point-like in F444W. We did attempt a two-component fit to the F444W image, but the second component was completely unconstrained \citep[see also][]{Chen:2024}. As we will argue in \S \ref{sec:contfit}, the spectrum is dominated by Balmer break feature in the F200W band, and therefore we derive the structural measurements from this band. We measure a size at F200W of $r_e = 0.010 \pm 0.001$\arcsec and a similar size for F300M. F300M is useful as it avoids strong emission lines and provides a clean measurement of the rest-frame optical continuum 0.55\micron. However, from extensive testing of size limits with stars in \citet{Labbe:2023uncover}, we find that we cannot reliably measure sizes below a third of a pixel. Given 0.04\arcsec\ pixels, we find a magnification-corrected size of $r_e \lesssim 70 \pm 10$~pc for the primary compact component to the South.

The second extended component towards the North is both fainter and bluer than the primary component. At F070W, the component is quite diffuse, but in the F090W and F115W images the second component is clearly visible (Figure \ref{fig:thedata}). Fitting this component at F070W, we find an extended blue component with $n \approx 1$ and $r_e \approx 0.16\arcsec$; the Northern component is asymmetric and only falls partially within the MSA. The angular size corresponds to $r_e \approx 700$~pc (accounting for lensing), if it is present at the same redshift as the main source. The Northern source is barely detected in F200W, and becomes undetectable redward. Unfortunately, the grism data are not sensitive enough in the blue to detect the extended component. The resolution is also growing worse redward of F200W and the very prominent point-source dominates the total light. 

The presence of extended blue emission on hundreds of parsec scales seems to argue for star-light dominating the UV continuum. With a surface brightness of 23.5 mag~arcsec$^{-2}$, the level is much brighter than would be expected from scattered light, if scaling from lower-redshift analogs at similar intrinsic luminosity \citep{Zakamska:2006}. On the other hand, it is possible that the extended emission comes from lines like Ly$\alpha$, photoionized by the same source that ionizes the H$\alpha$. Taking the integrated flux from the extended component, we find a luminosity of $L \approx 6 \times 10^{43}$~erg/s, which is very similar to the H$\alpha$ luminosity (see \S \ref{sec:altha}). Thus, assuming Case B recombination, perhaps $\sim 10\%$ of the Ly$\alpha$ has scattered into this extended component. 
Future integral-field data could determine the nature of the Northern component, and particularly whether \ion{Mg}{2} and other resonant lines also contribute.

\section{Emission Lines}
\label{sec:lines}

\subsection{Far-UV Emission Line Fits}

\begin{figure}
\hspace{-3mm}
\includegraphics[width=0.45\textwidth]{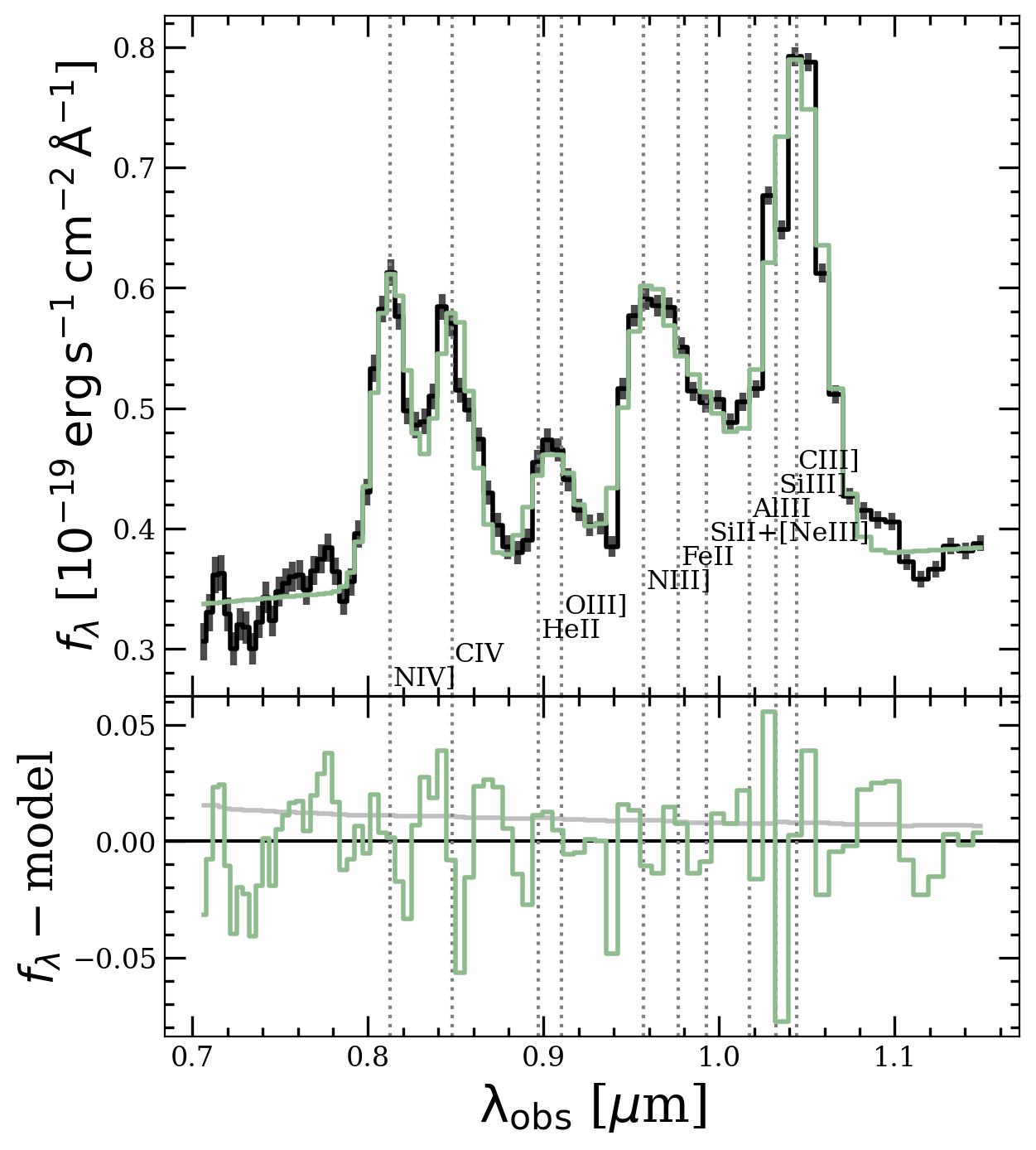}
\caption{Rest-frame far-UV fit to \source. We see prominent 
\ion{N}{4}]$\lambda 1483$+\ion{N}{4}]$\lambda 1487$~\AA, CIV $\lambda\lambda$1548,1551, HeII $\lambda$1640, OIII] $\lambda$1661,1666, \ion{N}{3}] $\lambda 1750$, and [CIII]$\lambda$1907+CIII]$\lambda$1909. 
Note that SiIII] and CIII] are blended, as are HeII and OIII].
Also needed to explain the spectrum is the blend of \ion{Al}{3}$\lambda 1860$, [\ion{Ne}{3}]+\ion{Si}{2}$\lambda 1815$ and \ion{Fe}{2}$\lambda 1786$, which are lines also found in the \cite{Shen:2019} z$\gtrsim$5.7 quasar composite. This broad-line fit is marginally preferred over fits constrained to narrower FWHM. The CIV profile drives this preference, which may be because of the blending in most other lines or because CIV alone is broad. We show the broad fit because of this fit preference and the evidence from Fe~II of broad-line emission in the rest-frame UV.}
\label{fig:uvfit}
\end{figure}

The NIRSpec/PRISM spectra cover from $0.7-5 \, \micron$ observed, which probes from Ly$\alpha$ to beyond H$\alpha$ at $z=4.47$. Note that the PRISM has high sensitivity but low dispersion, with $R=50-400$ across the wavelength coverage for a point source \citep{DeGraaff:2024LSF}. In this section we present new fits to the rest-UV lines \citep{Treiber:2024}. 

We fit the region from 1280-2110\AA\ rest-frame with the combination of a power-law ($f_{\lambda} \propto \lambda^{\beta}$) and emission lines: \ion{N}{4}]~$\lambda 1488$, \ion{C}{4}~$\lambda 1549$, \ion{He}{2}~$\lambda 1640$, \ion{O}{3}]~$\lambda 1666$, \ion{N}{3}]~$\lambda 1750$, \ion{Si}{3}]~$\lambda 1888$, and \ion{C}{3}]~$\lambda 1909$ (Figure \ref{fig:uvfit}). We also include the blend of AlIII$\lambda 1860$, [\ion{Ne}{3}]+\ion{Si}{2}$\lambda 1815$ and \ion{Fe}{2}$\lambda 1786$ next to \ion{N}{3}]. See Table \ref{tab:lines} for rest wavelengths and fluxes. In the UV, where $R\sim 50-100$, we have weak constraints on the line widths. Nominally the fits prefer line-widths greater than 2000~km/s. 

With the instrument-model thus constructed, we then create an oversampled but variable wavelength grid in order to handle the variable resolution. We convolve the model Gaussians with the instrument model as described in \S2.1. Our fit is presented in Figure \ref{fig:uvfit}. All strong UV lines are well-detected, including the blend of AlIII, [\ion{Ne}{3}]+\ion{Si}{2} and \ion{Fe}{2} next to \ion{N}{3}] \citep{Shen:2019}. Without these lines, there was a clear excess in the fit. The line equivalent widths are among the highest observed in UV emission lines in the UNCOVER/PRISM UV-line emitter sample at $z>3$ \citep{Treiber:2024}. The fluxes are reported in Table \ref{tab:uv}.

\cite{Treiber:2024} also investigate the location of \source\ in UV diagnostic diagrams, finding that the line ratios are in the region of these diagrams occupied both by broad UV emission lines in quasars and by star-forming galaxies. Based on the detection of broad Fe~II presented in \S \ref{sec:nearuv}, we favor the hypothesis that the UV emission lines also emerge from the broad-line region. First, we discuss the strong Nitrogen lines also detected in the rest-frame UV.

\subsubsection{Strong Nitrogen emitter}

We detect \ion{N}{4}]$\lambda \lambda 1483, 1486$ with an equivalent width of $25$~\AA\ and \ion{N}{3}]$\lambda 1750$ with an equivalent width of 30~\AA\ in the \source\ spectrum. \ion{N}{4}] is seen in narrow-line AGN selected in the UV and optical \citep[see the stacks in][]{Hainline:2011,Alexandroff:2018}, while \citet{LeFevre:2019} and \citet{Sobral:2018} present AGN stacks with Ly$\alpha$ and CIII] emission that show \ion{N}{3}]. In contrast, these two lines are rarely seen in broad-line AGN \citep[$<1\%$ in SDSS][]{Bentz:2004,Glikman:2007}. In the rare SDSS quasars with strong \ion{N}{4}] and \ion{N}{3}], the elevated equivalent widths seem to suggest super-solar abundance ratios of N/O \citep[e.g.,][]{Jiang:2008}, although \citet{Batra:2014} present an alternate interpretation of high overall metallicity. Interestingly, the incidence of luminous radio activity is much higher among the N-emitting quasars \citep[e.g.,][]{Humphrey:2008,Jiang:2008}, perhaps suggesting either that shock excitation is part of the strong Nitrogen emission, or that there is an evolutionary link between radio activity and Nitrogen emission; a search for radio emission from \source\ could be quite interesting, although in a stack of hundreds of photometrically selected little red dots, no radio continuum is detected \citep[][]{Akins:2024}. Finally, we note that \ion{N}{4}] is also detected in the prototypical narrow-line AGN NGC~1068 \citep{Kraemer:2000}. In that case, photoionization modeling does a very good job reproducing nearly all the UV/optical emission lines, but models with supersolar N/O tuned to reproduce \ion{N}{5} tend to overproduce \ion{N}{4}]. In other words, it can be challenging to model the Nitrogen lines in even nearby AGN.

Before \jw, very few star-forming galaxies had been seen to have the \ion{N}{4}] line, either local dwarf starbursts \citep{Berg:2022} or high-redshift lensed systems \citep{Fosbury:2003,Raiter:2010}. Now, more such galaxies are emerging at high redshift \citep[e.g.,][]{Bunker:2023,Topping:2023,Isobe:2023,Barchiesi:2023,Marques-Chaves:2024}. It is not yet clear in individual cases whether an AGN is needed to supply the hard ionizing radiation needed to excite the \ion{N}{4}]. Other explanations relate more directly to the massive stars associated with the observed starbursts. For instance, \citet{Cameron:2023} explore scenarios in which stellar collisions or tidal disruption events might pollute the ISM with extra Nitrogen produced in the CNO cycle \citep[see also][]{Senchyna:2024}. Higher spectral resolution data are needed to decouple broad and narrow lines, as well as to measure line ratios of the N lines themselves. These data will allow for detailed photoionization modeling and shed new light on the origin of the Nitrogen lines in \source.

\subsection{Near-UV Emission line fits}
\label{sec:nearuv}
The rest-frame near UV $2000-3200$\angstrom shows a complex aggregate of emission lines around \MgII$\lambda2798$ which we identify as broad \feii line emission. \feii emission is common in broad-line AGN \citep{Osterbrock:1977} and can be a prominent contributor to the UV, optical, and NIR emission line spectrum in AGN \citep{Wills:1985,BorosonGreen:1992,rodriguezetal:2002}. Iron emission is thought to originate in the outer parts of the broad line region (BLR) \citep{Kovacevic:2010}, with a shape and equivalent width reflecting the physical conditions of BLR clouds \citep{Baldwin:2004}, and the equivalent width is suggested to correlate with the relative black hole accretion rate \citep{BorosonGreen:1992}.   

\begin{figure}
\vspace{-10mm}
\includegraphics[width=0.48\textwidth]{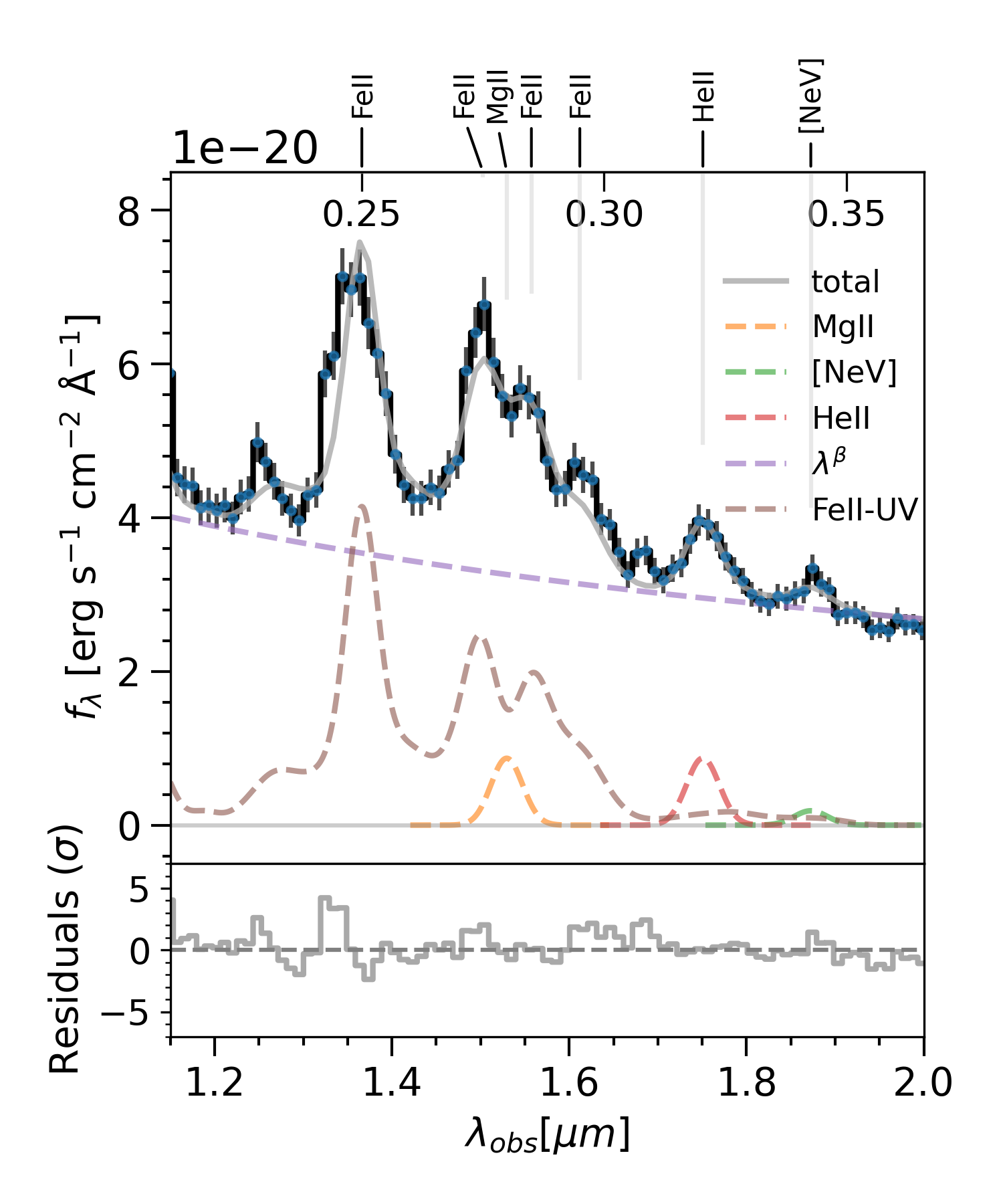}
\caption{Model fits to the rest-frame near-UV. Shown are the observations (black solid line), the local AGN powerlaw continuum (purple dashed), strong broad-line \feii pseudo-continuum emission (brown dashed) and the broad \MgII$\lambda2798$ (yellow), broad \HeII$\lambda3203$ (red), and narrow \nev$\lambda3426$ (green) emission lines. }
\label{fig:agn_lines_feii_uv}
\end{figure}

The challenge in modeling the \feii emission complex is that the \ion{Fe$^{+}$}{0}ion emits through thousands of emission line multiplets that blend together to form a pseudo continuum. Typically, the \feii\ is modeled using empirical templates from low-redshift, well-studied sources \citep[e.g.,][]{VestergaardWilkes:2001}.
Here we use empirical templates based on the Seyfert 1 galaxy I Zwicky 1, divided into sections to allow more flexibility in reproducing the shape of the observed emission: below 2000\angstrom the template of \citet{VestergaardWilkes:2001} is used, at $2200-2400$\angstrom, $2400-2600$\angstrom, and $2650-3090$\angstrom the template of \citet{Salviander:2006}, which models the flux underneath \MgII, and from $3090-3600$\angstrom the template of \citet{Tsuzuki:2006}. The  \feii line templates are convolved with a Gaussian to match the width of the broad line \halpha (see \S \ref{sec:altha}).

To model the near-UV from $2000-3600$\angstrom, we fit a combination of a power law, the \feii templates, and the \MgII$\lambda2798$, \HeII$\lambda3203$, and \nev$\lambda3426$ emission lines, assuming Gaussian profiles, with NIRSpec PRISM observations modeled as above. At 1.5\micron\ the effective resolution of NIRSpec is only $R\sim40$, so individual lines are unresolved. As shown in Figure {\ref{fig:agn_lines_feii_uv}} all lines are clearly detected. The permitted \FeII, \MgII, and \HeII are common BLR lines, and the narrow line [\ion{Ne}{5}]~$\lambda3426$ is a strong indicator of AGN accretion due to its high ionization potential \citep[e.g.,][]{Abel:2008,Chisholm:2024}. We estimate the equivalent width of \FeII, by integrating the flux between $2200-3050$\angstrom, relative to the powerlaw continuum. The measurements are presented in Table \ref{tab:feii}. The \feii-UV emission is exceptionally strong, with rest-frame equivalent width $W_\mathrm{FeII-UV}\sim340$\angstrom. For comparison, the largest EW measured in a sample of 4178 Seyfert 1 galaxies and QSOs in SDSS at $z<0.8$ is $W_\mathrm{FeII-UV}\sim400$\angstrom \citep{dongetal:2011}, with only 0.07\% of AGN exceeding 300\angstrom. 

We also present fits to the rest-frame optical and NIR lines from the PRISM data in Appendix \ref{appendix:lines}. Some noteworthy findings include broad \ion{He}{2}$\lambda4686$, broad Balmer, broad \ion{O}{1}$\lambda8646$, tentative broad \ion{Ca}{2} triplet lines, and the broad \ion{Fe}{2}$\lambda9200$ bump in emission, indicating widespread high equivalent width BLR line emission throughout the UV-optical-NIR spectrum.

\subsection{ALT Optical Emission Line Fits}
\label{sec:altha}

\begin{figure}
\hspace{1mm}
\includegraphics[width=0.45\textwidth]{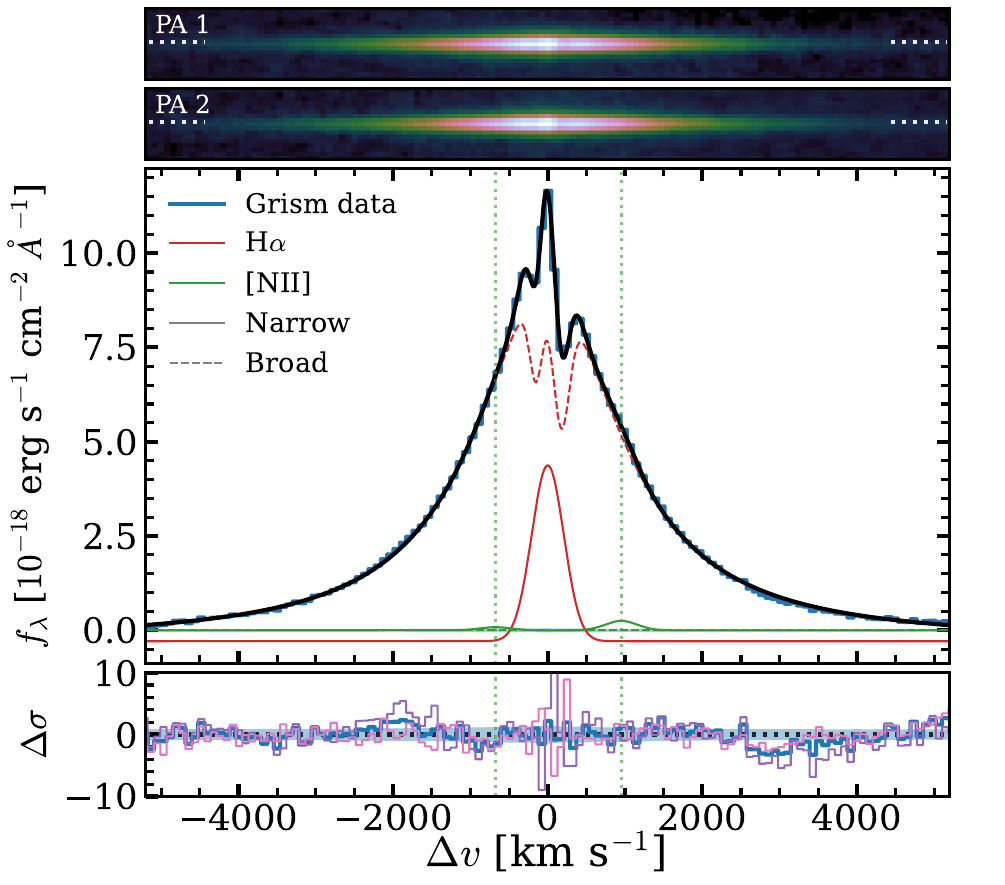}
\caption{The total H$\alpha$+\nii\ emission line profile as seen by the JWST/F356W GrismR with best-fit line emission and absorption complex. The profile is centered on the central velocity of the broad H$\alpha$ component. Green dashed lines mark the location of the \nii\ doublet. The top panels shows the two-dimensional grism spectrum from each roll angle with a $\gamma=0.5$ power-law normalisation to highlight the dynamic range in the data. The bottom panel shows the residuals of the best fit, normalised by the noise level. In pink and purple we show the residuals from individual roll angles, while in blue we show the residuals from the average of both roll angles.}
\label{fig:hafit}  
\end{figure}  

We fit the one-dimensional H$\alpha$ profile measured in the NIRCam grism data with the same methodology as used in \cite{Matthee:2023}, which means that we simultaneously fit for a narrow component for H$\alpha$ and \nii\ (fixing the \nii\ $\lambda6549,6585$ doublet ratio to 1:2.94 as set by quantum mechanics), a broad H$\alpha$ wing, and H$\alpha$ absorption components. We include the data from $\lambda_0=6420 - 6680$ {\AA} in the fit and use a least squares minimizer implemented in the {\tt lmfit} package.

The H$\alpha$+\nii\ line profile is quite complex (Figure \ref{fig:hafit}). To describe the wings of the H$\alpha$ line, we use a Lorentzian function with a FWHM~$=4540 \pm 50$~km/s which is strongly preferred over a Gaussian profile. Then, in order to model the narrow H$\alpha$ and \nii components, we include a Gaussian component at the position of H$\alpha$ and the \nii doublet with best-fit FWHM$=460 \pm 30$~km/s. The widths of the narrow components are fixed to each other. There is no significant velocity offset between the narrow and broad component. The best-fit \nii/H$\alpha$ ratio of the narrow component is $0.02\pm0.02$, i.e. no significant \nii emission is detected. The very low ratio of \nii/H$\alpha$ is not dependent on the absorber velocities or strengths (see below), as they are far from the \nii\ in velocity. Thus, we find a genuine low ratio of \nii/H$\alpha$, which typically implies low gas-phase metallicities \citep[e.g.,][]{grovesetal2006}.
 
After trying a number of configurations, we find that the best overall fit is achieved with two additional absorption components at $-143$ and $+172$ km/s from the systemic velocity set by the broad H$\alpha$ component. The absorption lines are barely spectroscopically resolved (FWHMs of 240-330 km s$^{-1}$) and their rest-frame equivalent widths are 1.2-2.0 and 2.3-4.4 {\AA}. The fit is significantly preferred compared to a fit with no absorbers ($\Delta \chi^2 = 11$) and with a single absorber ($\Delta \chi^2 = 4.4$). However, there is significant degeneracy between the width and strength of the narrow H$\alpha$ and the width and equivalent width of the absorbers. If we had high-resolution coverage of \oiii we would be able to break this degeneracy. For now, we simply argue that absorption features are needed, and defer quantitative analysis of these lines until multiple transitions can be observed at high resolution.

The best-fit profile has a reduced $\chi^2=2.4$. The residuals, shown in the bottom panel of Fig. $\ref{fig:hafit}$, are relatively free of correlated structure, in particular after the addition of the absorbing components and using a Lorentzian broad component. Some large-scale skewness can be identified over the $-$2000 to 2000 km s$^{-1}$ range, which could either be indicative of a third line-profile component (such as an outflow), or could be a result of a degeneracy with the centroiding of the broad component with the complex absorption near line-center. A higher resolution spectrum is required for even more complex modeling. We will revisit the implications of the Balmer absorption lines in \S \ref{sec:discussion}.

Finally, we note that data in separate roll angles yield fully consistent profiles. The complexity therefore is truly from the line-profile and not due to the spatial-spectral degeneracy of grism data. Multiple narrow or broad components without absorption are far less successful at yielding a good fit to the data, in particular due the strong symmetry of the broad wings of the line. Thus, despite significant degeneracy in fitting the narrow components on top of the absorption, the broad line width and flux are quite robust, as well as reassuringly consistent with the original PRISM fit \citep{Greene:2024}.

\section{Spectral energy Distribution Fits}\label{sec:sed}
\begin{figure*}
\hspace{10mm}
\centering 
\includegraphics[width=0.85\textwidth]{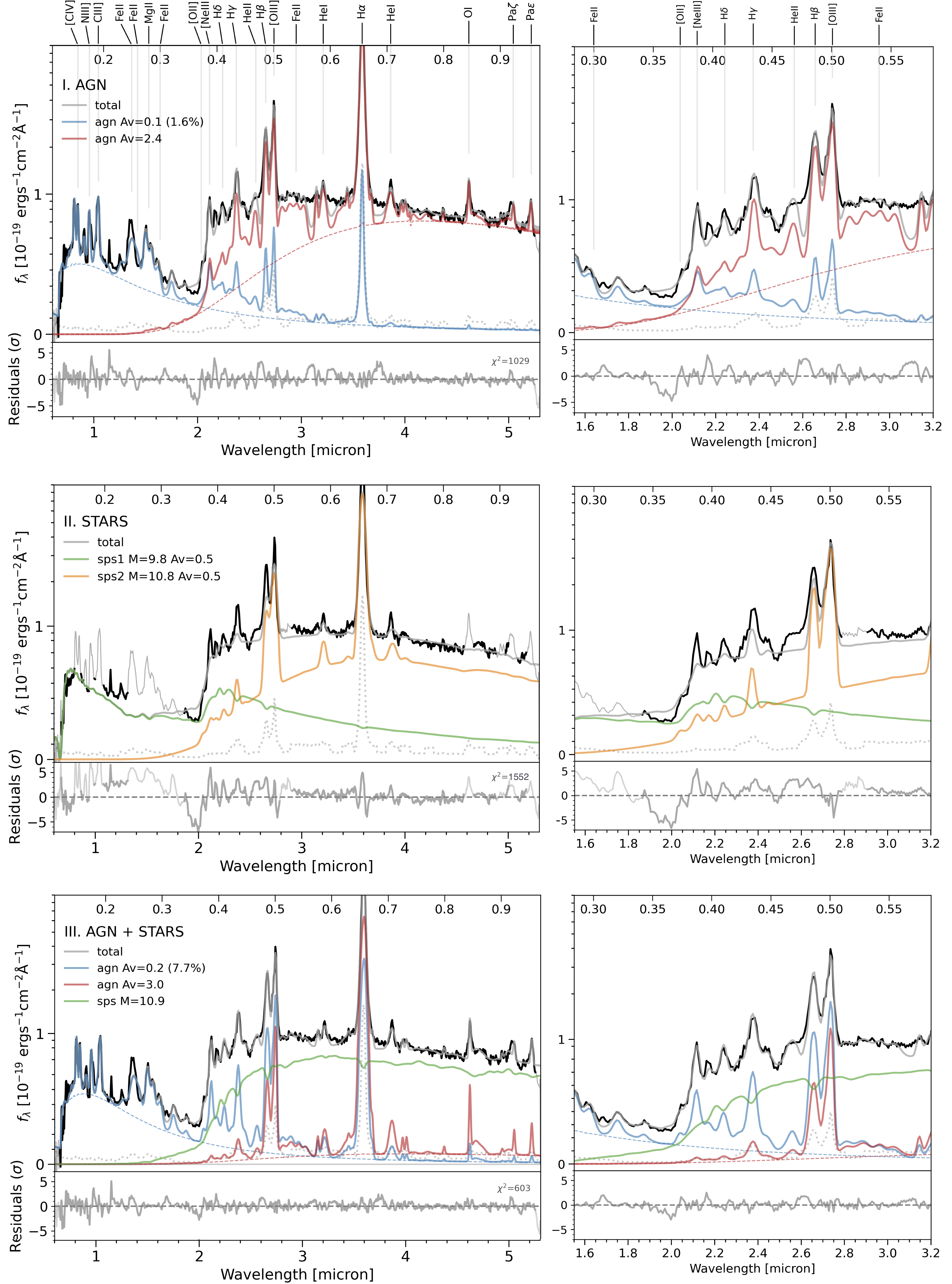}
\caption{Model fitting results using only AGN components (model I, top panels), only stellar components (model II, middle), and a mix of AGN and stars (model III, bottom). NIRSpec/PRISM observations (black), uncertainties (dotted), model fits (gray), and normalized residuals are shown. Model I consists of a dust-reddened AGN component (red) and a blue component (here parameterized as a fraction of dust-free scattered light; blue), which dominates at $<2 \micron$ despite being only a small percentage of the red luminosity. The model can produce an apparent sharp break at $\sim3500$\AA\ due to a blend of strong broad emission lines and broad \ion{Fe}{2} pseudo-continuum at $4000-5500\AA$, but leaves significant residuals throughout the rest-frame optical. Model II consists of two FSPS population components, each velocity-broadened according to the stellar mass and effective radius at F300W and where the broad lines are stellar in origin. AGN specific lines are masked. Model III is a mix of AGN and a stellar component. This fit prefers an evolved stellar population to dominating the rest-frame optical to match the observed Balmer break, with significant contribution from AGN emission lines and \ion{Fe}{2}. Between these models, model III is favored by the data.
}
\label{fig:agn_detailed_fit}
\end{figure*}

A defining property of the little red dots is the ``v-shaped'' spectrum, consisting of a red rest-frame optical continuum and a blue rest-frame UV continuum \citep[e.g.,][]{Labbe:2023uncover,Kocevski:2024}. However, as more and more LRD spectra become available, it has become increasingly clear that the transition from red to blue continuum often occurs at $\lambda_{\rm break} \approx 3600$~\AA, and in some cases there is a defined break in the continuum shape that resembles a Balmer break \citep{Wang:2024brd,Wang:2024UB,Setton:2024break}.  

In the case of \source, there is clear evidence of a sharp and large Balmer Break at $3650$\AA, suggesting the presence of an evolved stellar population. At the same time, the broad and high EW \ion{Fe}{2} in the UV and NIR, and broad H$\alpha$ in the optical suggest that AGN continuum might contribute significantly at wavelengths blueward and redward of the break. If the Balmer Break is stellar in origin (but see \citet{Inayoshi:2024} for an alternate explanation), any fit to the full spectrum using current models thus requires similar contributions from an AGN continuum with strong emission lines and an evolved stellar population. Joint modeling of a composite AGN, along with a stellar component is a delicate procedure \citep{Akins:2023,Wang:2024brd}. Because of the many independent components, it is hard to be sure that the solutions are unique. The SNR of \source\ is by far the highest amongst the population of little red dots with spectroscopic follow-up, offering the opportunity to disentangle the AGN and stellar components.
 
\subsection{AGN and Stellar SED models}

We first describe a flexible AGN model with ingredients based on known UV-selected AGN, and including a range of BLR emission line components seen in the spectrum of \source. The model has the following components:

\begin{itemize}
    \item A power-law continuum, $\lambda^\beta$, parameterized by slope $\beta$ and luminosity.
    
    \item Balmer series emission lines, with relative emission line strengths fixed to Case B line ratios\footnote{Fixing the Balmer line ratios helps mitigate degeneracies at $<4100$\AA, where the low resolution of NIRSpec/PRISM blends lines into a pseudo-continuum ending at $3650$\AA, resembling a Balmer Break.}:
    \begin{itemize}
        \item a broad Balmer series, set to the FWHM of the broad H$\alpha$ line. The strengths of the broad Balmer lines are tied to the AGN power-law continuum via the equivalent width of H$\alpha$.
        \item a secondary narrow Balmer series component, set to the FWHM of \oiii, with free normalization.
    \end{itemize}
    
    \item {} [\ion{O}{2}], [\ion{Ne}{3}]$\lambda3867$, and \oiii$\lambda\lambda4960,5008$ doublet emission lines, each set to the FWHM of \oiii\ and modeled as free parameters. The \oiii doublet line ratio is set to 1:3.
    
    \item \ion{Fe}{2} lines using a combination of templates. Originating in the broad-line region, these lines are numerous enough to form a “pseudo-continuum” after convolution to the FWHM of \halpha: 
    \begin{itemize}
        \item empirical templates, based on the Seyfert 1 galaxy I Zwicky 1 \citep[e.g.,][]{VestergaardWilkes:2001, Tsuzuki:2006, BorosonGreen:1992, veroncetty:2004}. The relatively narrow broad lines of I Zwicky 1 facilitate matching to the observed FWHM after convolution.
        \item flexible theoretical templates from \citet{Kovacevic:2010}, which combines the strongest \feii multiplets in three groups.
    \end{itemize}
    
    \item Additional far-UV, optical, and near-IR broad and narrow lines specific to AGN, modeled as Gaussians (full list provided in Table \ref{tab:lines}).
    
    \item AGN dust attenuation, modeled as a power law, with the index shared with the stellar population component.
\end{itemize}

The AGN model has the following free parameters: continuum slope $\beta$, continuum luminosity, EW H$\alpha_{\text{broad}}$, $L_{\text{H}\alpha_{\text{narrow}}}$, $A_{\rm V,AGN}$ and dust index, UV-optical-NIR broad and narrow lines, and Fe~II width and amplitude. 

To model the stellar continuum, we use FSPS stellar population models \citep{Conroy:2010} based on a \citet{Chabrier:2003} IMF, $0.1-2.0 Z\sun$ metallicity, and delayed-$\tau$ star formation histories (SFH). Note that strong constraints on metallicity and SFH are difficult to achieve in the presence of significant AGN contributions at all wavelengths. Some additional flexibility in SFH can be achieved by allowing a two component stellar population model (e.g., young burst and older population), but we do not explore more freedom in these parameters, e.g., ``non-parametric'' star formation histories \citep{Leja:2019} or attempt to interpret the best-fit ages.  Dust attenuation is modeled as a power law with a variable slope with the same index as the AGN dust. 
Each stellar population component has six free parameters: stellar mass, metallicity, $\tau$, age, $A_{\rm V,sps}$, and ionization parameter $U$.

We note that our approach shares many similarities with other works attempting to model the rest-frame UV/optical spectra of little red dots with prominent apparent breaks, in particular \citet{Wang:2024brd,Wang:2024UB} and \citet{Ma:2024lens}. The AGN models, and the treatment of dust attenuation, are quite similar across these works. As in \citet{Ma:2024lens}, when tying the dust slopes for the AGN and the galaxy together, we find quite steep slopes ($\delta = -1.8$), and as in both Ma et al.\ and Wang et al.\ we find it difficult to reproduce the break with smooth power-law AGN components alone.

The observed spectrum has a very high SNR, but the effective SNR is limited by calibration and other systematic uncertainties. We therefore set a systematic uncertainty floor of $5\%$ of the flux, added in quadrature to the Poisson noise, comparable to systematic uncertainties in the photometry. At the location of strong lines we increase the value to $10\%$ to account for additional systematics in the line modeling (e.g., line spread function, wavelength calibration, line center and shape etc, as these details are not explicitly modeled). In cases that emission lines are not modeled, we mask them. Posterior distributions are sampled with {\tt pymultinest} \citep{buchneretal2019, Feroz:2010}. Model parameters, priors, and posteriors are in Table \ref{tab:contfits}.

\subsection{Stellar and AGN Continuum fits}
\label{sec:contfit}

\begin{deluxetable}{lccc}
\tabletypesize{\footnotesize}
\tablecolumns{14}
\tablewidth{0pt}
\tablecaption{ Continuum Fits \label{tab:contfits}}
\tablehead{
\colhead{Var.} & \colhead{Model I}  & \colhead{Model II} & \colhead{Model III}
 }
\startdata
\hline
$\chi2$ & 1029 & 1552 & 603 \\ 
n$_\text{DOF}$  & 649 & 520 & 644 \\
ln$Z$ & $-588 \pm 0.5$ & $-931 \pm 0.5$ & $-400 \pm 0.8$  \\ 
\multicolumn{4}{c}{dust} \\
\hline
dust index & $-1.79 \pm 0.01$ & $-1.79 \pm 0.01$ &  $-1.79 \pm 0.01$ \\
\multicolumn{4}{c}{AGN} \\
\hline
$\rm{log} \, L_0$ & $12.56 \pm 0.01$ & \nodata & $12.23 \pm 0.06$ \\
$\rm{log} \, L_s$ & $10.82 \pm 0.01$ & \nodata & $11.02 \pm 0.02$ \\
$\beta$ & $-2.2 \pm 0.01$ & \nodata & $-2.90 \pm 0.05$ \\
$\rm{log \, \, EW_b} $ & $2.99 \pm 0.02$ & \nodata & $3.75 \pm 0.04$ \\
$\rm{log \, \, EW_n} $ & $2.82 \pm 0.03$ & \nodata & $3.48 \pm 0.04$ \\
$A_V$ & $2.40 \pm 0.02$ & \nodata & $3.0 \pm 0.2$ \\
$A_V,s$ & $0.1 \pm 0.01$ & \nodata & $0.2 \pm 0.01$ \\
\multicolumn{4}{c}{Gal (primary)} \\
\hline
$\rm{log} \, M_*$ & \nodata & $10.8 \pm 0.1$ & $10.9 \pm 0.02$ \\
$\rm{log} \, \tau$ & \nodata & $7.3 \pm 0.3$ & $6.9 \pm 0.3$  \\
$\rm{log \, t}$ &  \nodata & $7.4 \pm 0.5$ & $8.7 \pm 0.01$ \\
$[Z/\mathrm{H}]$ & \nodata & $-1 \pm 0.01$ & $0.28 \pm 0.02$ \\
$A_V$ & \nodata & $2.5 \pm 0.02$ & $1.17 \pm 0.04$ \\
\multicolumn{4}{c}{Gal (blue)} \\
\hline
$\rm{log} \, M_*$ & \nodata & $9.83 \pm 0.03$ & \nodata \\
$\rm{log \, t}$ &  \nodata & $8.74 \pm 0.04$ & \nodata \\
$A_V$ & \nodata & $0.005 \pm 0.007$ & \nodata \\
\enddata
\tablecomments{Luminosities and stellar masses are corrected for the lensing magnification $\mu=1.7\pm0.2$. The AGN \halpha rest-frame equivalent widths are computed relative to the AGN powerlaw continuum. Note: the uncertainties on dust index and age are artificially small because they run into the limits of the prior. In general, all quoted uncertainties should be considered underestimates and much smaller than the expected systematics.}
\end{deluxetable}

We  proceed with fitting three models, one with only AGN components (I),  one with only stellar components (II), and one with a mix of stars and AGN (III). 

\subsubsection{I. AGN only}
From the spectrum of \source it is apparent two AGN components are needed: a blue AGN continuum to explain the blue UV spectrum, and a reddened AGN to represent the red part of the spectrum (with a cross-over point near the Balmer break region). To keep the number of free parameters small, we couple the AGN models so that the blue AGN is simply some dust-free fraction of the reddened AGN model, effectively introducing one extra free parameter $f_{blue}$. In this model the blue AGN can be thought of as a fraction of dust-scattered light escaping without reddening. The fit is presented in the top row of Figure \ref{fig:agn_detailed_fit}. While the model is able to reproduce certain features well, including the complex emission complex in the UV and the overall shape of the SED, two clear discrepancies are visible: 1) a large mismatch around the Balmer break, indicating that the underlying power-law AGN continuum model is too smooth, and related 2)  large oscillating residuals at $0.36-0.7\micron$, due to strong emission lines and \ion{Fe}{2} pseudo-continuum attempting to make up for the continuum model mismatch. The mismatch is especially apparent in regions where \ion{Fe}{2} is known to be weak or absent; here the residuals are largest. Another notable aspect of the model is the need for a large negative dust law index $\alpha=-1.8$ reflecting large reddening around the break, but low reddening at longer wavelengths. This is likely another expression of the fundamental mismatch between a powerlaw continuum of the model and the large observed break in the observed spectrum. Despite the reasonably conservative errors of $5-10\%$ per pixel, the formal fit is rather poor $\chi_{red}^2=1.6$. %($\chi^2=1029$ for $649$ degrees of freedom n$_\text{DOF}$), giving a fit probability $P(\text{AGN}) = 0.0$.

\subsubsection{II. Stars only}
Next we model the spectrum with only stellar population models. Like above, we include two stellar models, a low-dust model $A_V<1$ to reflect the UV light, and $A_V>1$ model for the red. The star formation histories are independent, allowing significantly more freedom in the combined SFH (e.g., a young burst on top of an older stellar population). Strong \ion{Fe}{2} features are masked, as these are not produced in the FSPS models. Line widths are assumed to be due to virial motion of the stars. The stellar model is convolved with a velocity dispersion parameterized by the stellar mass and a size $r_e=70$pc (see \S\ref{sec:morphology}) following the relation by \citep{vandokkum:2015}. As shown by \citet{Baggen:2024}, extremely compact sizes and large stellar masses $\log(M/M_{\sun})>10$ can produce line widths of 1000s km/s, similar to observed. The results are shown in the middle row of Figure \ref{fig:agn_detailed_fit}. As with the AGN only model, some aspects of the observations can be reproduced, including the overall shape and broad lines, but the fit is poor $\chi_{red}^2=3.0$
The model does a better job at matching the continuum at $0.5-0.7$\micron, but shows large residuals around the Balmer break and at $0.4-0.5$\micron, where \feii lines contribute. Some of the mismatch could be due to restrictions in the combination of SFH and dust (e.g., a dust-reddened post-starburst population without emission lines could be added). But this exercise is meant to show that the most straightforward and reasonably simple models do not reproduce the full spectrum of \source. Obviously, the stellar-only model is incapable of producing AGN-specific BLR lines. 

\subsubsection{III. Stars + AGN}
Finally, we attempt a model containing both AGN continuum (two components as in model I) and a stellar population. The motivation is that the high EW of the \ion{Fe}{2}$\lambda$2500, \ion{Fe}{2}$\lambda$9200, and \halpha suggest that the rest-frame $0.1-0.3$\micron\ and $0.6-1.0$\micron\ are AGN dominated, while the rest-frame optical $0.3-0.6$\micron\ could be dominated by a stellar population producing the Balmer break \citep[e.g.,][]{Wang:2024UB,Baggen:2024}. As shown in Figure \ref{fig:agn_detailed_fit} ({\em bottom panels}) this model indeed is a much better fit $\chi_{red}^2=0.9$.
To see which of the three models is favored by the data we can use the evidence produced by the multinest fitting procedure. Comparing with the simpler AGN-only model I, we compute the Bayes Factor (the ratio of the logarithm of the evidence $Z$) to see if the added complexity of model III is justified by the data. $Z$ is the probability of the data given the model, marginalized over the model parameter space. It accounts for model complexity, favoring simpler models unless the data strongly supports the added complexity. We find $ln(B_{I,III}) = ln(Z_I/Z_{III}) = -81$, indicating strong evidence for model III. The conclusion from the SED fitting is that AGN power-law continua do not provide a good description of the data, and that a model component is required that produces a strong optical continuum break. In the context of the models considered here, this suggests a stellar-dominated optical continuum, with inferred $M_* \sim 8\times10^{10}$~\msun\ in stars, an age of $\sim500$~Myr and significant attenuation $A_V \sim 1$. In  Appendix \ref{appendix:sed}, we explore the sensitivity of the fit to reducing the Balmer Break component. 

Unfortunately, none of the fits are fully satisfying. As will be discussed more fully in \S \ref{sec:discussion}, while the fit is acceptable in a $\chi^2$ sense, the interpretation of the parameters are less straightforward. The best-fit model leads to rather extreme properties for both the stars (extremely high central stellar densities) and the AGN (extremely high EW H$\alpha \, >$3000~\AA\ measured relative to the AGN continuum), neither of which have been observed in any other galaxy or AGN. So we consider it likely that we are missing a critical physical ingredient or insight in our modeling. For context, the SDSS AGN EW(H$\alpha$) is $\sim 200-300$~\AA\ \citep{Stern:2012a} and the EW distribution of broad-line AGN found with JWST at these redshifts is $\sim500\pm300$\angstrom\citep{Maiolino:2024Xray}, so the EW \halpha of \source is already quite high (relative to the observed total continuum). We note that a key assumption the AGN modeling performed here is the powerlaw shape of the AGN continuum. If the AGN continuum has a different shape, or itself features a Balmer break \citep{Inayoshi:2024}, a massive stellar component would not be needed.

\section{X-ray emission}
It has become clear that little red dots, and perhaps broad-line selected AGN more generally at $z>4$, are very X-ray weak compared with lower-redshift sources \citep{Furtak:2023,Furtak:2023nature,Yue:2024,Maiolino:2024Xray}. With UNCOVER, we get the benefit of 2.14 Ms of \emph{Chandra} imaging (Cycle 23; ACIS-I; Proposal ID 23700107; PI: A.\ Bogdan) and the small boost of lensing ($\mu = 1.7$). Our best constraint comes from the observed 2-7 keV, corresponding to a rest-frame 10-40 keV luminosity upper-limit of $<2 \times 10^{43}$~erg/s. At rest-frame 10-40 keV, hard X-ray photons are relatively insensitive to absorption up to and including extreme Compton-thick column densities of $N_{\rm H}\sim10^{25}$~cm$^{-2}$. Relative to a standard X-ray bolometric correction of 80-90 \citep{Duras:2020}, \source\ is nominally 10 times weaker in the X-rays than would be expected assuming the broad H$\alpha$ is photoionized by a normal UV-emitting AGN (see \S \ref{sec:discussion}). Any reddening correction would only make \source\ intrinsically weaker in the X-ray. Nominally, this level of X-ray weakness could be explained purely by absorption if $N_H > 10^{25}$~cm$^{-2}$. Although such a high value is quite extreme, the stellar and gas densities in these objects are also extreme. Alternatively, our limit is consistent with the higher end of the bolometric correction ($L_X/L_{\rm bol}= 100-1000$) that is derived for the most luminous quasars \citep[e.g.,][]{Martocchia:2017}. 

\section{Environment}
\label{sec:environ}

\begin{figure}
\hspace{-1mm}
\includegraphics[width=0.45\textwidth]{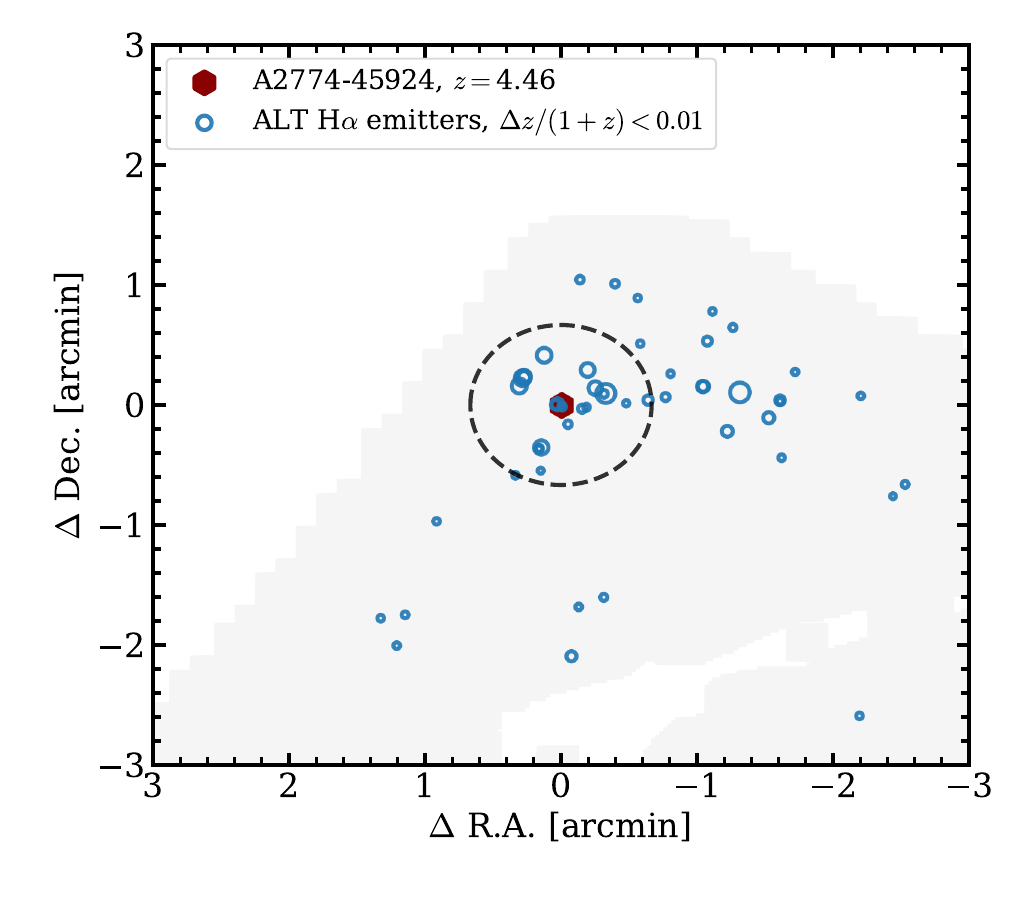}
\caption{ The overdensity around \source\ as revealed by ALT spectroscopy. \source\ is the red hexagon, and those of 40 galaxies with $4.45 < z < 4.49$ as blue circles whose size correlates with their stellar mass. These were all identified via H$\alpha$ emission in ALT. For references the circle represents 1 cMpc, while the grey region shows the ALT area with $\mu < 3$. There is a clear overdensity around \source\ at this redshift.}
\label{fig:env}
\end{figure}

\source\ is found in a large galaxy overdensity. We have already seen hints that some LRDs are preferentially associated with overdensities \citep{Matthee:2023,Fujimoto:2023overdense}, but the extensive data in the Abell 2744 field enables the most comprehensive look at the environment of an LRD afforded to date.  With the ALT grism redshifts \citep{Naidu24}, it is very clear that many galaxies are associated with this overdensity (Figure \ref{fig:env}). There are 40 galaxies within $\delta z = 0.02$ of \source\ in the full 27.5 arcmin$^2$ coverage in this redshift interval. \source\ is about 50 arcsec away from the edge of the field of view of the ALT data. In the full $z=4-5$ coverage, ALT detects 17.6 galaxies per $\Delta z=0.04$ on average. This suggests that \source\ is in a mild over-density of $\delta=N/\langle N\rangle-1=1.3$.  However, as seen in Fig. $\ref{fig:env}$, most are at relatively small angular separations. Half are within 30 arcsec, with an over-density $\delta\approx40$. This over-density is similar to some of the most UV luminous QSOs at $z\sim6-7$ \citep{Eilers:2024}, although as a class the high number densities of little red dots argues that they cannot occupy the same massive halos as UV-selected QSOs \citep{Pizzati:2024}. 

\begin{figure*}
\hspace{1mm}
\includegraphics[width=0.38\textwidth]{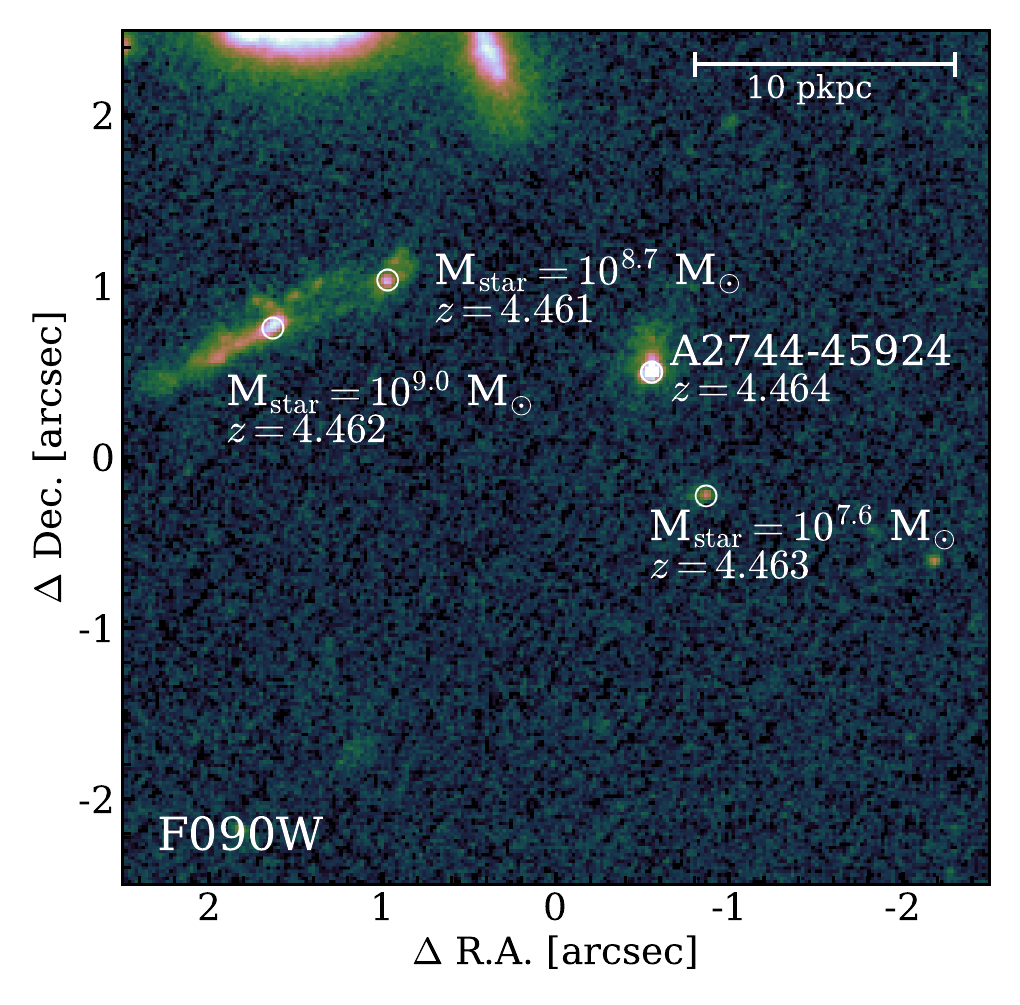}
\includegraphics[width=0.52\textwidth]{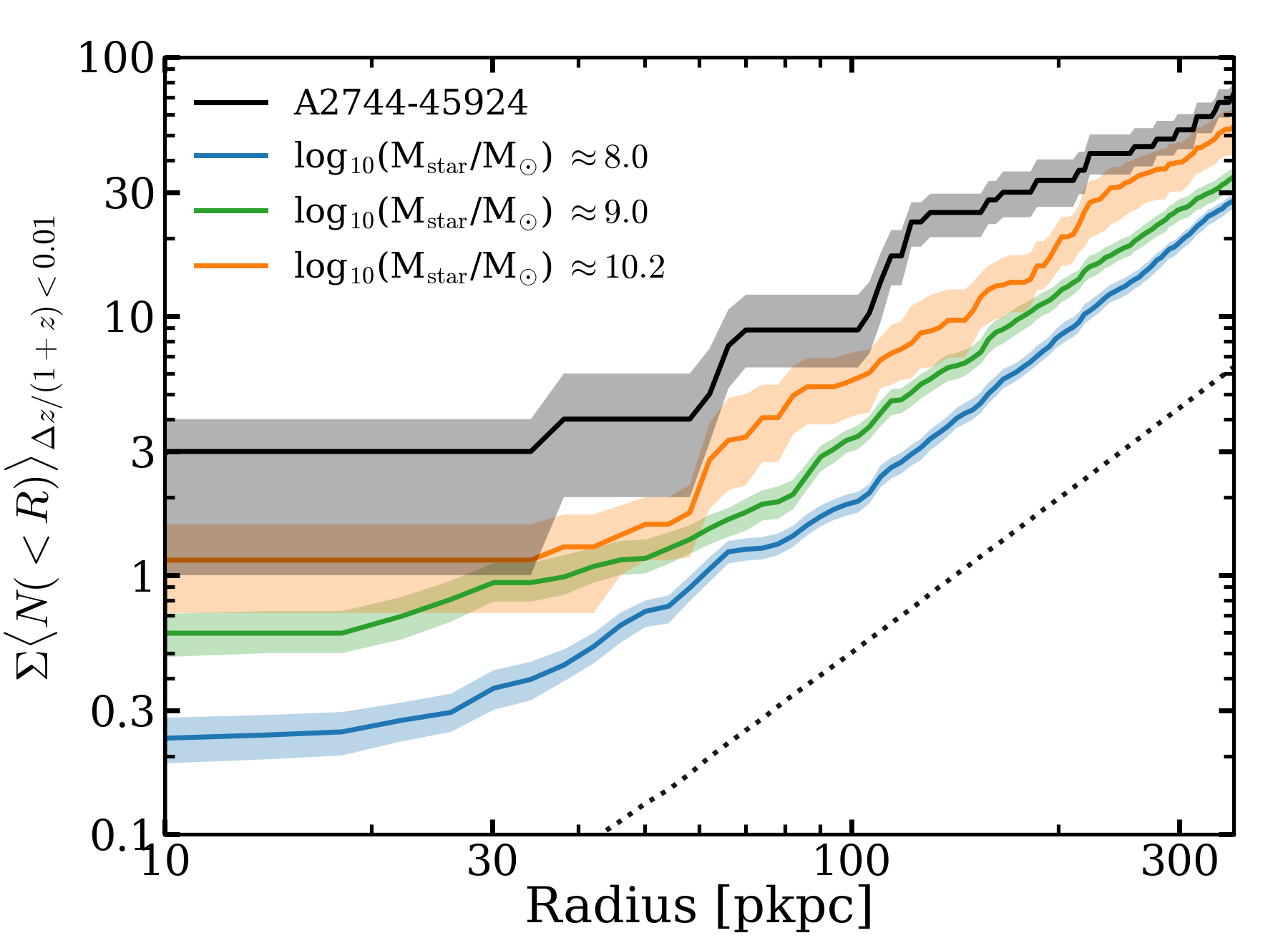}
\caption{
{\it Left}: The neighboring galaxies to \source\ in the NIRCam F090W image. We highlight the masses and redshifts of other galaxies for which ALT detected H$\alpha$ emission.
{\it Right}:
The average number of neighbours within $
\Delta z/(1+z) <0.01$ as a function of increasing projected radii in the source-plane around \source\ and around normal star forming galaxies at $z=4-$ from the ALT survey for various stellar masses, compared to the random expectation. Errors are from bootstrapping the pairs (in the case of \source) or the central galaxies (in the stellar mass bins). The dashed line (normalised to the blue curve) shows the expectation around a random position.  }
\label{fig:galaxycounts} 
\end{figure*}

The left panel of Figure $\ref{fig:galaxycounts}$ shows the near environment of \source, in which a total of four galaxies are detected in H$\alpha$ emission in the ALT data. None of the three neighbours shows evidence for AGN activity.  These galaxies are all found with velocity differences within 80 km s$^{-1}$, roughly the precision of the grism redshifts \citep[e.g.][]{Torralba:2024}. One companion is found at a distance of 0.7 arcsec to \source\ and has a stellar mass $M_* = 4 \times 10^7$~\msun, a second is 1.5 arcsec away and has $M_* = 5 \times 10^8$~\msun, while the third at 2.3 arcsec has $M_* = 10^9$~\msun. The latter two are interacting with each other as indicated by diffuse emission. These systems are very likely all in the virial radius of the same halo (2.5 arcsec is the virial radius of a $3\times10^{10}$ M$_{\odot}$ halo at $z=4.5$; e.g. \citealt{Behroozi:2019}) and bound to the same system.

We empirically compare the environment of \source\ to the environments of other galaxies in the same dataset. Both as a comparison set and a tracer of the galaxy density we used other ALT galaxies with spectroscopically measured redshifts at $z=4-5$ with a magnification $\mu<3$ and L$_{\rm H\alpha}>2\times10^{41}$ erg s$^{-1}$ (see Matthee et al. in prep for details). We count the number of neighbors with increasing distance (along the projected direction, in the source-plane modeled using the \citealt{Furtak:2022b_SLmodel} magnification model), with a redshift difference of $\Delta z/(1+z)<0.01$ (i.e. 3000 km s$^{-1}$). For the comparison set, we count the average number of neighbors around galaxies in a range of stellar masses. The rationale of this empirical comparison is that the galaxy and galaxy-pair selection functions are identical and that the completeness of these selections only mildly vary with redshift and across the field of view, especially with our luminosity cut. When calculating average neighbouring source density, we calculate the effective covered area taking the edges of the field of view and the masked $\mu>3$ regions into account.

Fig. $\ref{fig:galaxycounts}$ shows the measurement for \source\, compared to the average neighbouring galaxy counts around 149, 62 and 7 sources with $M_* \approx 10^{8, 9, 10.2}$~\msun\, respectively. We estimate the errors in the neighbouring counts around \source\ by bootstrapping the neighbours with replacement. We bootstrap the comparison set to also capture the variation from source to source among that set. We detect that galaxy over-densities are dependent on mass, and that on all scales the overdensity around \source\ is comparable or even higher than the most massive, non-LRD galaxies in the reference sample. There is an apparent flattening in the galaxy pair counts that occurs at scales 50 pkpc ($\sim 8$ arcsec) likely due to satellites within the same dark matter halo. 

In summary, \source\ is found in one of the most overdense regions in ALT at $z=4-5$, and the most overdense on $\sim1-2$ cMpc scales. The overdensity exceeds that found around galaxies with $M_* \sim 10^{10}$~\msun, which in itself is higher than around galaxies with lower masses.  While we will not attempt to quantify the stellar mass from this overdensity, the richness provides indirect evidence for a high stellar mass \citep{Matthee:2024} 
At this point it remains unclear whether either the recent merger history (i.e. the relatively nearby environment) or the very large-scale environment is important in either forming or triggering the black holes powering the little red dots. A clustering analysis of a larger sample that would allow us to control for BH activity would be very valuable to explore the diversity in their environments quantitatively.

\section{An extreme little red dot}
\label{sec:discussion}

\source\ is an extreme system. It is one of the brightest and most luminous little red dots known, even over wider areas \citep{Kokorev:2024,Kocevski:2024,Akins:2024,Wang:2024brd}. 

The broad observed \halpha line width FWHM(H$\alpha$)$=4500$~km/s and luminosity $L_{\mathrm{H}\alpha} = 10^{44}$~erg/s are among the highest observed in any LRD. Meanwhile, the source also exhibits one of the strongest Balmer breaks seen in any source at $z>4$ \citep[see][for a compilation]{Weibel:2024}. Regardless of the source of the continuum, \source\ has a very high  H$\alpha$ rest-frame EW ($W_{\mathrm{H}\alpha,0}\sim800~\AA$), but this means that any significant contribution from a stellar continuum would cause the intrinsic EW (relative to AGN continuum only) to be much higher (\S \ref{sec:contfit}).

In the following subsection, we explore a range of possible origins both for the broad emission lines and the strong Balmer continuum to see if there is a way to reconcile the peculiar extreme properties of \source. As of now there is no physical picture that naturally explains the high number densities, the apparent Balmer break, the implied extraordinary stellar density, the remarkable H$\alpha$ (width, flux, and EW), and the lack of commensurate X-ray \citep{Yue:2024} and mid-infrared emission \citep{Williams:2024,Akins:2024}.

\subsection{Broad \ion{Fe}{2} Emission: Unambiguous Evidence for the Broad-line Region}

One of the novel aspects of our analysis here is the detection of broad Fe~II multiples throughout the UV (Figure \ref{fig:agn_lines_feii_uv}), optical, and near-infrared spectrum (see Figure \ref{fig:appendix:lines}). These Fe multiplets are commonly seen in AGN continuum \citep[e.g.,][]{Wills:1985} and are a tell-tale signature that we are seeing emission from the broad-line region although we cannot spectrally resolve the lines in the UV. They are thought to be strongest in sources accreted close to their Eddington limit \citep[e.g.,][]{BorosonGreen:1992}. A persistent question about little red dots has been the origin of the blue and faint UV continuum. In this one case, we have clear evidence that the emission lines arise from the broad-line region of an accreting black hole. In principle, this emission could arise from scattered light, as seen in some lower-redshift AGN \citep[e.g.,][]{Glikman:2023}. However, we do not favor an explanation where a reddened AGN  (with some unreddened scattered light) produces the entire spectrum, because the inflection point in the spectrum occurs right at $3600$~\AA, which is the energy required to ionize $n=2$ hydrogen. As shown by \citet{Setton:2024break}, the ensemble of little red dot spectra in the \jwst\ archive always show an inflection point at the same wavelength. This is inconsistent with a combination of reddened directly-transmitted AGN continuum with scattered light, which would produce inflections at different wavelengths depending on the dust law and fraction of scattered light.

In summary, the detection of \ion{Fe}{2} in this spectrum strongly implies that the emission lines throughout the spectrum emerge from broad-line region conditions and are photoionized by an accreting massive black hole. We caution that other little red dot spectra have quite different spectral morphologies \citep[e.g., the triply imaged source has very weak metal lines,][]{Furtak:2023nature}, so we cannot necessarily draw more general conclusions about other objects yet.

\subsection{Implications of a Pure Stellar Continuum}
\label{sec:alternatives}

\begin{figure}
\hspace{5mm}
\includegraphics[width=0.45\textwidth]{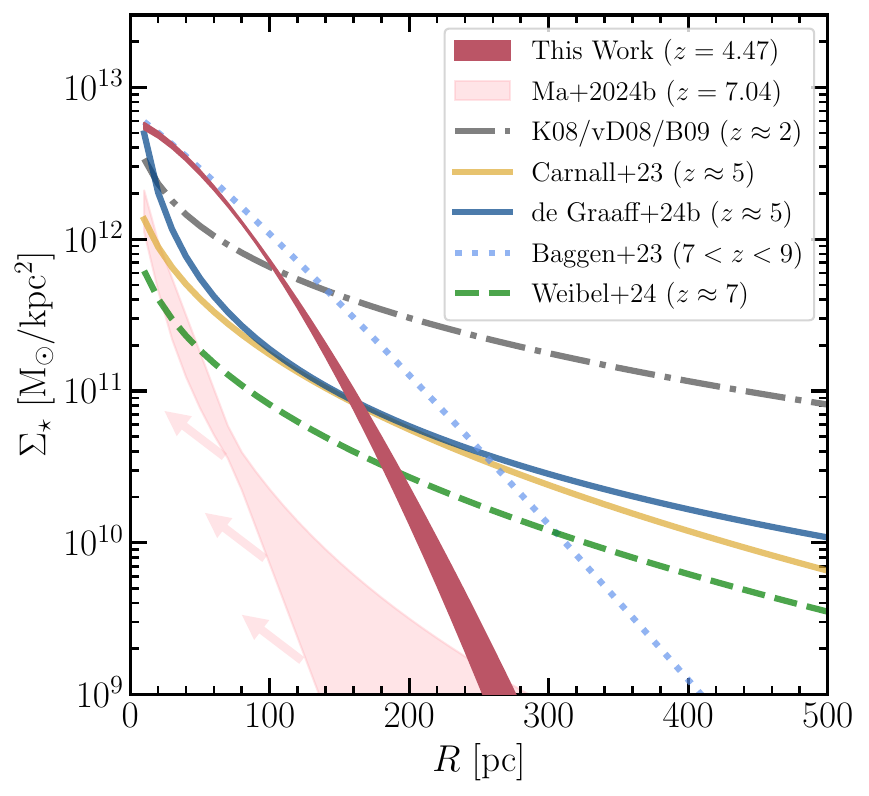}
\caption{ We compare the projected stellar density of \source\ with \jw-detected objects at similar redshift from \citet{degraaff:2024} and \citep{Carnall:2023}, with higher-redshift little red dots from \citet[][see also \citealt{Wang:2024UB}]{Baggen:2023}, and $z \sim 2.3$ compact red galaxies from \citet[][see also \citealt{vanDokkum:2008,Bezanson:2009}]{Kriek:2008}. In the stellar-only case, the density is high, but not much higher ($\sim 2\times$), than an extrapolation of the cosmic noon sources.
}
\label{fig:density} 
\end{figure}

Regardless of the origin of the emission lines, our best-fit models in \S \ref{sec:contfit} strongly prefer a galaxy-dominated continuum at rest-frame $>0.4$\micron. A stellar-only picture for \source\ would also explain the lack of X-ray detection \citep[see also][]{Yue:2024}, as well as the lack of a hot dust upturn in the mid-infrared seen in general in little red dots \citep{Williams:2024}, and potentially the lack of variability \citep{Kokubo:2024}.

In Figure \ref{fig:density}, we contextualize the implied stellar density of \source\ compared to other little red dots \citep{Baggen:2023, Baggen:2024, Ma:2024lens}, contemporary massive and passive galaxies \citep{Carnall:2023,degraaff:2024}, and lower-redshift massive and compact systems \citep{Kriek:2008,vanDokkum:2008,Bezanson:2009}. These densities are high, but given the uncertainties they are not unprecedented \citep[e.g.,][]{Norris:2014}. If we confirm the implied stellar densities, then it will have important implications for star formation at high density and the role of feedback \citep[e.g.,][]{Hopkins:2010,Grudic:2019} and perhaps the role of dark matter in driving star formation \citep{Boylan-Kolchin:2024}.

Perhaps the hardest thing to explain about \source\ in the absence of an AGN is the H$\alpha$ luminosity. With an observed (magnification but not reddening-corrected) H$\alpha$ luminosity of $L_{\rm H\alpha} = 10^{44}$~erg/s and a FWHM$=4500$~km/s, this line is hard to explain given the observed continuum. The luminosity would require $500-1000$~\msun/yr of star formation to produce sufficient ionizing photons \citep{KennicuttEvans:2012,Wilkins:2019}. Such a star-formation rate within a rest-frame optical size of $<70$~pc (see \S \ref{sec:morphology}) would likely exceed the Eddington limit for star formation \citep[e.g.,][]{Murray:2005,Thompson:2005,Hopkins:2010}. The faint UV fluxes imply that the star formation needs to be highly dust-reddened, further exacerbating the situation. The other possibility is that the broad H$\alpha$ may be powered by supernovae \citep[e.g.,][]{Terlevich:1985,Filippenko:1997}. However, given that the number of ionizing photons from the supernovae is roughly 100 times lower than that from young stars at a fixed star formation rate \citep{Johnson:2011}, it seems very unlikely that supernovas could power this broad line \citep[see also arguments in][]{Maiolino:2024Xray}, unless the IMF is extremely top heavy.

\subsection{Nature of the AGN}
\label{sec:alternatives}

Taking the AGN presence as a given due to the presence of \ion{Fe}{2}, we discuss the nature of the AGN. It is hard to avoid the conclusion that the AGN has properties that have not been observed in AGN before. First of all, the spectrum cannot be described by the combination of power-laws that constitute our current AGN models (\S \ref{sec:contfit}). Furthermore, the broad-line EWs are quite high, and potentially much higher if stellar continuum dominates at 6564\angstrom. Finally, we observe that \source\ is at least ten times fainter in the UV than expected from the H$\alpha$ line assuming typical type 1 QSOs\citep{VandenBerk:2001}, even with no reddening correction. 

It is possible that the dearth of UV/optical and X-ray continuum from the AGN could be explained by a much higher covering factor of broad-line region clouds. The observed narrow associated Balmer absorption lines we see in \source\ and other little red dots \citep[20-30\%][]{Matthee:2023,Wang:2024brd,Kocevski:2024,Juodzbalis:2024} also may indicate an excess of high-density clouds. Only a small number of associated Balmer absorbers have been observed in lower-redshift broad-line quasars to date, almost exclusively in systems with known absorption systems in UV lines, known as Broad Absorption Line quasars \citep[BALs; for an overview, see][]{Hutchings:2002,Schulze:2018}. The velocity distribution of the absorbers, falling within $<500$~km/s, strongly suggests that they sit at the outer part of the broad-line region, and the fact that there is sufficient hydrogen in an excited state to absorb H$\alpha$ requires densities $n_e > 10^9$~cm$^{-3}$. It is possible that these dense clouds are so numerous that they can mostly block the light from the physically smaller accretion disk, while still allowing broad-line region emission to escape \citep{Kocevski:2024,Maiolino:2024Xray}, as has been seen in the X-ray before \citep[e.g.,][]{Ricci:2023}. It is also possible that lower-metallicity gas can boost the observed EWs, since there are fewer metal lines available for cooling. This metallicity effect is seen in star-forming galaxies \citep[e.g.,][]{Reddy:2018}. The presence of such dense clouds beyond the sublimation radius could also cause additional absorption even into the mid-infrared, leading to colder torus models \citep[e.g.,][]{Williams:2024,Wang:2024brd}.

The other challenge in an AGN-dominated scenario is how to explain the strong Balmer break. In our current models, a relatively old stellar population is required to explain the observed break. We use the revised index from \citet{Wang:2024UB} to measure a strength of $f\lambda4100/f\lambda3670=$2.4 in \source. This Balmer break strength is higher than is reported by Wang et al.\ for similar compact red sources at $z=7-8$. It is tempting to wonder whether the continuum could be coming from the accretion disk \citep{Laor:2011} or the very dense gas in the broad-line region \citep[e.g.,][]{Baldwin:2004, Inayoshi:2024, Setton:2024break}. However, this has yet to be observed in other AGN and more modeling work is needed to determine how and where such a continuum might form.

\subsection{Implications for galaxy evolution and black hole growth}

\begin{figure}
\hspace{-3mm}
\includegraphics[width=0.48\textwidth]{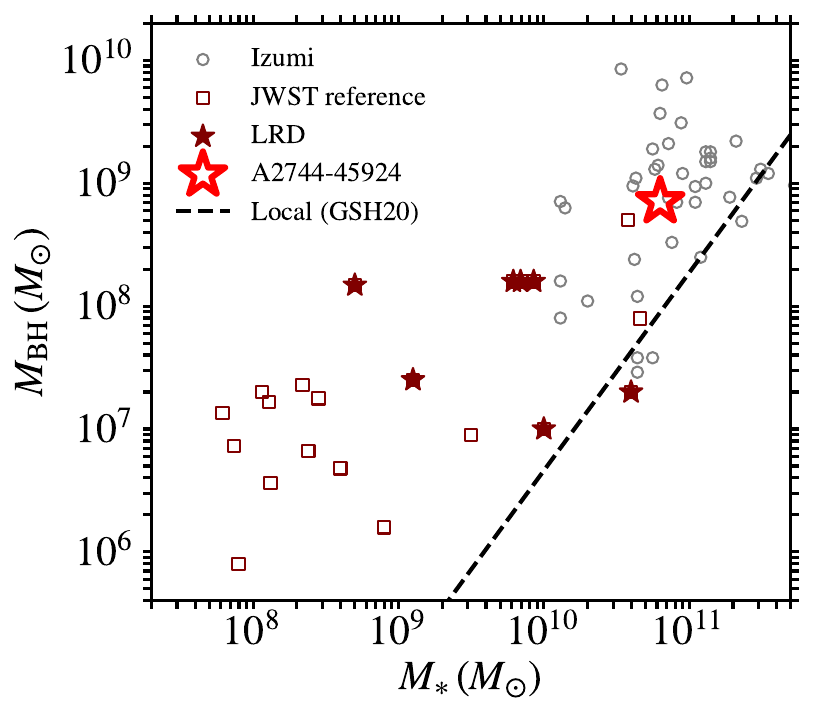}
\caption{ The position of \source\ in a black hole mass-galaxy mass scaling relation plot. Since the inferred $M_*$ is similar across Model II and III (and unconstrained in Model I), we simply show one value. In terms of $M_{\rm BH}$, we note that the formal errors on all \mbh\ values are $\sim 0.5$~dex \citep{Shen:2013}. However, in the case of little red dots, given the high H$\alpha$ EWs and unknown bolometric (or UV) luminosities, we caution that the \mbh\ values are particularly systematics dominated. When taken at face value, \source\ is not such a big outlier from the local scaling relations. For comparison, we show other broad-line samples, including little red dots from \citet{Fujimoto:2022}, \citet{Furtak:2023nature}, \citet{Kokorev:2023}, and \citet{Wang:2024UB} as filled stars, along with other AGN selections from \citet{Maiolino:2023}, \citet{Larson:2023}, and \citet{Chisholm:2024} as open squares. The scaling relation line is taken from \citet{Greene:2020}.}
\label{fig:mbhmgal} 
\end{figure}

Little red dots are common, accounting for a few percent of the galaxy population at $4<z<6$. It is therefore important to consider the implications for galaxy and black hole growth under different possible interpretations of their SEDs. Granting that in our estimation the broad Balmer lines are likely powered by accretion onto a supermassive black hole, with some substantial contribution from an evolved stellar population, and granting that no scenario satisfactorily explains all of the observed phenomena, we muse about possible implications for the coevolution of black holes and galaxies.

The black hole to galaxy mass ratio appears to be quite high across the board in little red dots \citep[e.g.,][]{Furtak:2023nature,Kokorev:2023}. Given their high number density, a high ratio of \mbh/$M_{\rm gal}$ may have important implications for seeding mechanisms \citep[e.g.,][]{Pacucci:2023}. However, we cannot be confident in either the black hole mass or the stellar mass. 

The measured H$\alpha$ flux is $f_{\rm H \alpha} = 7.3 \times 10^-16$~erg/s/cm$^2$/\AA, corresponding to a magnification-corrected $L_{\rm H\alpha} \sim 10^{44}$~erg/s with no reddening correction. %Taking this luminosity as a conservative lower-limit we infer a bolometric luminosity of $L_{\rm bol} = 130 \times L_{\halpha} = 2 \times 10^{46}$~erg/s.
We thus infer a black hole mass of \mbh$= 7 \times 10^8$~\msun\ using the single-epoch scaling from \citet{GreeneHo:2005}. The dominant sources of error include the uncertain reddening correction (factors of several) and unusual spectral energy distribution precluding the unambiguous use of low-redshift scaling relations. Simply put, we do not know how to relate the observed H$\alpha$ luminosity to the radius of the broad-line region \citep[e.g.,][]{krolik2001}. 
We therefore consider the black hole mass an order of magnitude estimate at best. Our best fit host decomposition likewise yields $M_* \sim 10^{11}$~\msun, consistent with the finding of significant clustering of nearby galaxies (Fig.\ \ref{fig:galaxycounts}). In that case, we find \mbh/$M_{\rm gal} \sim 1\%$ (Figure \ref{fig:mbhmgal}), which is certainly high compared to local relations \citep[e.g.,][]{mcconnellma2013} but not as  dramatic of an outlier as other sources \citep{Kokorev:2023,Bogdan:2023}.

In this galaxy, many signs are pointing to very rapid assembly within an environment characterized by high gas density. Most of the red light in the source emerges from within $r_e < 70$~pc, and we have argued above that the Balmer absorption requires very dense gas. Furthermore, the high-EW UV Nitrogen lines also seem to point to a rapid star formation event, although the presence of these lines has been used to argue for the presence of an AGN \citep{Bunker:2023,Ubler:2023,Maiolino:2024gnz11}. Indeed, \ion{N}{3}] is seen in stacks of narrow-line AGN \citep{Hainline:2011} and radio-selected narrow-line AGN \citep{DeBreuck:2000}. On the other hand, these lines are only rarely seen in broad-line quasars \citep{Bentz:2004,Glikman:2007,Jiang:2008,Batra:2014}. To date, a number of explanations for these Nitrogen lines have emerged that relate to the products of very massive stars, globular cluster formation \citep[e.g.,][]{Senchyna:2024}, or even a tidal disruption event \citep[e.g.,][]{Kochanek:2016}. Regardless of the details, the current explanations of these lines all point to an event involving recent star formation \citep[e.g.,][]{Cameron:2023, Shi:2024}. The overdense environment and complicated UV morphology also are suggestive that some of the activity may be triggered by the environment.

\section{Summary}

We present high SNR NIRSpec/PRISM, G395M grating, and NIRCam imaging for \source, one of the optically brightest little red dots known. Discovered from the UNCOVER survey \citep{Bezanson:2022}, this object was photometrically identified \citep{Labbe:2023uncover} and then observed for 16.3 hours with NIRSpec PRISM \citep{Price:2024}. 

\source\ is unresolved in the rest-frame optical, with an upper limit to its size of $r_e < 70 \pm 10$~pc. There is very blue $n=1$ asymmetric fluff detected in the rest-frame UV whose nature is yet unknown that is not included in the PRISM spectrum. From detailed modeling of the emission lines in the PRISM and grating spectra, we find clear evidence for a broad Fe~II pseudo-continuum that argues strongly for an AGN to power the observed emission lines. Explaining the observed continuum proves more challenging. Our modeling strongly prefers that the rest-frame optical continuum be dominated by emission from a $\sim 500$~Myr old stellar population with moderate reddening $A_V \sim 1$. The implied lower limit to the central stellar density of $\sim 5 \times 10^{12}$~$M_{\odot}$/kpc$^2$ appears as high as any observed stellar system. Furthermore, we have no mechanism to excite the $L_{\rm H\alpha}>10^{44}$~erg/s broad emission line, except AGN activity. Other clues to the nature of the source include high-EW UV Nitrogen lines \ion{N}{3}] and \ion{N}{4}], and narrow associated Balmer absorption of the broad H$\alpha$ line. The implied rapid star-formation event that may have led to high Nitrogen enrichment may be related to the overdensity of galaxies found in the 100 kpc around \source. 

This exploration of the properties of one luminous little red dot strongly suggests that AGN power at least some of these sources. To make more definitive progress on the composition of the host galaxy, high-resolution NIRSpec spectroscopy of the Balmer break region would provide qualitatively new information about the origin of the continuum. Photometric monitoring, to search for variability, could provide another clue as to whether there is emission from accretion disk scales \citep[e.g.,][]{Kokubo:2024,Zhang:2024}. We will also observe this source with ALMA and MIRI, in an attempt to understand the bolometric luminosity and dust temperature (Setton in prep). These studies, hopefully combined with deep UV spectroscopy of a larger sample, may help us put together a more complete picture of this new and enigmatic class of objects.

\acknowledgements

We acknowledge funding from JWST-GO-02561 and JWST-GO-04111, provided through a grant from the STScI under NASA contract NAS5-03127. I.L. acknowledges support from Australian Research Council Future Fellowship FT220100798. K.G. and T.N. acknowledge support from Australian Research Council Laureate Fellowship FL180100060. AZ acknowledges support by Grant No. 2020750 from the United States-Israel Binational Science Foundation (BSF) and Grant No. 2109066 from the United States National Science Foundation (NSF); by the Ministry of Science \& Technology, Israel; and by the Israel Science Foundation Grant No. 864/23. JM and IK are funded by the European Union (ERC, AGENTS, 101076224). Views and opinions expressed are however those of the author(s) only and do not necessarily reflect those of the European Union or the European Research Council. Neither the European Union nor the granting authority can be held responsible for them.  YF ackowledges supports from JSPS KAKENHI Grant Number JSPS KAKENHI Grant Numbers JP22K21349 and JP23K13149.

\bibliographystyle{aasjournal}
%\bibliography{imbh.bib}

\begin{thebibliography}{}
\expandafter\ifx\csname natexlab\endcsname\relax\def\natexlab#1{#1}\fi
\providecommand{\url}[1]{\href{#1}{#1}}
\providecommand{\dodoi}[1]{doi:~\href{http://doi.org/#1}{\nolinkurl{#1}}}
\providecommand{\doeprint}[1]{\href{http://ascl.net/#1}{\nolinkurl{http://ascl.net/#1}}}
\providecommand{\doarXiv}[1]{\href{https://arxiv.org/abs/#1}{\nolinkurl{https://arxiv.org/abs/#1}}}

\bibitem[{{Abel} \& {Satyapal}(2008)}]{Abel:2008}
{Abel}, N.~P., \& {Satyapal}, S. 2008, \apj, 678, 686, \dodoi{10.1086/529013}

\bibitem[{{Akins} {et~al.}(2023){Akins}, {Casey}, {Allen}, {Bagley}, {Dickinson}, {Finkelstein}, {Franco}, {Harish}, {Arrabal Haro}, {Ilbert}, {Kartaltepe}, {Koekemoer}, {Liu}, {Long}, {McCracken}, {Paquereau}, {Papovich}, {Pirzkal}, {Rhodes}, {Robertson}, {Shuntov}, {Toft}, {Yang}, {Barro}, {Bisigello}, {Buat}, {Champagne}, {Cooper}, {Costantin}, {de La Vega}, {Drakos}, {Faisst}, {Fontana}, {Fujimoto}, {Gillman}, {G{\'o}mez-Guijarro}, {Gozaliasl}, {Hathi}, {Hayward}, {Hirschmann}, {Holwerda}, {Jin}, {Kocevski}, {Kokorev}, {Lambrides}, {Lucas}, {Magdis}, {Magnelli}, {McKinney}, {Mobasher}, {P{\'e}rez-Gonz{\'a}lez}, {Rich}, {Seill{\'e}}, {Talia}, {Urry}, {Valentino}, {Whitaker}, {Yung}, {Zavala}, {Cosmos-Web Team}, \& {Ceers Team}}]{Akins:2023}
{Akins}, H.~B., {Casey}, C.~M., {Allen}, N., {et~al.} 2023, \apj, 956, 61, \dodoi{10.3847/1538-4357/acef21}

\bibitem[{{Akins} {et~al.}(2024){Akins}, {Casey}, {Lambrides}, {Allen}, {Andika}, {Brinch}, {Champagne}, {Cooper}, {Ding}, {Drakos}, {Faisst}, {Finkelstein}, {Franco}, {Fujimoto}, {Gentile}, {Gillman}, {Gozaliasl}, {Harish}, {Hayward}, {Hirschmann}, {Ilbert}, {Kartaltepe}, {Kocevski}, {Koekemoer}, {Kokorev}, {Liu}, {Long}, {McCracken}, {McKinney}, {Onoue}, {Paquereau}, {Renzini}, {Rhodes}, {Robertson}, {Shuntov}, {Silverman}, {Tanaka}, {Toft}, {Trakhtenbrot}, {Valentino}, \& {Zavala}}]{Akins:2024}
{Akins}, H.~B., {Casey}, C.~M., {Lambrides}, E., {et~al.} 2024, arXiv e-prints, arXiv:2406.10341, \dodoi{10.48550/arXiv.2406.10341}

\bibitem[{{Alexandroff} {et~al.}(2018){Alexandroff}, {Zakamska}, {Barth}, {Hamann}, {Strauss}, {Krolik}, {Greene}, {P{\^a}ris}, \& {Ross}}]{Alexandroff:2018}
{Alexandroff}, R.~M., {Zakamska}, N.~L., {Barth}, A.~J., {et~al.} 2018, \mnras, 479, 4936, \dodoi{10.1093/mnras/sty1685}

\bibitem[{{Atek} {et~al.}(2024){Atek}, {Labb{\'e}}, {Furtak}, {Chemerynska}, {Fujimoto}, {Setton}, {Miller}, {Oesch}, {Bezanson}, {Price}, {Dayal}, {Zitrin}, {Kokorev}, {Weaver}, {Brammer}, {Dokkum}, {Williams}, {Cutler}, {Feldmann}, {Fudamoto}, {Greene}, {Leja}, {Maseda}, {Muzzin}, {Pan}, {Papovich}, {Nelson}, {Nanayakkara}, {Stark}, {Stefanon}, {Suess}, {Wang}, \& {Whitaker}}]{Atek:2023eor}
{Atek}, H., {Labb{\'e}}, I., {Furtak}, L.~J., {et~al.} 2024, \nat, 626, 975, \dodoi{10.1038/s41586-024-07043-6}

\bibitem[{{Baggen} {et~al.}(2023){Baggen}, {van Dokkum}, {Labb{\'e}}, {Brammer}, {Miller}, {Bezanson}, {Leja}, {Wang}, {Whitaker}, {Suess}, \& {Nelson}}]{Baggen:2023}
{Baggen}, J. F.~W., {van Dokkum}, P., {Labb{\'e}}, I., {et~al.} 2023, \apjl, 955, L12, \dodoi{10.3847/2041-8213/acf5ef}

\bibitem[{{Baggen} {et~al.}(2024){Baggen}, {van Dokkum}, {Brammer}, {de Graaff}, {Franx}, {Greene}, {Labb{\'e}}, {Leja}, {Maseda}, {Nelson}, {Rix}, {Wang}, \& {Weibel}}]{Baggen:2024}
{Baggen}, J. F.~W., {van Dokkum}, P., {Brammer}, G., {et~al.} 2024, arXiv e-prints, arXiv:2408.07745, \dodoi{10.48550/arXiv.2408.07745}

\bibitem[{{Baldwin} {et~al.}(2004){Baldwin}, {Ferland}, {Korista}, {Hamann}, \& {LaCluyz{\'e}}}]{Baldwin:2004}
{Baldwin}, J.~A., {Ferland}, G.~J., {Korista}, K.~T., {Hamann}, F., \& {LaCluyz{\'e}}, A. 2004, \apj, 615, 610, \dodoi{10.1086/424683}

\bibitem[{{Banerji} {et~al.}(2015){Banerji}, {Alaghband-Zadeh}, {Hewett}, \& {McMahon}}]{Banerji:2015}
{Banerji}, M., {Alaghband-Zadeh}, S., {Hewett}, P.~C., \& {McMahon}, R.~G. 2015, \mnras, 447, 3368, \dodoi{10.1093/mnras/stu2649}

\bibitem[{{Barchiesi} {et~al.}(2023){Barchiesi}, {Dessauges-Zavadsky}, {Vignali}, {Pozzi}, {Marques-Chaves}, {Feltre}, {Faisst}, {B{\'e}thermin}, {Cassata}, {Charlot}, {Fudamoto}, {Ginolfi}, {Ibar}, {Jones}, {Romano}, {Schaerer}, {Vallini}, {Vanzella}, \& {Yan}}]{Barchiesi:2023}
{Barchiesi}, L., {Dessauges-Zavadsky}, M., {Vignali}, C., {et~al.} 2023, \aap, 675, A30, \dodoi{10.1051/0004-6361/202244838}

\bibitem[{{Barro} {et~al.}(2024){Barro}, {P{\'e}rez-Gonz{\'a}lez}, {Kocevski}, {McGrath}, {Trump}, {Simons}, {Somerville}, {Yung}, {Arrabal Haro}, {Akins}, {Bagley}, {Cleri}, {Costantin}, {Davis}, {Dickinson}, {Finkelstein}, {Giavalisco}, {G{\'o}mez-Guijarro}, {Hathi}, {Hirschmann}, {Holwerda}, {Huertas-Company}, {Kartaltepe}, {Koekemoer}, {Lucas}, {Papovich}, {Pirzkal}, {Seill{\'e}}, {Tacchella}, {Wuyts}, {Wilkins}, {de la Vega}, {Yang}, \& {Zavala}}]{Barro:2023}
{Barro}, G., {P{\'e}rez-Gonz{\'a}lez}, P.~G., {Kocevski}, D.~D., {et~al.} 2024, \apj, 963, 128, \dodoi{10.3847/1538-4357/ad167e}

\bibitem[{{Batra} \& {Baldwin}(2014)}]{Batra:2014}
{Batra}, N.~D., \& {Baldwin}, J.~A. 2014, \mnras, 439, 771, \dodoi{10.1093/mnras/stu007}

\bibitem[{{Behroozi} {et~al.}(2019){Behroozi}, {Wechsler}, {Hearin}, \& {Conroy}}]{Behroozi:2019}
{Behroozi}, P., {Wechsler}, R.~H., {Hearin}, A.~P., \& {Conroy}, C. 2019, \mnras, 488, 3143, \dodoi{10.1093/mnras/stz1182}

\bibitem[{{Bentz} {et~al.}(2004){Bentz}, {Hall}, \& {Osmer}}]{Bentz:2004}
{Bentz}, M.~C., {Hall}, P.~B., \& {Osmer}, P.~S. 2004, \aj, 128, 561, \dodoi{10.1086/422346}

\bibitem[{{Berg} {et~al.}(2022){Berg}, {James}, {King}, {McDonald}, {Chen}, {Chisholm}, {Heckman}, {Martin}, {Stark}, {Aloisi}, {Amor{\'\i}n}, {Arellano-C{\'o}rdova}, {Bayliss}, {Bordoloi}, {Brinchmann}, {Charlot}, {Chevallard}, {Clark}, {Erb}, {Feltre}, {Gronke}, {Hayes}, {Henry}, {Hernandez}, {Jaskot}, {Jones}, {Kewley}, {Kumari}, {Leitherer}, {Llerena}, {Maseda}, {Mingozzi}, {Nanayakkara}, {Ouchi}, {Plat}, {Pogge}, {Ravindranath}, {Rigby}, {Sanders}, {Scarlata}, {Senchyna}, {Skillman}, {Steidel}, {Strom}, {Sugahara}, {Wilkins}, {Wofford}, {Xu}, \& {Classy Team}}]{Berg:2022}
{Berg}, D.~A., {James}, B.~L., {King}, T., {et~al.} 2022, \apjs, 261, 31, \dodoi{10.3847/1538-4365/ac6c03}

\bibitem[{{Bezanson} {et~al.}(2009){Bezanson}, {van Dokkum}, {Tal}, {Marchesini}, {Kriek}, {Franx}, \& {Coppi}}]{Bezanson:2009}
{Bezanson}, R., {van Dokkum}, P.~G., {Tal}, T., {et~al.} 2009, \apj, 697, 1290, \dodoi{10.1088/0004-637X/697/2/1290}

\bibitem[{{Bezanson} {et~al.}(2024){Bezanson}, {Labbe}, {Whitaker}, {Leja}, {Price}, {Franx}, {Brammer}, {Marchesini}, {Zitrin}, {Wang}, {Weaver}, {Furtak}, {Atek}, {Coe}, {Cutler}, {Dayal}, {van Dokkum}, {Feldmann}, {F{\"o}rster Schreiber}, {Fujimoto}, {Geha}, {Glazebrook}, {de Graaff}, {Greene}, {Juneau}, {Kassin}, {Kriek}, {Khullar}, {Maseda}, {Mowla}, {Muzzin}, {Nanayakkara}, {Nelson}, {Oesch}, {Pacifici}, {Pan}, {Papovich}, {Setton}, {Shapley}, {Smit}, {Stefanon}, {Taylor}, \& {Williams}}]{Bezanson:2022}
{Bezanson}, R., {Labbe}, I., {Whitaker}, K.~E., {et~al.} 2024, \apj, 974, 92, \dodoi{10.3847/1538-4357/ad66cf}

\bibitem[{{Bogd{\'a}n} {et~al.}(2024){Bogd{\'a}n}, {Goulding}, {Natarajan}, {Kov{\'a}cs}, {Tremblay}, {Chadayammuri}, {Volonteri}, {Kraft}, {Forman}, {Jones}, {Churazov}, \& {Zhuravleva}}]{Bogdan:2023}
{Bogd{\'a}n}, {\'A}., {Goulding}, A.~D., {Natarajan}, P., {et~al.} 2024, Nature Astronomy, 8, 126, \dodoi{10.1038/s41550-023-02111-9}

\bibitem[{{Boroson} \& {Green}(1992)}]{BorosonGreen:1992}
{Boroson}, T.~A., \& {Green}, R.~F. 1992, \apjs, 80, 109, \dodoi{10.1086/191661}

\bibitem[{{Boylan-Kolchin}(2024)}]{Boylan-Kolchin:2024}
{Boylan-Kolchin}, M. 2024, arXiv e-prints, arXiv:2407.10900, \dodoi{10.48550/arXiv.2407.10900}

\bibitem[{{Brammer}(2022)}]{Brammer:22}
{Brammer}, G. 2022, {msaexp: NIRSpec analyis tools}, 0.3.4, Zenodo,  Zenodo, \dodoi{10.5281/zenodo.7299500}

\bibitem[{{Buchner} {et~al.}(2019){Buchner}, {Treister}, {Bauer}, {Sartori}, \& {Schawinski}}]{buchneretal2019}
{Buchner}, J., {Treister}, E., {Bauer}, F.~E., {Sartori}, L.~F., \& {Schawinski}, K. 2019, \apj, 874, 117, \dodoi{10.3847/1538-4357/aafd32}

\bibitem[{{Bunker} {et~al.}(2023){Bunker}, {Saxena}, {Cameron}, {Willott}, {Curtis-Lake}, {Jakobsen}, {Carniani}, {Smit}, {Maiolino}, {Witstok}, {Curti}, {D'Eugenio}, {Jones}, {Ferruit}, {Arribas}, {Charlot}, {Chevallard}, {Giardino}, {de Graaff}, {Looser}, {L{\"u}tzgendorf}, {Maseda}, {Rawle}, {Rix}, {Del Pino}, {Alberts}, {Egami}, {Eisenstein}, {Endsley}, {Hainline}, {Hausen}, {Johnson}, {Rieke}, {Rieke}, {Robertson}, {Shivaei}, {Stark}, {Sun}, {Tacchella}, {Tang}, {Williams}, {Willmer}, {Baker}, {Baum}, {Bhatawdekar}, {Bowler}, {Boyett}, {Chen}, {Circosta}, {Helton}, {Ji}, {Kumari}, {Lyu}, {Nelson}, {Parlanti}, {Perna}, {Sandles}, {Scholtz}, {Suess}, {Topping}, {{\"U}bler}, {Wallace}, \& {Whitler}}]{Bunker:2023}
{Bunker}, A.~J., {Saxena}, A., {Cameron}, A.~J., {et~al.} 2023, \aap, 677, A88, \dodoi{10.1051/0004-6361/202346159}

\bibitem[{{Burgasser} {et~al.}(2024){Burgasser}, {Bezanson}, {Labbe}, {Brammer}, {Cutler}, {Furtak}, {Greene}, {Gerasimov}, {Leja}, {Pan}, {Price}, {Wang}, {Weaver}, {Whitaker}, {Fujimoto}, {Kokorev}, {Dayal}, {Nanayakkara}, {Williams}, {Marchesini}, {Zitrin}, \& {van Dokkum}}]{Burgasser:2023}
{Burgasser}, A.~J., {Bezanson}, R., {Labbe}, I., {et~al.} 2024, \apj, 962, 177, \dodoi{10.3847/1538-4357/ad206f}

\bibitem[{{Cameron} {et~al.}(2023){Cameron}, {Katz}, {Rey}, \& {Saxena}}]{Cameron:2023}
{Cameron}, A.~J., {Katz}, H., {Rey}, M.~P., \& {Saxena}, A. 2023, \mnras, 523, 3516, \dodoi{10.1093/mnras/stad1579}

\bibitem[{{Carnall} {et~al.}(2023){Carnall}, {McLure}, {Dunlop}, {McLeod}, {Wild}, {Cullen}, {Magee}, {Begley}, {Cimatti}, {Donnan}, {Hamadouche}, {Jewell}, \& {Walker}}]{Carnall:2023}
{Carnall}, A.~C., {McLure}, R.~J., {Dunlop}, J.~S., {et~al.} 2023, \nat, 619, 716, \dodoi{10.1038/s41586-023-06158-6}

\bibitem[{{Chabrier}(2003)}]{Chabrier:2003}
{Chabrier}, G. 2003, \pasp, 115, 763, \dodoi{10.1086/376392}

\bibitem[{{Chemerynska} {et~al.}(2024){Chemerynska}, {Atek}, {Dayal}, {Furtak}, {Feldmann}, {Greene}, {Maseda}, {Nanayakkara}, {Oesch}, {Labbe}, {Bezanson}, {Brammer}, {Cutler}, {Leja}, {Pan}, {Price}, {Wang}, {Weaver}, \& {Whitaker}}]{Chemerynska:2024}
{Chemerynska}, I., {Atek}, H., {Dayal}, P., {et~al.} 2024, arXiv e-prints, arXiv:2407.17110, \dodoi{10.48550/arXiv.2407.17110}

\bibitem[{{Chen} {et~al.}(2024){Chen}, {Ho}, {Li}, \& {Zhuang}}]{Chen:2024}
{Chen}, C.-H., {Ho}, L.~C., {Li}, R., \& {Zhuang}, M.-Y. 2024, arXiv e-prints, arXiv:2411.04446, \dodoi{10.48550/arXiv.2411.04446}

\bibitem[{{Chisholm} {et~al.}(2024){Chisholm}, {Berg}, {Endsley}, {Gazagnes}, {Richardson}, {Lambrides}, {Greene}, {Finkelstein}, {Flury}, {Guseva}, {Henry}, {Hutchison}, {Izotov}, {Marques-Chaves}, {Oesch}, {Papovich}, {Saldana-Lopez}, {Schaerer}, \& {Stephenson}}]{Chisholm:2024}
{Chisholm}, J., {Berg}, D.~A., {Endsley}, R., {et~al.} 2024, \mnras, 534, 2633, \dodoi{10.1093/mnras/stae2199}

\bibitem[{{Conroy} \& {Gunn}(2010)}]{Conroy:2010}
{Conroy}, C., \& {Gunn}, J.~E. 2010, \apj, 712, 833, \dodoi{10.1088/0004-637X/712/2/833}

\bibitem[{{Dayal}(2024)}]{Dayal:2024}
{Dayal}, P. 2024, \aap, 690, A182, \dodoi{10.1051/0004-6361/202451481}

\bibitem[{{De Breuck} {et~al.}(2000){De Breuck}, {R{\"o}ttgering}, {Miley}, {van Breugel}, \& {Best}}]{DeBreuck:2000}
{De Breuck}, C., {R{\"o}ttgering}, H., {Miley}, G., {van Breugel}, W., \& {Best}, P. 2000, \aap, 362, 519, \dodoi{10.48550/arXiv.astro-ph/0008264}

\bibitem[{{de Graaff} {et~al.}(2024{\natexlab{a}}){de Graaff}, {Brammer}, {Weibel}, {Lewis}, {Maseda}, {Oesch}, {Bezanson}, {Boogaard}, {Cleri}, {Cooper}, {Gottumukkala}, {Greene}, {Hirschmann}, {Hviding}, {Katz}, {Labb{\'e}}, {Leja}, {Matthee}, {McConachie}, {Miller}, {Naidu}, {Price}, {Rix}, {Setton}, {Suess}, {Wang}, {Whitaker}, \& {Williams}}]{Degraaff:2024RUBIES}
{de Graaff}, A., {Brammer}, G., {Weibel}, A., {et~al.} 2024{\natexlab{a}}, arXiv e-prints, arXiv:2409.05948, \dodoi{10.48550/arXiv.2409.05948}

\bibitem[{{de Graaff} {et~al.}(2024{\natexlab{b}}){de Graaff}, {Rix}, {Carniani}, {Suess}, {Charlot}, {Curtis-Lake}, {Arribas}, {Baker}, {Boyett}, {Bunker}, {Cameron}, {Chevallard}, {Curti}, {Eisenstein}, {Franx}, {Hainline}, {Hausen}, {Ji}, {Johnson}, {Jones}, {Maiolino}, {Maseda}, {Nelson}, {Parlanti}, {Rawle}, {Robertson}, {Tacchella}, {{\"U}bler}, {Williams}, {Willmer}, \& {Willott}}]{DeGraaff:2024LSF}
{de Graaff}, A., {Rix}, H.-W., {Carniani}, S., {et~al.} 2024{\natexlab{b}}, \aap, 684, A87, \dodoi{10.1051/0004-6361/202347755}

\bibitem[{{de Graaff} {et~al.}(2024{\natexlab{c}}){de Graaff}, {Setton}, {Brammer}, {Cutler}, {Suess}, {Labbe}, {Leja}, {Weibel}, {Maseda}, {Whitaker}, {Bezanson}, {Boogaard}, {Cleri}, {De Lucia}, {Franx}, {Greene}, {Hirschmann}, {Matthee}, {McConachie}, {Naidu}, {Oesch}, {Price}, {Rix}, {Valentino}, {Wang}, \& {Williams}}]{degraaff:2024}
{de Graaff}, A., {Setton}, D.~J., {Brammer}, G., {et~al.} 2024{\natexlab{c}}, arXiv e-prints, arXiv:2404.05683, \dodoi{10.48550/arXiv.2404.05683}

\bibitem[{{D'Eugenio} {et~al.}(2024){D'Eugenio}, {Cameron}, {Scholtz}, {Carniani}, {Willott}, {Curtis-Lake}, {Bunker}, {Parlanti}, {Maiolino}, {Willmer}, {Jakobsen}, {Robertson}, {Johnson}, {Tacchella}, {Cargile}, {Rawle}, {Arribas}, {Chevallard}, {Curti}, {Egami}, {Eisenstein}, {Kumari}, {Looser}, {Rieke}, {Rodr{\'\i}guez Del Pino}, {Saxena}, {{\"U}bler}, {Venturi}, {Witstok}, {Baker}, {Bhatawdekar}, {Bonaventura}, {Boyett}, {Charlot}, {Danhaive}, {Hainline}, {Hausen}, {Helton}, {Ji}, {Ji}, {Jones}, {Joud{\v{z}}balis}, {Maseda}, {P{\'e}rez-Gonz{\'a}lez}, {Perna}, {Pusk{\'a}s}, {Shivaei}, {Silcock}, {Simmonds}, {Smit}, {Sun}, {Villanueva}, {Williams}, \& {Zhu}}]{DEugenio:2024}
{D'Eugenio}, F., {Cameron}, A.~J., {Scholtz}, J., {et~al.} 2024, arXiv e-prints, arXiv:2404.06531, \dodoi{10.48550/arXiv.2404.06531}

\bibitem[{{Dong} {et~al.}(2011){Dong}, {Wang}, {Ho}, {Wang}, {Fan}, {Wang}, {Zhou}, \& {Yuan}}]{dongetal:2011}
{Dong}, X.-B., {Wang}, J.-G., {Ho}, L.~C., {et~al.} 2011, \apj, 736, 86, \dodoi{10.1088/0004-637X/736/2/86}

\bibitem[{{Duras} {et~al.}(2020){Duras}, {Bongiorno}, {Ricci}, {Piconcelli}, {Shankar}, {Lusso}, {Bianchi}, {Fiore}, {Maiolino}, {Marconi}, {Onori}, {Sani}, {Schneider}, {Vignali}, \& {La Franca}}]{Duras:2020}
{Duras}, F., {Bongiorno}, A., {Ricci}, F., {et~al.} 2020, \aap, 636, A73, \dodoi{10.1051/0004-6361/201936817}

\bibitem[{{Eilers} {et~al.}(2024){Eilers}, {Mackenzie}, {Pizzati}, {Matthee}, {Hennawi}, {Zhang}, {Bordoloi}, {Kashino}, {Lilly}, {Naidu}, {Simcoe}, {Yue}, {Frenk}, {Helly}, {Schaller}, \& {Schaye}}]{Eilers:2024}
{Eilers}, A.-C., {Mackenzie}, R., {Pizzati}, E., {et~al.} 2024, \apj, 974, 275, \dodoi{10.3847/1538-4357/ad778b}

\bibitem[{{Fan} {et~al.}(2023){Fan}, {Ba{\~n}ados}, \& {Simcoe}}]{Fan:2023}
{Fan}, X., {Ba{\~n}ados}, E., \& {Simcoe}, R.~A. 2023, \araa, 61, 373, \dodoi{10.1146/annurev-astro-052920-102455}

\bibitem[{{Feroz} {et~al.}(2010){Feroz}, {Hobson}, \& {Trotta}}]{Feroz:2010}
{Feroz}, F., {Hobson}, M.~P., \& {Trotta}, R. 2010, arXiv e-prints, arXiv:1001.0719, \dodoi{10.48550/arXiv.1001.0719}

\bibitem[{{Ferruit} {et~al.}(2022){Ferruit}, {Jakobsen}, {Giardino}, {Rawle}, {Alves de Oliveira}, {Arribas}, {Beck}, {Birkmann}, {B{\"o}ker}, {Bunker}, {Charlot}, {de Marchi}, {Franx}, {Henry}, {Karakla}, {Kassin}, {Kumari}, {L{\'o}pez-Caniego}, {L{\"u}tzgendorf}, {Maiolino}, {Manjavacas}, {Marston}, {Moseley}, {Muzerolle}, {Pirzkal}, {Rauscher}, {Rix}, {Sabbi}, {Sirianni}, {te Plate}, {Valenti}, {Willott}, \& {Zeidler}}]{Ferruit:2022}
{Ferruit}, P., {Jakobsen}, P., {Giardino}, G., {et~al.} 2022, \aap, 661, A81, \dodoi{10.1051/0004-6361/202142673}

\bibitem[{{Filippenko}(1997)}]{Filippenko:1997}
{Filippenko}, A.~V. 1997, \araa, 35, 309, \dodoi{10.1146/annurev.astro.35.1.309}

\bibitem[{{Fosbury} {et~al.}(2003){Fosbury}, {Villar-Mart{\'\i}n}, {Humphrey}, {Lombardi}, {Rosati}, {Stern}, {Hook}, {Holden}, {Stanford}, {Squires}, {Rauch}, \& {Sargent}}]{Fosbury:2003}
{Fosbury}, R.~A.~E., {Villar-Mart{\'\i}n}, M., {Humphrey}, A., {et~al.} 2003, \apj, 596, 797, \dodoi{10.1086/378228}

\bibitem[{{Fujimoto} {et~al.}(2022){Fujimoto}, {Brammer}, {Watson}, {Magdis}, {Kokorev}, {Greve}, {Toft}, {Walter}, {Valiante}, {Ginolfi}, {Schneider}, {Valentino}, {Colina}, {Vestergaard}, {Marques-Chaves}, {Fynbo}, {Krips}, {Steinhardt}, {Cortzen}, {Rizzo}, \& {Oesch}}]{Fujimoto:2022}
{Fujimoto}, S., {Brammer}, G.~B., {Watson}, D., {et~al.} 2022, \nat, 604, 261, \dodoi{10.1038/s41586-022-04454-1}

\bibitem[{{Fujimoto} {et~al.}(2023{\natexlab{a}}){Fujimoto}, {Wang}, {Weaver}, {Kokorev}, {Atek}, {Bezanson}, {Labbe}, {Brammer}, {Greene}, {Chemerynska}, {Dayal}, {de Graaff}, {Furtak}, {Oesch}, {Setton}, {Price}, {Miller}, {Williams}, {Whitaker}, {Zitrin}, {Cutler}, {Leja}, {Pan}, {Coe}, {van Dokkum}, {Feldmann}, {Fudamoto}, {Goulding}, {Khullar}, {Marchesini}, {Maseda}, {Nanayakkara}, {Nelson}, {Smit}, {Stefanon}, \& {Weibel}}]{Fujimoto:2023overdense}
{Fujimoto}, S., {Wang}, B., {Weaver}, J., {et~al.} 2023{\natexlab{a}}, arXiv e-prints, arXiv:2308.11609, \dodoi{10.48550/arXiv.2308.11609}

\bibitem[{{Fujimoto} {et~al.}(2023{\natexlab{b}}){Fujimoto}, {Kohno}, {Ouchi}, {Oguri}, {Kokorev}, {Brammer}, {Sun}, {Gonzalez-Lopez}, {Bauer}, {Caminha}, {Hatsukade}, {Richard}, {Smail}, {Tsujita}, {Ueda}, {Uematsu}, {Zitrin}, {Coe}, {Kneib}, {Postman}, {Umetsu}, {Lagos}, {Popping}, {Ao}, {Bradley}, {Caputi}, {Dessauges-Zavadsky}, {Egami}, {Espada}, {Ivison}, {Jauzac}, {Knudsen}, {Koekemoer}, {Magdis}, {Mahler}, {Munoz Arancibia}, {Rawle}, {Shimasaku}, {Toft}, {Umehata}, {Valentino}, {Wang}, \& {Wang}}]{Fujimoto:2023}
{Fujimoto}, S., {Kohno}, K., {Ouchi}, M., {et~al.} 2023{\natexlab{b}}, arXiv e-prints, arXiv:2303.01658, \dodoi{10.48550/arXiv.2303.01658}

\bibitem[{{Furtak} {et~al.}(2023{\natexlab{a}}){Furtak}, {Zitrin}, {Plat}, {Fujimoto}, {Wang}, {Nelson}, {Labb{\'e}}, {Bezanson}, {Brammer}, {van Dokkum}, {Endsley}, {Glazebrook}, {Greene}, {Leja}, {Price}, {Smit}, {Stark}, {Weaver}, {Whitaker}, {Atek}, {Chevallard}, {Curtis-Lake}, {Dayal}, {Feltre}, {Franx}, {Fudamoto}, {Marchesini}, {Mowla}, {Pan}, {Suess}, {Vidal-Garc{\'\i}a}, \& {Williams}}]{Furtak:2022}
{Furtak}, L.~J., {Zitrin}, A., {Plat}, A., {et~al.} 2023{\natexlab{a}}, \apj, 952, 142, \dodoi{10.3847/1538-4357/acdc9d}

\bibitem[{{Furtak} {et~al.}(2023{\natexlab{b}}){Furtak}, {Zitrin}, {Plat}, {Fujimoto}, {Wang}, {Nelson}, {Labb{\'e}}, {Bezanson}, {Brammer}, {van Dokkum}, {Endsley}, {Glazebrook}, {Greene}, {Leja}, {Price}, {Smit}, {Stark}, {Weaver}, {Whitaker}, {Atek}, {Chevallard}, {Curtis-Lake}, {Dayal}, {Feltre}, {Franx}, {Fudamoto}, {Marchesini}, {Mowla}, {Pan}, {Suess}, {Vidal-Garc{\'\i}a}, \& {Williams}}]{Furtak:2023}
---. 2023{\natexlab{b}}, \apj, 952, 142, \dodoi{10.3847/1538-4357/acdc9d}

\bibitem[{{Furtak} {et~al.}(2023{\natexlab{c}}){Furtak}, {Zitrin}, {Weaver}, {Atek}, {Bezanson}, {Labb{\'e}}, {Whitaker}, {Leja}, {Price}, {Brammer}, {Wang}, {Marchesini}, {Pan}, {Dayal}, {van Dokkum}, {Feldmann}, {Fujimoto}, {Franx}, {Khullar}, {Nelson}, \& {Mowla}}]{Furtak:2022b_SLmodel}
{Furtak}, L.~J., {Zitrin}, A., {Weaver}, J.~R., {et~al.} 2023{\natexlab{c}}, \mnras, 523, 4568, \dodoi{10.1093/mnras/stad1627}

\bibitem[{{Furtak} {et~al.}(2024){Furtak}, {Labb{\'e}}, {Zitrin}, {Greene}, {Dayal}, {Chemerynska}, {Kokorev}, {Miller}, {Goulding}, {de Graaff}, {Bezanson}, {Brammer}, {Cutler}, {Leja}, {Pan}, {Price}, {Wang}, {Weaver}, {Whitaker}, {Atek}, {Bogd{\'a}n}, {Charlot}, {Curtis-Lake}, {van Dokkum}, {Endsley}, {Feldmann}, {Fudamoto}, {Fujimoto}, {Glazebrook}, {Juneau}, {Marchesini}, {Maseda}, {Nelson}, {Oesch}, {Plat}, {Setton}, {Stark}, \& {Williams}}]{Furtak:2023nature}
{Furtak}, L.~J., {Labb{\'e}}, I., {Zitrin}, A., {et~al.} 2024, \nat, 628, 57, \dodoi{10.1038/s41586-024-07184-8}

\bibitem[{{Glikman} {et~al.}(2007){Glikman}, {Djorgovski}, {Stern}, {Bogosavljevi{\'c}}, \& {Mahabal}}]{Glikman:2007}
{Glikman}, E., {Djorgovski}, S.~G., {Stern}, D., {Bogosavljevi{\'c}}, M., \& {Mahabal}, A. 2007, \apjl, 663, L73, \dodoi{10.1086/520085}

\bibitem[{{Glikman} {et~al.}(2012){Glikman}, {Urrutia}, {Lacy}, {Djorgovski}, {Mahabal}, {Myers}, {Ross}, {Petitjean}, {Ge}, {Schneider}, \& {York}}]{Glikman:2012}
{Glikman}, E., {Urrutia}, T., {Lacy}, M., {et~al.} 2012, \apj, 757, 51, \dodoi{10.1088/0004-637X/757/1/51}

\bibitem[{{Glikman} {et~al.}(2023){Glikman}, {Rusu}, {Chen}, {Chan}, {Spingola}, {Stacey}, {McKean}, {Berghea}, {Djorgovski}, {Graham}, {Stern}, {Urrutia}, {Lacy}, {Secrest}, \& {O'Meara}}]{Glikman:2023}
{Glikman}, E., {Rusu}, C.~E., {Chen}, G. C.~F., {et~al.} 2023, \apj, 943, 25, \dodoi{10.3847/1538-4357/aca093}

\bibitem[{{Goulding} {et~al.}(2023){Goulding}, {Greene}, {Setton}, {Labbe}, {Bezanson}, {Miller}, {Atek}, {Bogd{\'a}n}, {Brammer}, {Chemerynska}, {Cutler}, {Dayal}, {Fudamoto}, {Fujimoto}, {Furtak}, {Kokorev}, {Khullar}, {Leja}, {Marchesini}, {Natarajan}, {Nelson}, {Oesch}, {Pan}, {Papovich}, {Price}, {van Dokkum}, {Wang}, {Weaver}, {Whitaker}, \& {Zitrin}}]{Goulding:2023}
{Goulding}, A.~D., {Greene}, J.~E., {Setton}, D.~J., {et~al.} 2023, \apjl, 955, L24, \dodoi{10.3847/2041-8213/acf7c5}

\bibitem[{{Greene} \& {Ho}(2005)}]{GreeneHo:2005}
{Greene}, J.~E., \& {Ho}, L.~C. 2005, \apj, 630, 122, \dodoi{10.1086/431897}

\bibitem[{{Greene} {et~al.}(2020){Greene}, {Strader}, \& {Ho}}]{Greene:2020}
{Greene}, J.~E., {Strader}, J., \& {Ho}, L.~C. 2020, \araa, 58, 257, \dodoi{10.1146/annurev-astro-032620-021835}

\bibitem[{{Greene} {et~al.}(2024){Greene}, {Labbe}, {Goulding}, {Furtak}, {Chemerynska}, {Kokorev}, {Dayal}, {Volonteri}, {Williams}, {Wang}, {Setton}, {Burgasser}, {Bezanson}, {Atek}, {Brammer}, {Cutler}, {Feldmann}, {Fujimoto}, {Glazebrook}, {de Graaff}, {Khullar}, {Leja}, {Marchesini}, {Maseda}, {Matthee}, {Miller}, {Naidu}, {Nanayakkara}, {Oesch}, {Pan}, {Papovich}, {Price}, {van Dokkum}, {Weaver}, {Whitaker}, \& {Zitrin}}]{Greene:2024}
{Greene}, J.~E., {Labbe}, I., {Goulding}, A.~D., {et~al.} 2024, \apj, 964, 39, \dodoi{10.3847/1538-4357/ad1e5f}

\bibitem[{{Groves} {et~al.}(2006){Groves}, {Heckman}, \& {Kauffmann}}]{grovesetal2006}
{Groves}, B.~A., {Heckman}, T.~M., \& {Kauffmann}, G. 2006, \mnras, 371, 1559, \dodoi{10.1111/j.1365-2966.2006.10812.x}

\bibitem[{{Grudi{\'c}} {et~al.}(2019){Grudi{\'c}}, {Hopkins}, {Quataert}, \& {Murray}}]{Grudic:2019}
{Grudi{\'c}}, M.~Y., {Hopkins}, P.~F., {Quataert}, E., \& {Murray}, N. 2019, \mnras, 483, 5548, \dodoi{10.1093/mnras/sty3386}

\bibitem[{{Habouzit} {et~al.}(2022){Habouzit}, {Onoue}, {Ba{\~n}ados}, {Neeleman}, {Angl{\'e}s-Alc{\'a}zar}, {Walter}, {Pillepich}, {Dav{\'e}}, {Jahnke}, \& {Dubois}}]{Habouzit:2022}
{Habouzit}, M., {Onoue}, M., {Ba{\~n}ados}, E., {et~al.} 2022, \mnras, 511, 3751, \dodoi{10.1093/mnras/stac225}

\bibitem[{{Hainline} {et~al.}(2011){Hainline}, {Shapley}, {Greene}, \& {Steidel}}]{Hainline:2011}
{Hainline}, K.~N., {Shapley}, A.~E., {Greene}, J.~E., \& {Steidel}, C.~C. 2011, \apj, 733, 31, \dodoi{10.1088/0004-637X/733/1/31}

\bibitem[{{Harikane} {et~al.}(2022){Harikane}, {Ono}, {Ouchi}, {Liu}, {Sawicki}, {Shibuya}, {Behroozi}, {He}, {Shimasaku}, {Arnouts}, {Coupon}, {Fujimoto}, {Gwyn}, {Huang}, {Inoue}, {Kashikawa}, {Komiyama}, {Matsuoka}, \& {Willott}}]{Harikane:2022}
{Harikane}, Y., {Ono}, Y., {Ouchi}, M., {et~al.} 2022, \apjs, 259, 20, \dodoi{10.3847/1538-4365/ac3dfc}

\bibitem[{Hoffman {et~al.}(2019)Hoffman, Sountsov, Dillon, Langmore, Tran, \& Vasudevan}]{Hoffman2019}
Hoffman, M., Sountsov, P., Dillon, J.~V., {et~al.} 2019, arXiv preprint arXiv:1903.03704

\bibitem[{Hoffman {et~al.}(2014)Hoffman, Gelman, {et~al.}}]{Hoffman2014}
Hoffman, M.~D., Gelman, A., {et~al.} 2014, J. Mach. Learn. Res., 15, 1593

\bibitem[{{Hopkins} {et~al.}(2010){Hopkins}, {Murray}, {Quataert}, \& {Thompson}}]{Hopkins:2010}
{Hopkins}, P.~F., {Murray}, N., {Quataert}, E., \& {Thompson}, T.~A. 2010, \mnras, 401, L19, \dodoi{10.1111/j.1745-3933.2009.00777.x}

\bibitem[{{Horne}(1986)}]{Horne:1986}
{Horne}, K. 1986, \pasp, 98, 609, \dodoi{10.1086/131801}

\bibitem[{{Humphrey} {et~al.}(2008){Humphrey}, {Villar-Mart{\'\i}n}, {Vernet}, {Fosbury}, {di Serego Alighieri}, \& {Binette}}]{Humphrey:2008}
{Humphrey}, A., {Villar-Mart{\'\i}n}, M., {Vernet}, J., {et~al.} 2008, \mnras, 383, 11, \dodoi{10.1111/j.1365-2966.2007.12506.x}

\bibitem[{{Hutchings} {et~al.}(2002){Hutchings}, {Crenshaw}, {Kraemer}, {Gabel}, {Kaiser}, {Weistrop}, \& {Gull}}]{Hutchings:2002}
{Hutchings}, J.~B., {Crenshaw}, D.~M., {Kraemer}, S.~B., {et~al.} 2002, \aj, 124, 2543, \dodoi{10.1086/344080}

\bibitem[{{Inayoshi} \& {Maiolino}(2024)}]{Inayoshi:2024}
{Inayoshi}, K., \& {Maiolino}, R. 2024, arXiv e-prints, arXiv:2409.07805, \dodoi{10.48550/arXiv.2409.07805}

\bibitem[{{Isobe} {et~al.}(2023){Isobe}, {Ouchi}, {Tominaga}, {Watanabe}, {Nakajima}, {Umeda}, {Yajima}, {Harikane}, {Fukushima}, {Xu}, {Ono}, \& {Zhang}}]{Isobe:2023}
{Isobe}, Y., {Ouchi}, M., {Tominaga}, N., {et~al.} 2023, \apj, 959, 100, \dodoi{10.3847/1538-4357/ad09be}

\bibitem[{{Jakobsen} {et~al.}(2022){Jakobsen}, {Ferruit}, {Alves de Oliveira}, {Arribas}, {Bagnasco}, {Barho}, {Beck}, {Birkmann}, {B{\"o}ker}, {Bunker}, {Charlot}, {de Jong}, {de Marchi}, {Ehrenwinkler}, {Falcolini}, {Fels}, {Franx}, {Franz}, {Funke}, {Giardino}, {Gnata}, {Holota}, {Honnen}, {Jensen}, {Jentsch}, {Johnson}, {Jollet}, {Karl}, {Kling}, {K{\"o}hler}, {Kolm}, {Kumari}, {Lander}, {Lemke}, {L{\'o}pez-Caniego}, {L{\"u}tzgendorf}, {Maiolino}, {Manjavacas}, {Marston}, {Maschmann}, {Maurer}, {Messerschmidt}, {Moseley}, {Mosner}, {Mott}, {Muzerolle}, {Pirzkal}, {Pittet}, {Plitzke}, {Posselt}, {Rapp}, {Rauscher}, {Rawle}, {Rix}, {R{\"o}del}, {Rumler}, {Sabbi}, {Salvignol}, {Schmid}, {Sirianni}, {Smith}, {Strada}, {te Plate}, {Valenti}, {Wettemann}, {Wiehe}, {Wiesmayer}, {Willott}, {Wright}, {Zeidler}, \& {Zincke}}]{Jakobsen:2022}
{Jakobsen}, P., {Ferruit}, P., {Alves de Oliveira}, C., {et~al.} 2022, \aap, 661, A80, \dodoi{10.1051/0004-6361/202142663}

\bibitem[{{Jiang} {et~al.}(2008){Jiang}, {Fan}, \& {Vestergaard}}]{Jiang:2008}
{Jiang}, L., {Fan}, X., \& {Vestergaard}, M. 2008, \apj, 679, 962, \dodoi{10.1086/587868}

\bibitem[{{Johnson} \& {Khochfar}(2011)}]{Johnson:2011}
{Johnson}, J.~L., \& {Khochfar}, S. 2011, \apj, 743, 126, \dodoi{10.1088/0004-637X/743/2/126}

\bibitem[{{Juod{\v{z}}balis} {et~al.}(2024){Juod{\v{z}}balis}, {Ji}, {Maiolino}, {D'Eugenio}, {Scholtz}, {Risaliti}, {Fabian}, {Mazzolari}, {Gilli}, {Prandoni}, {Arribas}, {Bunker}, {Carniani}, {Charlot}, {Curtis-Lake}, {de Graaff}, {Hainline}, {Parlanti}, {Perna}, {P{\'e}rez-Gonz{\'a}lez}, {Robertson}, {Tacchella}, {{\"U}bler}, {Williams}, {Willott}, \& {Witstok}}]{Juodzbalis:2024}
{Juod{\v{z}}balis}, I., {Ji}, X., {Maiolino}, R., {et~al.} 2024, arXiv e-prints, arXiv:2407.08643, \dodoi{10.48550/arXiv.2407.08643}

\bibitem[{{Kashino} {et~al.}(2023){Kashino}, {Lilly}, {Matthee}, {Eilers}, {Mackenzie}, {Bordoloi}, \& {Simcoe}}]{Kashino:2023}
{Kashino}, D., {Lilly}, S.~J., {Matthee}, J., {et~al.} 2023, \apj, 950, 66, \dodoi{10.3847/1538-4357/acc588}

\bibitem[{{Kennicutt} \& {Evans}(2012)}]{KennicuttEvans:2012}
{Kennicutt}, R.~C., \& {Evans}, N.~J. 2012, \araa, 50, 531, \dodoi{10.1146/annurev-astro-081811-125610}

\bibitem[{{Killi} {et~al.}(2024){Killi}, {Watson}, {Brammer}, {McPartland}, {Antwi-Danso}, {Newshore}, {Coe}, {Allen}, {Fynbo}, {Gould}, {Heintz}, {Rusakov}, \& {Vejlgaard}}]{Killi:2024}
{Killi}, M., {Watson}, D., {Brammer}, G., {et~al.} 2024, \aap, 691, A52, \dodoi{10.1051/0004-6361/202348857}

\bibitem[{{Kocevski} {et~al.}(2023){Kocevski}, {Onoue}, {Inayoshi}, {Trump}, {Arrabal Haro}, {Grazian}, {Dickinson}, {Finkelstein}, {Kartaltepe}, {Hirschmann}, {Aird}, {Holwerda}, {Fujimoto}, {Juneau}, {Amor{\'\i}n}, {Backhaus}, {Bagley}, {Barro}, {Bell}, {Bisigello}, {Calabr{\`o}}, {Cleri}, {Cooper}, {Ding}, {Grogin}, {Ho}, {Hutchison}, {Inoue}, {Jiang}, {Jones}, {Koekemoer}, {Li}, {Li}, {McGrath}, {Molina}, {Papovich}, {P{\'e}rez-Gonz{\'a}lez}, {Pirzkal}, {Wilkins}, {Yang}, \& {Yung}}]{Kocevski:2023}
{Kocevski}, D.~D., {Onoue}, M., {Inayoshi}, K., {et~al.} 2023, \apjl, 954, L4, \dodoi{10.3847/2041-8213/ace5a0}

\bibitem[{{Kocevski} {et~al.}(2024){Kocevski}, {Finkelstein}, {Barro}, {Taylor}, {Calabr{\`o}}, {Laloux}, {Buchner}, {Trump}, {Leung}, {Yang}, {Dickinson}, {P{\'e}rez-Gonz{\'a}lez}, {Pacucci}, {Inayoshi}, {Somerville}, {McGrath}, {Akins}, {Bagley}, {Bisigello}, {Bowler}, {Carnall}, {Casey}, {Cheng}, {Cleri}, {Costantin}, {Cullen}, {Davis}, {Donnan}, {Dunlop}, {Ellis}, {Ferguson}, {Fujimoto}, {Fontana}, {Giavalisco}, {Grazian}, {Grogin}, {Hathi}, {Hirschmann}, {Huertas-Company}, {Holwerda}, {Illingworth}, {Juneau}, {Kartaltepe}, {Koekemoer}, {Li}, {Lucas}, {Magee}, {Mason}, {McLeod}, {McLure}, {Napolitano}, {Papovich}, {Pirzkal}, {Rodighiero}, {Santini}, {Wilkins}, \& {Yung}}]{Kocevski:2024}
{Kocevski}, D.~D., {Finkelstein}, S.~L., {Barro}, G., {et~al.} 2024, arXiv e-prints, arXiv:2404.03576, \dodoi{10.48550/arXiv.2404.03576}

\bibitem[{{Kochanek}(2016)}]{Kochanek:2016}
{Kochanek}, C.~S. 2016, \mnras, 458, 127, \dodoi{10.1093/mnras/stw267}

\bibitem[{{Kokorev} {et~al.}(2023){Kokorev}, {Fujimoto}, {Labbe}, {Greene}, {Bezanson}, {Dayal}, {Nelson}, {Atek}, {Brammer}, {Caputi}, {Chemerynska}, {Cutler}, {Feldmann}, {Fudamoto}, {Furtak}, {Goulding}, {de Graaff}, {Leja}, {Marchesini}, {Miller}, {Nanayakkara}, {Oesch}, {Pan}, {Price}, {Setton}, {Smit}, {Stefanon}, {Wang}, {Weaver}, {Whitaker}, {Williams}, \& {Zitrin}}]{Kokorev:2023}
{Kokorev}, V., {Fujimoto}, S., {Labbe}, I., {et~al.} 2023, \apjl, 957, L7, \dodoi{10.3847/2041-8213/ad037a}

\bibitem[{{Kokorev} {et~al.}(2024){Kokorev}, {Caputi}, {Greene}, {Dayal}, {Trebitsch}, {Cutler}, {Fujimoto}, {Labb{\'e}}, {Miller}, {Iani}, {Navarro-Carrera}, \& {Rinaldi}}]{Kokorev:2024}
{Kokorev}, V., {Caputi}, K.~I., {Greene}, J.~E., {et~al.} 2024, \apj, 968, 38, \dodoi{10.3847/1538-4357/ad4265}

\bibitem[{{Kokubo} \& {Harikane}(2024)}]{Kokubo:2024}
{Kokubo}, M., \& {Harikane}, Y. 2024, arXiv e-prints, arXiv:2407.04777, \dodoi{10.48550/arXiv.2407.04777}

\bibitem[{{Kova{\v{c}}evi{\'c}} {et~al.}(2010){Kova{\v{c}}evi{\'c}}, {Popovi{\'c}}, \& {Dimitrijevi{\'c}}}]{Kovacevic:2010}
{Kova{\v{c}}evi{\'c}}, J., {Popovi{\'c}}, L.~{\v{C}}., \& {Dimitrijevi{\'c}}, M.~S. 2010, \apjs, 189, 15, \dodoi{10.1088/0067-0049/189/1/15}

\bibitem[{{Kraemer} \& {Crenshaw}(2000)}]{Kraemer:2000}
{Kraemer}, S.~B., \& {Crenshaw}, D.~M. 2000, \apj, 532, 256, \dodoi{10.1086/308572}

\bibitem[{{Kriek} {et~al.}(2008){Kriek}, {van Dokkum}, {Franx}, {Illingworth}, {Marchesini}, {Quadri}, {Rudnick}, {Taylor}, {F{\"o}rster Schreiber}, {Gawiser}, {Labb{\'e}}, {Lira}, \& {Wuyts}}]{Kriek:2008}
{Kriek}, M., {van Dokkum}, P.~G., {Franx}, M., {et~al.} 2008, \apj, 677, 219, \dodoi{10.1086/528945}

\bibitem[{{Krolik}(2001)}]{krolik2001}
{Krolik}, J.~H. 2001, \apj, 551, 72, \dodoi{10.1086/320091}

\bibitem[{{Labb{\'e}} {et~al.}(2023){Labb{\'e}}, {Greene}, {Bezanson}, {Fujimoto}, {Furtak}, {Goulding}, {Matthee}, {Naidu}, {Oesch}, {Atek}, {Brammer}, {Chemerynska}, {Coe}, {Cutler}, {Dayal}, {Feldmann}, {Franx}, {Glazebrook}, {Leja}, {Marchesini}, {Maseda}, {Nanayakkara}, {Nelson}, {Pan}, {Papovich}, {Price}, {Suess}, {Wang}, {Whitaker}, {Williams}, \& {Zitrin}}]{Labbe:2023uncover}
{Labb{\'e}}, I., {Greene}, J.~E., {Bezanson}, R., {et~al.} 2023, arXiv e-prints, arXiv:2306.07320, \dodoi{10.48550/arXiv.2306.07320}

\bibitem[{{Langeroodi} \& {Hjorth}(2023)}]{Langeroodi:2023}
{Langeroodi}, D., \& {Hjorth}, J. 2023, \apjl, 957, L27, \dodoi{10.3847/2041-8213/acfeec}

\bibitem[{{Laor} \& {Davis}(2011)}]{Laor:2011}
{Laor}, A., \& {Davis}, S.~W. 2011, \mnras, 417, 681, \dodoi{10.1111/j.1365-2966.2011.19310.x}

\bibitem[{{Larson} {et~al.}(2023){Larson}, {Finkelstein}, {Kocevski}, {Hutchison}, {Trump}, {Arrabal Haro}, {Bromm}, {Cleri}, {Dickinson}, {Fujimoto}, {Kartaltepe}, {Koekemoer}, {Papovich}, {Pirzkal}, {Tacchella}, {Zavala}, {Bagley}, {Behroozi}, {Champagne}, {Cole}, {Jung}, {Morales}, {Yang}, {Zhang}, {Zitrin}, {Amor{\'\i}n}, {Burgarella}, {Casey}, {Ch{\'a}vez Ortiz}, {Cox}, {Chworowsky}, {Fontana}, {Gawiser}, {Grazian}, {Grogin}, {Harish}, {Hathi}, {Hirschmann}, {Holwerda}, {Juneau}, {Leung}, {Lucas}, {McGrath}, {P{\'e}rez-Gonz{\'a}lez}, {Rigby}, {Seill{\'e}}, {Simons}, {de La Vega}, {Weiner}, {Wilkins}, {Yung}, \& {Ceers Team}}]{Larson:2023}
{Larson}, R.~L., {Finkelstein}, S.~L., {Kocevski}, D.~D., {et~al.} 2023, \apjl, 953, L29, \dodoi{10.3847/2041-8213/ace619}

\bibitem[{{Le F{\`e}vre} {et~al.}(2019){Le F{\`e}vre}, {Lemaux}, {Nakajima}, {Schaerer}, {Talia}, {Zamorani}, {Cassata}, {Garilli}, {Maccagni}, {Pentericci}, {Tasca}, {Zucca}, {Amorin}, {Bardelli}, {Cimatti}, {Giavalisco}, {Guaita}, {Hathi}, {Marchi}, {Vanzella}, {Vergani}, \& {Dunlop}}]{LeFevre:2019}
{Le F{\`e}vre}, O., {Lemaux}, B.~C., {Nakajima}, K., {et~al.} 2019, \aap, 625, A51, \dodoi{10.1051/0004-6361/201732197}

\bibitem[{{Leja} {et~al.}(2019){Leja}, {Carnall}, {Johnson}, {Conroy}, \& {Speagle}}]{Leja:2019}
{Leja}, J., {Carnall}, A.~C., {Johnson}, B.~D., {Conroy}, C., \& {Speagle}, J.~S. 2019, \apj, 876, 3, \dodoi{10.3847/1538-4357/ab133c}

\bibitem[{{Lotz} {et~al.}(2017){Lotz}, {Koekemoer}, {Coe}, {Grogin}, {Capak}, {Mack}, {Anderson}, {Avila}, {Barker}, {Borncamp}, {Brammer}, {Durbin}, {Gunning}, {Hilbert}, {Jenkner}, {Khandrika}, {Levay}, {Lucas}, {MacKenty}, {Ogaz}, {Porterfield}, {Reid}, {Robberto}, {Royle}, {Smith}, {Storrie-Lombardi}, {Sunnquist}, {Surace}, {Taylor}, {Williams}, {Bullock}, {Dickinson}, {Finkelstein}, {Natarajan}, {Richard}, {Robertson}, {Tumlinson}, {Zitrin}, {Flanagan}, {Sembach}, {Soifer}, \& {Mountain}}]{Lotz:2017}
{Lotz}, J.~M., {Koekemoer}, A., {Coe}, D., {et~al.} 2017, \apj, 837, 97, \dodoi{10.3847/1538-4357/837/1/97}

\bibitem[{{Ma} {et~al.}(2024){Ma}, {Greene}, {Setton}, {Volonteri}, {Leja}, {Wang}, {Bezanson}, {Brammer}, {Cutler}, {Dayal}, {van Dokkum}, {Furtak}, {Glazebrook}, {Goulding}, {de Graaff}, {Kokorev}, {Labbe}, {Pan}, {Price}, {Weaver}, {Williams}, {Whitaker}, \& {Zitrin}}]{Ma:2024lens}
{Ma}, Y., {Greene}, J.~E., {Setton}, D.~J., {et~al.} 2024, arXiv e-prints, arXiv:2410.06257, \dodoi{10.48550/arXiv.2410.06257}

\bibitem[{{Maiolino} {et~al.}(2023){Maiolino}, {Scholtz}, {Curtis-Lake}, {Carniani}, {Baker}, {de Graaff}, {Tacchella}, {{\"U}bler}, {D'Eugenio}, {Witstok}, {Curti}, {Arribas}, {Bunker}, {Charlot}, {Chevallard}, {Eisenstein}, {Egami}, {Ji}, {Jones}, {Lyu}, {Rawle}, {Robertson}, {Rujopakarn}, {Perna}, {Sun}, {Venturi}, {Williams}, \& {Willott}}]{Maiolino:2023}
{Maiolino}, R., {Scholtz}, J., {Curtis-Lake}, E., {et~al.} 2023, arXiv e-prints, arXiv:2308.01230, \dodoi{10.48550/arXiv.2308.01230}

\bibitem[{{Maiolino} {et~al.}(2024{\natexlab{a}}){Maiolino}, {Risaliti}, {Signorini}, {Trefoloni}, {Juodzbalis}, {Scholtz}, {Uebler}, {D'Eugenio}, {Carniani}, {Fabian}, {Ji}, {Mazzolari}, {Bertola}, {Brusa}, {Bunker}, {Charlot}, {Comastri}, {Cresci}, {DeCoursey}, {Egami}, {Fiore}, {Gilli}, {Perna}, {Tacchella}, \& {Venturi}}]{Maiolino:2024Xray}
{Maiolino}, R., {Risaliti}, G., {Signorini}, M., {et~al.} 2024{\natexlab{a}}, arXiv e-prints, arXiv:2405.00504, \dodoi{10.48550/arXiv.2405.00504}

\bibitem[{{Maiolino} {et~al.}(2024{\natexlab{b}}){Maiolino}, {Scholtz}, {Witstok}, {Carniani}, {D'Eugenio}, {de Graaff}, {{\"U}bler}, {Tacchella}, {Curtis-Lake}, {Arribas}, {Bunker}, {Charlot}, {Chevallard}, {Curti}, {Looser}, {Maseda}, {Rawle}, {Rodr{\'\i}guez del Pino}, {Willott}, {Egami}, {Eisenstein}, {Hainline}, {Robertson}, {Williams}, {Willmer}, {Baker}, {Boyett}, {DeCoursey}, {Fabian}, {Helton}, {Ji}, {Jones}, {Kumari}, {Laporte}, {Nelson}, {Perna}, {Sandles}, {Shivaei}, \& {Sun}}]{Maiolino:2024gnz11}
{Maiolino}, R., {Scholtz}, J., {Witstok}, J., {et~al.} 2024{\natexlab{b}}, \nat, 627, 59, \dodoi{10.1038/s41586-024-07052-5}

\bibitem[{{Marques-Chaves} {et~al.}(2024){Marques-Chaves}, {Schaerer}, {Kuruvanthodi}, {Korber}, {Prantzos}, {Charbonnel}, {Weibel}, {Izotov}, {Messa}, {Brammer}, {Dessauges-Zavadsky}, \& {Oesch}}]{Marques-Chaves:2024}
{Marques-Chaves}, R., {Schaerer}, D., {Kuruvanthodi}, A., {et~al.} 2024, \aap, 681, A30, \dodoi{10.1051/0004-6361/202347411}

\bibitem[{{Martocchia} {et~al.}(2017){Martocchia}, {Piconcelli}, {Zappacosta}, {Duras}, {Vietri}, {Vignali}, {Bianchi}, {Bischetti}, {Bongiorno}, {Brusa}, {Lanzuisi}, {Marconi}, {Mathur}, {Miniutti}, {Nicastro}, {Bruni}, \& {Fiore}}]{Martocchia:2017}
{Martocchia}, S., {Piconcelli}, E., {Zappacosta}, L., {et~al.} 2017, \aap, 608, A51, \dodoi{10.1051/0004-6361/201731314}

\bibitem[{{Matthee} {et~al.}(2024{\natexlab{a}}){Matthee}, {Naidu}, {Brammer}, {Chisholm}, {Eilers}, {Goulding}, {Greene}, {Kashino}, {Labbe}, {Lilly}, {Mackenzie}, {Oesch}, {Weibel}, {Wuyts}, {Xiao}, {Bordoloi}, {Bouwens}, {van Dokkum}, {Illingworth}, {Kramarenko}, {Maseda}, {Mason}, {Meyer}, {Nelson}, {Reddy}, {Shivaei}, {Simcoe}, \& {Yue}}]{Matthee:2023}
{Matthee}, J., {Naidu}, R.~P., {Brammer}, G., {et~al.} 2024{\natexlab{a}}, \apj, 963, 129, \dodoi{10.3847/1538-4357/ad2345}

\bibitem[{{Matthee} {et~al.}(2024{\natexlab{b}}){Matthee}, {Brammer}, {Weibel}, {Lewis}, {Maseda}, {Oesch}, {Bezanson}, {Boogaard}, {Cleri}, {Cooper}, {Gottumukkala}, {Greene}, {Hirschmann}, {Hviding}, {Katz}, {Labb{\'e}}, {Leja}, {Matthee}, {McConachie}, {Miller}, {Naidu}, {Price}, {Rix}, {Setton}, {Suess}, {Wang}, {Whitaker}, \& {Williams}}]{Matthee:2024}
{Matthee}, J., {Brammer}, G., {Weibel}, A., {et~al.} 2024{\natexlab{b}}, arXiv e-prints, arXiv:2412.02846, \dodoi{10.48550/arXiv.2412.02846}

\bibitem[{{McConnell} \& {Ma}(2013)}]{mcconnellma2013}
{McConnell}, N.~J., \& {Ma}, C.-P. 2013, \apj, 764, 184, \dodoi{10.1088/0004-637X/764/2/184}

\bibitem[{{Murray} {et~al.}(2005){Murray}, {Quataert}, \& {Thompson}}]{Murray:2005}
{Murray}, N., {Quataert}, E., \& {Thompson}, T.~A. 2005, \apj, 618, 569, \dodoi{10.1086/426067}

\bibitem[{{Naidu} {et~al.}(2024){Naidu}, {Matthee}, {Kramarenko}, {Weibel}, {Brammer}, {Oesch}, {Lechner}, {Furtak}, {Di Cesare}, {Torralba}, {Kotiwale}, {Bezanson}, {Bouwens}, {Chandra}, {Claeyssens}, {Danhaive}, {Frebel}, {de Graaff}, {Greene}, {Heintz}, {Ji}, {Kashino}, {Katz}, {Labbe}, {Leja}, {Li}, {Maseda}, {Richard}, {Shivaei}, {Simcoe}, {Sobral}, {Suess}, {Tacchella}, \& {Williams}}]{Naidu24}
{Naidu}, R.~P., {Matthee}, J., {Kramarenko}, I., {et~al.} 2024, arXiv e-prints, arXiv:2410.01874, \dodoi{10.48550/arXiv.2410.01874}

\bibitem[{{Norris} {et~al.}(2014){Norris}, {Kannappan}, {Forbes}, {Romanowsky}, {Brodie}, {Faifer}, {Huxor}, {Maraston}, {Moffett}, {Penny}, {Pota}, {Smith-Castelli}, {Strader}, {Bradley}, {Eckert}, {Fohring}, {McBride}, {Stark}, \& {Vaduvescu}}]{Norris:2014}
{Norris}, M.~A., {Kannappan}, S.~J., {Forbes}, D.~A., {et~al.} 2014, \mnras, 443, 1151, \dodoi{10.1093/mnras/stu1186}

\bibitem[{{Oke} \& {Gunn}(1983)}]{oke1983}
{Oke}, J.~B., \& {Gunn}, J.~E. 1983, \apj, 266, 713, \dodoi{10.1086/160817}

\bibitem[{{Osterbrock}(1977)}]{Osterbrock:1977}
{Osterbrock}, D.~E. 1977, \apj, 215, 733, \dodoi{10.1086/155407}

\bibitem[{{Pacucci} {et~al.}(2023){Pacucci}, {Nguyen}, {Carniani}, {Maiolino}, \& {Fan}}]{Pacucci:2023}
{Pacucci}, F., {Nguyen}, B., {Carniani}, S., {Maiolino}, R., \& {Fan}, X. 2023, \apjl, 957, L3, \dodoi{10.3847/2041-8213/ad0158}

\bibitem[{{Pasha} \& {Miller}(2023)}]{Pasha2023}
{Pasha}, I., \& {Miller}, T.~B. 2023, The Journal of Open Source Software, 8, 5703, \dodoi{10.21105/joss.05703}

\bibitem[{{P{\'e}rez-Gonz{\'a}lez} {et~al.}(2024){P{\'e}rez-Gonz{\'a}lez}, {Barro}, {Rieke}, {Lyu}, {Rieke}, {Alberts}, {Williams}, {Hainline}, {Sun}, {Pusk{\'a}s}, {Annunziatella}, {Baker}, {Bunker}, {Egami}, {Ji}, {Johnson}, {Robertson}, {Rodr{\'\i}guez Del Pino}, {Rujopakarn}, {Shivaei}, {Tacchella}, {Willmer}, \& {Willott}}]{PerezGonzalez:2024}
{P{\'e}rez-Gonz{\'a}lez}, P.~G., {Barro}, G., {Rieke}, G.~H., {et~al.} 2024, \apj, 968, 4, \dodoi{10.3847/1538-4357/ad38bb}

\bibitem[{Phan {et~al.}(2019)Phan, Pradhan, \& Jankowiak}]{Phan2019}
Phan, D., Pradhan, N., \& Jankowiak, M. 2019, arXiv preprint arXiv:1912.11554

\bibitem[{{Pizzati} {et~al.}(2024){Pizzati}, {Hennawi}, {Schaye}, {Eilers}, {Huang}, {Schindler}, \& {Wang}}]{Pizzati:2024}
{Pizzati}, E., {Hennawi}, J.~F., {Schaye}, J., {et~al.} 2024, arXiv e-prints, arXiv:2409.18208, \dodoi{10.48550/arXiv.2409.18208}

\bibitem[{{Planck Collaboration} {et~al.}(2020){Planck Collaboration}, {Aghanim}, {Akrami}, {Ashdown}, {Aumont}, {Baccigalupi}, {Ballardini}, {Banday}, {Barreiro}, {Bartolo}, {Basak}, {Battye}, {Benabed}, {Bernard}, {Bersanelli}, {Bielewicz}, {Bock}, {Bond}, {Borrill}, {Bouchet}, {Boulanger}, {Bucher}, {Burigana}, {Butler}, {Calabrese}, {Cardoso}, {Carron}, {Challinor}, {Chiang}, {Chluba}, {Colombo}, {Combet}, {Contreras}, {Crill}, {Cuttaia}, {de Bernardis}, {de Zotti}, {Delabrouille}, {Delouis}, {Di Valentino}, {Diego}, {Dor{\'e}}, {Douspis}, {Ducout}, {Dupac}, {Dusini}, {Efstathiou}, {Elsner}, {En{\ss}lin}, {Eriksen}, {Fantaye}, {Farhang}, {Fergusson}, {Fernandez-Cobos}, {Finelli}, {Forastieri}, {Frailis}, {Fraisse}, {Franceschi}, {Frolov}, {Galeotta}, {Galli}, {Ganga}, {G{\'e}nova-Santos}, {Gerbino}, {Ghosh}, {Gonz{\'a}lez-Nuevo}, {G{\'o}rski}, {Gratton}, {Gruppuso}, {Gudmundsson}, {Hamann}, {Handley}, {Hansen}, {Herranz}, {Hildebrandt}, {Hivon}, {Huang}, {Jaffe}, {Jones}, {Karakci}, {Keih{\"a}nen},
  {Keskitalo}, {Kiiveri}, {Kim}, {Kisner}, {Knox}, {Krachmalnicoff}, {Kunz}, {Kurki-Suonio}, {Lagache}, {Lamarre}, {Lasenby}, {Lattanzi}, {Lawrence}, {Le Jeune}, {Lemos}, {Lesgourgues}, {Levrier}, {Lewis}, {Liguori}, {Lilje}, {Lilley}, {Lindholm}, {L{\'o}pez-Caniego}, {Lubin}, {Ma}, {Mac{\'\i}as-P{\'e}rez}, {Maggio}, {Maino}, {Mandolesi}, {Mangilli}, {Marcos-Caballero}, {Maris}, {Martin}, {Martinelli}, {Mart{\'\i}nez-Gonz{\'a}lez}, {Matarrese}, {Mauri}, {McEwen}, {Meinhold}, {Melchiorri}, {Mennella}, {Migliaccio}, {Millea}, {Mitra}, {Miville-Desch{\^e}nes}, {Molinari}, {Montier}, {Morgante}, {Moss}, {Natoli}, {N{\o}rgaard-Nielsen}, {Pagano}, {Paoletti}, {Partridge}, {Patanchon}, {Peiris}, {Perrotta}, {Pettorino}, {Piacentini}, {Polastri}, {Polenta}, {Puget}, {Rachen}, {Reinecke}, {Remazeilles}, {Renzi}, {Rocha}, {Rosset}, {Roudier}, {Rubi{\~n}o-Mart{\'\i}n}, {Ruiz-Granados}, {Salvati}, {Sandri}, {Savelainen}, {Scott}, {Shellard}, {Sirignano}, {Sirri}, {Spencer}, {Sunyaev}, {Suur-Uski}, {Tauber}, {Tavagnacco},
  {Tenti}, {Toffolatti}, {Tomasi}, {Trombetti}, {Valenziano}, {Valiviita}, {Van Tent}, {Vibert}, {Vielva}, {Villa}, {Vittorio}, {Wandelt}, {Wehus}, {White}, {White}, {Zacchei}, \& {Zonca}}]{planck20}
{Planck Collaboration}, {Aghanim}, N., {Akrami}, Y., {et~al.} 2020, \aap, 641, A6, \dodoi{10.1051/0004-6361/201833910}

\bibitem[{{Price} {et~al.}(2023){Price}, {Suess}, {Williams}, {Bezanson}, {Khullar}, {Nelson}, {Wang}, {Weaver}, {Fujimoto}, {Kokorev}, {Greene}, {Brammer}, {Cutler}, {Dayal}, {Furtak}, {Labbe}, {Leja}, {Miller}, {Nanayakkara}, {Pan}, \& {Whitaker}}]{Price:2023}
{Price}, S.~H., {Suess}, K.~A., {Williams}, C.~C., {et~al.} 2023, arXiv e-prints, arXiv:2310.02500, \dodoi{10.48550/arXiv.2310.02500}

\bibitem[{{Price} {et~al.}(2024){Price}, {Bezanson}, {Labbe}, {Furtak}, {de Graaff}, {Greene}, {Kokorev}, {Setton}, {Suess}, {Brammer}, {Cutler}, {Leja}, {Pan}, {Wang}, {Weaver}, {Whitaker}, {Atek}, {Burgasser}, {Chemerynska}, {Dayal}, {Feldmann}, {F{\"o}rster Schreiber}, {Fudamoto}, {Fujimoto}, {Glazebrook}, {Goulding}, {Khullar}, {Kriek}, {Marchesini}, {Maseda}, {Miller}, {Muzzin}, {Nanayakkara}, {Nelson}, {Oesch}, {Shipley}, {Smit}, {Taylor}, {van Dokkum}, {Williams}, \& {Zitrin}}]{Price:2024}
{Price}, S.~H., {Bezanson}, R., {Labbe}, I., {et~al.} 2024, arXiv e-prints, arXiv:2408.03920, \dodoi{10.48550/arXiv.2408.03920}

\bibitem[{{Raiter} {et~al.}(2010){Raiter}, {Fosbury}, \& {Teimoorinia}}]{Raiter:2010}
{Raiter}, A., {Fosbury}, R.~A.~E., \& {Teimoorinia}, H. 2010, \aap, 510, A109, \dodoi{10.1051/0004-6361/200912429}

\bibitem[{{Reddy} {et~al.}(2018){Reddy}, {Shapley}, {Sanders}, {Kriek}, {Coil}, {Shivaei}, {Freeman}, {Mobasher}, {Siana}, {Azadi}, {Fetherolf}, {Fornasini}, {Leung}, {Price}, {Zick}, \& {Barro}}]{Reddy:2018}
{Reddy}, N.~A., {Shapley}, A.~E., {Sanders}, R.~L., {et~al.} 2018, \apj, 869, 92, \dodoi{10.3847/1538-4357/aaed1e}

\bibitem[{{Ricci} \& {Trakhtenbrot}(2023)}]{Ricci:2023}
{Ricci}, C., \& {Trakhtenbrot}, B. 2023, Nature Astronomy, 7, 1282, \dodoi{10.1038/s41550-023-02108-4}

\bibitem[{{Rodriguez} {et~al.}(2002){Rodriguez}, {Varni{\`e}re}, {Tagger}, \& {Durouchoux}}]{rodriguezetal:2002}
{Rodriguez}, J., {Varni{\`e}re}, P., {Tagger}, M., \& {Durouchoux}, P. 2002, \aap, 387, 487, \dodoi{10.1051/0004-6361:20000524}

\bibitem[{{Salviander} {et~al.}(2006){Salviander}, {Shields}, {Gebhardt}, \& {Bonning}}]{Salviander:2006}
{Salviander}, S., {Shields}, G.~A., {Gebhardt}, K., \& {Bonning}, E.~W. 2006, \nar, 50, 803, \dodoi{10.1016/j.newar.2006.06.018}

\bibitem[{{Scholtz} {et~al.}(2023){Scholtz}, {Maiolino}, {D'Eugenio}, {Curtis-Lake}, {Carniani}, {Charlot}, {Curti}, {Silcock}, {Arribas}, {Baker}, {Bhatawdekar}, {Boyett}, {Bunker}, {Chevallard}, {Circosta}, {Eisenstein}, {Hainline}, {Hausen}, {Ji}, {Ji}, {Johnson}, {Kumari}, {Looser}, {Lyu}, {Maseda}, {Parlanti}, {Perna}, {Rieke}, {Robertson}, {Rodr{\'\i}guez Del Pino}, {Sun}, {Tacchella}, {{\"U}bler}, {Venturi}, {Williams}, {Willmer}, {Willott}, \& {Witstok}}]{Scholtz:2024}
{Scholtz}, J., {Maiolino}, R., {D'Eugenio}, F., {et~al.} 2023, arXiv e-prints, arXiv:2311.18731, \dodoi{10.48550/arXiv.2311.18731}

\bibitem[{{Schulze} {et~al.}(2018){Schulze}, {Misawa}, {Zuo}, \& {Wu}}]{Schulze:2018}
{Schulze}, A., {Misawa}, T., {Zuo}, W., \& {Wu}, X.-B. 2018, \apj, 853, 167, \dodoi{10.3847/1538-4357/aaa7f0}

\bibitem[{{Senchyna} {et~al.}(2024){Senchyna}, {Plat}, {Stark}, {Rudie}, {Berg}, {Charlot}, {James}, \& {Mingozzi}}]{Senchyna:2024}
{Senchyna}, P., {Plat}, A., {Stark}, D.~P., {et~al.} 2024, \apj, 966, 92, \dodoi{10.3847/1538-4357/ad235e}

\bibitem[{{Setton} {et~al.}(2024){Setton}, {Greene}, {de Graaff}, {Ma}, {Leja}, {Matthee}, {Bezanson}, {Boogaard}, {Cleri}, {Katz}, {Labbe}, {Maseda}, {McConachie}, {Miller}, {Price}, {Suess}, {van Dokkum}, {Wang}, {Weibel}, {Whitaker}, \& {Williams}}]{Setton:2024break}
{Setton}, D.~J., {Greene}, J.~E., {de Graaff}, A., {et~al.} 2024, arXiv e-prints, arXiv:2411.03424.
\newblock \doarXiv{2411.03424}

\bibitem[{{Shen}(2013)}]{Shen:2013}
{Shen}, Y. 2013, Bulletin of the Astronomical Society of India, 41, 61, \dodoi{10.48550/arXiv.1302.2643}

\bibitem[{{Shen} {et~al.}(2019){Shen}, {Wu}, {Jiang}, {Ba{\~n}ados}, {Fan}, {Ho}, {Riechers}, {Strauss}, {Venemans}, {Vestergaard}, {Walter}, {Wang}, {Willott}, {Wu}, \& {Yang}}]{Shen:2019}
{Shen}, Y., {Wu}, J., {Jiang}, L., {et~al.} 2019, \apj, 873, 35, \dodoi{10.3847/1538-4357/ab03d9}

\bibitem[{{Shi} {et~al.}(2024){Shi}, {Kremer}, \& {Hopkins}}]{Shi:2024}
{Shi}, Y., {Kremer}, K., \& {Hopkins}, P.~F. 2024, \aap, 691, A24, \dodoi{10.1051/0004-6361/202450964}

\bibitem[{{Sobral} {et~al.}(2018){Sobral}, {Matthee}, {Darvish}, {Smail}, {Best}, {Alegre}, {R{\"o}ttgering}, {Mobasher}, {Paulino-Afonso}, {Stroe}, \& {Oteo}}]{Sobral:2018}
{Sobral}, D., {Matthee}, J., {Darvish}, B., {et~al.} 2018, \mnras, 477, 2817, \dodoi{10.1093/mnras/sty782}

\bibitem[{{Stern} \& {Laor}(2012)}]{Stern:2012a}
{Stern}, J., \& {Laor}, A. 2012, \mnras, 423, 600, \dodoi{10.1111/j.1365-2966.2012.20901.x}

\bibitem[{{Suess} {et~al.}(2024){Suess}, {Weaver}, {Price}, {Pan}, {Wang}, {Bezanson}, {Brammer}, {Cutler}, {Labbe}, {Leja}, {Williams}, {Whitaker}, {Dayal}, {de Graaff}, {Feldmann}, {Franx}, {Fudamoto}, {Fujimoto}, {Furtak}, {Goulding}, {Greene}, {Khullar}, {Kokorev}, {Kriek}, {Lorenz}, {Marchesini}, {Maseda}, {Matthee}, {Miller}, {Mitsuhashi}, {Mowla}, {Muzzin}, {Naidu}, {Nanayakkara}, {Nelson}, {Oesch}, {Setton}, {Shipley}, {Smit}, {Spilker}, {van Dokkum}, \& {Zitrin}}]{Suess:2024}
{Suess}, K.~A., {Weaver}, J.~R., {Price}, S.~H., {et~al.} 2024, arXiv e-prints, arXiv:2404.13132, \dodoi{10.48550/arXiv.2404.13132}

\bibitem[{{Terlevich} \& {Melnick}(1985)}]{Terlevich:1985}
{Terlevich}, R., \& {Melnick}, J. 1985, \mnras, 213, 841, \dodoi{10.1093/mnras/213.4.841}

\bibitem[{{Thompson} {et~al.}(2005){Thompson}, {Quataert}, \& {Murray}}]{Thompson:2005}
{Thompson}, T.~A., {Quataert}, E., \& {Murray}, N. 2005, \apj, 630, 167, \dodoi{10.1086/431923}

\bibitem[{{Topping} {et~al.}(2024){Topping}, {Stark}, {Endsley}, {Whitler}, {Hainline}, {Johnson}, {Robertson}, {Tacchella}, {Chen}, {Alberts}, {Baker}, {Bunker}, {Carniani}, {Charlot}, {Chevallard}, {Curtis-Lake}, {DeCoursey}, {Egami}, {Eisenstein}, {Ji}, {Maiolino}, {Williams}, {Willmer}, {Willott}, \& {Witstok}}]{Topping:2023}
{Topping}, M.~W., {Stark}, D.~P., {Endsley}, R., {et~al.} 2024, \mnras, 529, 4087, \dodoi{10.1093/mnras/stae800}

\bibitem[{{Torralba-Torregrosa} {et~al.}(2024){Torralba-Torregrosa}, {Matthee}, {Naidu}, {Mackenzie}, {Pezzulli}, {Hutter}, {Arnalte-Mur}, {Gurung-L{\'o}pez}, {Tacchella}, {Oesch}, {Kashino}, {Conroy}, \& {Sobral}}]{Torralba:2024}
{Torralba-Torregrosa}, A., {Matthee}, J., {Naidu}, R.~P., {et~al.} 2024, \aap, 689, A44, \dodoi{10.1051/0004-6361/202450318}

\bibitem[{{Treiber} {et~al.}(2024){Treiber}, {Greene}, {Weaver}, {Miller}, {Furtak}, {Setton}, {Wang}, {de Graaff}, {Bezanson}, {Brammer}, {Cutler}, {Dayal}, {Feldmann}, {Fujimoto}, {Goulding}, {Kokorev}, {Labbe}, {Leja}, {Marchesini}, {Nanayakkara}, {Nelson}, {Pan}, {Price}, {Siegel}, {Suess}, \& {Whitaker}}]{Treiber:2024}
{Treiber}, H., {Greene}, J., {Weaver}, J.~R., {et~al.} 2024, arXiv e-prints, arXiv:2409.12232, \dodoi{10.48550/arXiv.2409.12232}

\bibitem[{{Tsuzuki} {et~al.}(2006){Tsuzuki}, {Kawara}, {Yoshii}, {Oyabu}, {Tanab{\'e}}, \& {Matsuoka}}]{Tsuzuki:2006}
{Tsuzuki}, Y., {Kawara}, K., {Yoshii}, Y., {et~al.} 2006, \apj, 650, 57, \dodoi{10.1086/506376}

\bibitem[{{{\"U}bler} {et~al.}(2023){{\"U}bler}, {Maiolino}, {Curtis-Lake}, {P{\'e}rez-Gonz{\'a}lez}, {Curti}, {Perna}, {Arribas}, {Charlot}, {Marshall}, {D'Eugenio}, {Scholtz}, {Bunker}, {Carniani}, {Ferruit}, {Jakobsen}, {Rix}, {Rodr{\'\i}guez Del Pino}, {Willott}, {Boeker}, {Cresci}, {Jones}, {Kumari}, \& {Rawle}}]{Ubler:2023}
{{\"U}bler}, H., {Maiolino}, R., {Curtis-Lake}, E., {et~al.} 2023, \aap, 677, A145, \dodoi{10.1051/0004-6361/202346137}

\bibitem[{{van Dokkum} {et~al.}(2008){van Dokkum}, {Franx}, {Kriek}, {Holden}, {Illingworth}, {Magee}, {Bouwens}, {Marchesini}, {Quadri}, {Rudnick}, {Taylor}, \& {Toft}}]{vanDokkum:2008}
{van Dokkum}, P.~G., {Franx}, M., {Kriek}, M., {et~al.} 2008, \apjl, 677, L5, \dodoi{10.1086/587874}

\bibitem[{{van Dokkum} {et~al.}(2015){van Dokkum}, {Nelson}, {Franx}, {Oesch}, {Momcheva}, {Brammer}, {F{\"o}rster Schreiber}, {Skelton}, {Whitaker}, {van der Wel}, {Bezanson}, {Fumagalli}, {Illingworth}, {Kriek}, {Leja}, \& {Wuyts}}]{vandokkum:2015}
{van Dokkum}, P.~G., {Nelson}, E.~J., {Franx}, M., {et~al.} 2015, \apj, 813, 23, \dodoi{10.1088/0004-637X/813/1/23}

\bibitem[{{Vanden Berk} {et~al.}(2001){Vanden Berk}, {Richards}, {Bauer}, {Strauss}, {Schneider}, {Heckman}, {York}, {Hall}, {Fan}, {Knapp}, {Anderson}, {Annis}, {Bahcall}, {Bernardi}, {Briggs}, {Brinkmann}, {Brunner}, {Burles}, {Carey}, {Castander}, {Connolly}, {Crocker}, {Csabai}, {Doi}, {Finkbeiner}, {Friedman}, {Frieman}, {Fukugita}, {Gunn}, {Hennessy}, {Ivezi{\'c}}, {Kent}, {Kunszt}, {Lamb}, {Leger}, {Long}, {Loveday}, {Lupton}, {Meiksin}, {Merelli}, {Munn}, {Newberg}, {Newcomb}, {Nichol}, {Owen}, {Pier}, {Pope}, {Rockosi}, {Schlegel}, {Siegmund}, {Smee}, {Snir}, {Stoughton}, {Stubbs}, {SubbaRao}, {Szalay}, {Szokoly}, {Tremonti}, {Uomoto}, {Waddell}, {Yanny}, \& {Zheng}}]{VandenBerk:2001}
{Vanden Berk}, D.~E., {Richards}, G.~T., {Bauer}, A., {et~al.} 2001, \aj, 122, 549, \dodoi{10.1086/321167}

\bibitem[{Vehtari {et~al.}(2021)Vehtari, Gelman, Simpson, Carpenter, \& B{\"u}rkner}]{Vehtari2021}
Vehtari, A., Gelman, A., Simpson, D., Carpenter, B., \& B{\"u}rkner, P.-C. 2021, Bayesian analysis, 16, 667

\bibitem[{{V{\'e}ron-Cetty} {et~al.}(2004){V{\'e}ron-Cetty}, {Joly}, \& {V{\'e}ron}}]{veroncetty:2004}
{V{\'e}ron-Cetty}, M.~P., {Joly}, M., \& {V{\'e}ron}, P. 2004, \aap, 417, 515, \dodoi{10.1051/0004-6361:20035714}

\bibitem[{{Vestergaard} \& {Wilkes}(2001)}]{VestergaardWilkes:2001}
{Vestergaard}, M., \& {Wilkes}, B.~J. 2001, \apjs, 134, 1, \dodoi{10.1086/320357}

\bibitem[{{Wang} {et~al.}(2023){Wang}, {Fujimoto}, {Labb{\'e}}, {Furtak}, {Miller}, {Setton}, {Zitrin}, {Atek}, {Bezanson}, {Brammer}, {Leja}, {Oesch}, {Price}, {Chemerynska}, {Cutler}, {Dayal}, {van Dokkum}, {Goulding}, {Greene}, {Fudamoto}, {Khullar}, {Kokorev}, {Marchesini}, {Pan}, {Weaver}, {Whitaker}, \& {Williams}}]{Wang:2023highz}
{Wang}, B., {Fujimoto}, S., {Labb{\'e}}, I., {et~al.} 2023, \apjl, 957, L34, \dodoi{10.3847/2041-8213/acfe07}

\bibitem[{{Wang} {et~al.}(2024{\natexlab{a}}){Wang}, {de Graaff}, {Davies}, {Greene}, {Leja}, {Goulding}, {Williams}, {Brammer}, {Suess}, {Weibel}, {Bezanson}, {Boogaard}, {Cleri}, {Hirschmann}, {Katz}, {Labbe}, {Maseda}, {Matthee}, {McConachie}, {Naidu}, {Oesch}, {Rix}, {Setton}, \& {Whitaker}}]{Wang:2024brd}
{Wang}, B., {de Graaff}, A., {Davies}, R.~L., {et~al.} 2024{\natexlab{a}}, arXiv e-prints, arXiv:2403.02304, \dodoi{10.48550/arXiv.2403.02304}

\bibitem[{{Wang} {et~al.}(2024{\natexlab{b}}){Wang}, {Leja}, {de Graaff}, {Brammer}, {Weibel}, {van Dokkum}, {Baggen}, {Suess}, {Greene}, {Bezanson}, {Cleri}, {Hirschmann}, {Labb{\'e}}, {Matthee}, {McConachie}, {Naidu}, {Nelson}, {Oesch}, {Setton}, \& {Williams}}]{Wang:2024UB}
{Wang}, B., {Leja}, J., {de Graaff}, A., {et~al.} 2024{\natexlab{b}}, \apjl, 969, L13, \dodoi{10.3847/2041-8213/ad55f7}

\bibitem[{{Wang} {et~al.}(2024{\natexlab{c}}){Wang}, {Leja}, {Labb{\'e}}, {Bezanson}, {Whitaker}, {Brammer}, {Furtak}, {Weaver}, {Price}, {Zitrin}, {Atek}, {Coe}, {Cutler}, {Dayal}, {van Dokkum}, {Feldmann}, {Marchesini}, {Franx}, {F{\"o}rster Schreiber}, {Fujimoto}, {Geha}, {Glazebrook}, {de Graaff}, {Greene}, {Juneau}, {Kassin}, {Kriek}, {Khullar}, {Maseda}, {Mowla}, {Muzzin}, {Nanayakkara}, {Nelson}, {Oesch}, {Pacifici}, {Pan}, {Papovich}, {Setton}, {Shapley}, {Smit}, {Stefanon}, {Suess}, {Taylor}, \& {Williams}}]{Wang:2023}
{Wang}, B., {Leja}, J., {Labb{\'e}}, I., {et~al.} 2024{\natexlab{c}}, \apjs, 270, 12, \dodoi{10.3847/1538-4365/ad0846}

\bibitem[{{Weaver} {et~al.}(2024){Weaver}, {Cutler}, {Pan}, {Whitaker}, {Labb{\'e}}, {Price}, {Bezanson}, {Brammer}, {Marchesini}, {Leja}, {Wang}, {Furtak}, {Zitrin}, {Atek}, {Chemerynska}, {Coe}, {Dayal}, {van Dokkum}, {Feldmann}, {F{\"o}rster Schreiber}, {Franx}, {Fujimoto}, {Fudamoto}, {Glazebrook}, {de Graaff}, {Greene}, {Juneau}, {Kassin}, {Kriek}, {Khullar}, {Maseda}, {Mowla}, {Muzzin}, {Nanayakkara}, {Nelson}, {Oesch}, {Pacifici}, {Papovich}, {Setton}, {Shapley}, {Shipley}, {Smit}, {Stefanon}, {Taylor}, {Weibel}, \& {Williams}}]{Weaver:2023}
{Weaver}, J.~R., {Cutler}, S.~E., {Pan}, R., {et~al.} 2024, \apjs, 270, 7, \dodoi{10.3847/1538-4365/ad07e0}

\bibitem[{{Weibel} {et~al.}(2024){Weibel}, {de Graaff}, {Setton}, {Miller}, {Oesch}, {Brammer}, {Lagos}, {Whitaker}, {Williams}, {Baggen}, {Bezanson}, {Boogaard}, {Cleri}, {Greene}, {Hirschmann}, {Hviding}, {Kuruvanthodi}, {Labb{\'e}}, {Leja}, {Maseda}, {Matthee}, {McConachie}, {Naidu}, {Roberts-Borsani}, {Schaerer}, {Suess}, {Valentino}, {van Dokkum}, \& {Wang}}]{Weibel:2024}
{Weibel}, A., {de Graaff}, A., {Setton}, D.~J., {et~al.} 2024, arXiv e-prints, arXiv:2409.03829, \dodoi{10.48550/arXiv.2409.03829}

\bibitem[{{Wilkins} {et~al.}(2019){Wilkins}, {Lovell}, \& {Stanway}}]{Wilkins:2019}
{Wilkins}, S.~M., {Lovell}, C.~C., \& {Stanway}, E.~R. 2019, \mnras, 490, 5359, \dodoi{10.1093/mnras/stz2894}

\bibitem[{{Williams} {et~al.}(2024){Williams}, {Alberts}, {Ji}, {Hainline}, {Lyu}, {Rieke}, {Endsley}, {Suess}, {Sun}, {Johnson}, {Florian}, {Shivaei}, {Rujopakarn}, {Baker}, {Bhatawdekar}, {Boyett}, {Bunker}, {Cameron}, {Carniani}, {Charlot}, {Curtis-Lake}, {DeCoursey}, {de Graaff}, {Egami}, {Eisenstein}, {Gibson}, {Hausen}, {Helton}, {Maiolino}, {Maseda}, {Nelson}, {P{\'e}rez-Gonz{\'a}lez}, {Rieke}, {Robertson}, {Saxena}, {Tacchella}, {Willmer}, \& {Willott}}]{Williams:2024}
{Williams}, C.~C., {Alberts}, S., {Ji}, Z., {et~al.} 2024, \apj, 968, 34, \dodoi{10.3847/1538-4357/ad3f17}

\bibitem[{{Wills} {et~al.}(1985){Wills}, {Netzer}, \& {Wills}}]{Wills:1985}
{Wills}, B.~J., {Netzer}, H., \& {Wills}, D. 1985, \apj, 288, 94, \dodoi{10.1086/162767}

\bibitem[{{Yue} {et~al.}(2024){Yue}, {Eilers}, {Ananna}, {Panagiotou}, {Kara}, \& {Miyaji}}]{Yue:2024}
{Yue}, M., {Eilers}, A.-C., {Ananna}, T.~T., {et~al.} 2024, \apjl, 974, L26, \dodoi{10.3847/2041-8213/ad7eba}

\bibitem[{{Zakamska} {et~al.}(2006){Zakamska}, {Strauss}, {Krolik}, {Ridgway}, {Schmidt}, {Smith}, {Heckman}, {Schneider}, {Hao}, \& {Brinkmann}}]{Zakamska:2006}
{Zakamska}, N.~L., {Strauss}, M.~A., {Krolik}, J.~H., {et~al.} 2006, \aj, 132, 1496, \dodoi{10.1086/506986}

\bibitem[{{Zhang} {et~al.}(2024){Zhang}, {Jiang}, {Liu}, \& {Ho}}]{Zhang:2024}
{Zhang}, Z., {Jiang}, L., {Liu}, W., \& {Ho}, L.~C. 2024, arXiv e-prints, arXiv:2411.02729, \dodoi{10.48550/arXiv.2411.02729}

\end{thebibliography}

\appendix

\begin{deluxetable}{lcccc}[]
\tabcolsep=2mm
\tablecaption{\label{tab:uv} UV Emission Line Fluxes}
\tablehead{
\colhead{Line} & \colhead{Component} & \colhead{$\lambda_{obs}$ (\AA)} & \colhead{Flux ($10^{-19}~\mathrm{erg~s^{-1}~cm^{-2}}$)} & \colhead{I/I(\hbeta)} 
}
\startdata
N IV] & Broad & 1488 & 64.9$\pm$1.7 & 0.13 \\
C IV & Broad & 1549 & 60.2$\pm$1.7 & 0.12 \\
He II & Broad & 1640 & 15.8$\pm$2.4 & 0.03 \\
O III] & Broad & 1666 & 19.4$\pm$2.3 & 0.04 \\
N III] & Broad & 1750 & 67.8$\pm$2.4 & 0.13 \\
Fe II & Broad & 1786 & 21.2$\pm$3.2 & 0.04 \\
$[$Ne III$]$+Si II & Broad & 1815 & 31.8$\pm$3.1 & 0.06 \\
Al III & Broad & 1860 & 17.0$\pm$3.0 & 0.03 \\
Si III] & Broad & 1888 & 1.6$\pm$1.1 & 0.00 \\
C III] & Broad & 1909 & 146.3$\pm$3.0 & 0.28 \\
\enddata
\tablecomments{Note that at shorter wavelengths than $H\beta$, broad and narrow components can not be reliably distinguished at NIRSpec PRISM resolution. The Fluxes are corrected for magnification $\mu=1.7\pm0.2$. Relative intensities are to the broad component of \hbeta.}
\end{deluxetable}
\begin{table}[ht]
\centering
\caption{FeII measurements}
\begin{tabular}{lcccc}
\hline
Ion  & $\lambda$[\angstrom] & Flux [1e-19 erg/s/cm$^2$] & EW & I/I(\hbeta) \\
\hline
Fe II-UV & (2200-2600) & $295.2$ & 151.2 & 0.40 \\
Fe II-UV & (2600-3050) & $332.1$ & 186.5 & 0.45 \\
Fe II-UV & (2200-3050) & $628.0$ & 338.2 & 0.84 \\
Fe II-opt & (4000-4800) & $233.9$ & 59.0 & 0.46 \\
Fe II-opt & (5000-5500) & $441.0$ & 103.3 & 0.87 \\
Fe II-opt & (4000-5500) & $705.9$ & 168.1 & 1.39 \\
Fe II-NIR & (9100-9300) & $34.3$ & 9.0 & 0.07 \\
\hline
\end{tabular} 
\begin{tabular}{lcccc}
Ion  & $\lambda$[\angstrom] & Ratio &  &  \\
\hline
Fe II-UV/Mg II & (2200-3050) & $10.6$  &  \\
Fe II-UV/Fe II-opt & (2200-3050) & 0.9  &  \\
\hline
\end{tabular} 
\label{tab:feii}
\end{table}

\begin{deluxetable}{lcccc}[]
\tabcolsep=2mm
\tablecaption{\label{tab:balmer} Balmer Line Series Fluxes}
\tablehead{
\colhead{Line} & \colhead{Component} & \colhead{$\lambda_{obs}$ $[$\AA$]$} & \colhead{Flux $[10^{-19}~\mathrm{erg~s^{-1}~cm^{-2}}]$} 
}
\startdata
H10 & narrow & 3799.014 & $5.3\pm0.6$ \\
H10 & broad & 3799.014 & $11.7\pm1.2$ \\
H9 & broad & 3836.511 & $16.3\pm1.7$ \\
H9 & narrow & 3836.511 & $7.4\pm0.8$ \\
H$\zeta$ & narrow & 3890.191 & $10.8\pm1.2$ \\
H$\zeta$ & broad & 3890.191 & $23.9\pm2.5$ \\
H$\epsilon$ & broad & 3971.236 & $37.4\pm3.9$ \\
H$\epsilon$ & narrow & 3971.236 & $16.9\pm1.9$ \\
H$\delta$ & broad & 4102.936 & $64.1\pm6.7$ \\
H$\delta$ & narrow & 4102.936 & $28.9\pm3.2$ \\
H$\gamma$ & narrow & 4341.731 & $57.9\pm6.4$ \\
H$\gamma$ & broad & 4341.731 & $128.3\pm13.4$ \\
H$\beta$ & narrow & 4862.738 & $157.8\pm17.6$ \\
H$\beta$ & broad & 4862.738 & $349.9\pm36.6$ \\
H$\alpha$ & narrow & 6564.697 & $964.3\pm107.5$ \\
H$\alpha$ & broad & 6564.697 & $2137.8\pm123.4$ \\
\enddata
\tablecomments{Balmer Series fluxes for fiducial model III. These are the sum of the attenuated blue and red AGN Balmer series components. Note that at shorter wavelengths than $H\beta$, broad and narrow components can not be reliably distinguished at NIRSpec PRISM resolution. The Fluxes are corrected for magnification $\mu=1.7\pm0.2$.}
\end{deluxetable}
\begin{table}[h!]
\centering
\caption{Other Emission Line Measurements}.
\label{tab:lines}
\begin{tabular}{lclcr}
\toprule
Ion & width & $\lambda$[\angstrom] & Flux [1e-19 erg/s/cm$^2$] & I/I(\hbeta) \\
\midrule
Ly$\alpha$ & narrow & 1215.4 & $4.1\pm0.4$ & 0.01 \\
Mg II & broad & 2799.1 & $59.5\pm9.1$ &   0.08 \\
He I & broad & 3188.7 & $5.1\pm2.2$ & 0.01 \\
He II & broad &  3204.0 & $23.7\pm4.3$ &   0.03 \\
$[$Ne V$]$ & narrow & 3426.8 & $5.2\pm3.6$ & 0.01 \\
$[$Ne V$]$ & narrow & 3426.8 & $6.3\pm1.0$ & 0.01 \\
$[$O II$]$ & narrow & 3727.1 & $17.0\pm1.1$ & 0.02 \\
$[$O II$]$ & narrow & 3729.9 & $17.0\pm1.1$ & 0.02 \\
$[$Ne III$]$ & narrow & 3869.9 & $50.3\pm1.8$ & 0.07 \\
He I & broad & 3889.8 & $13.1\pm4.9$ & 0.02 \\
He I & broad & 4027.3 & $14.8\pm4.1$ & 0.02 \\
$[$O III$]$ & narrow & 4364.4 & $21.2\pm3.7$ & 0.03 \\
He I & broad & 4472.7 & $23.5\pm4.0$ & 0.03 \\
He II & broad & 4685.7 & $67.6\pm2.3$ & 0.09 \\
$[$O III$]$ & narrow & 4960.3 & $196.6\pm0.9$ & 0.26 \\
$[$O III$]$ & narrow & 5008.2 & $586.0\pm2.7$ & 0.79 \\
$[$N II$]$ & narrow & 5756.2 & $34.2\pm1.6$ & 0.05 \\
He I & broad & 5877.2 & $110.0\pm3.8$ & 0.15 \\
$[$O I$]$ & narrow & 6302.0 & $11.5\pm1.8$ & 0.02 \\
$[$N II$]$ & narrow & 6549.9 & $82.3\pm36.4$ & 0.09 \\
$[$N II$]$ & narrow & 6585.3 & $246.8\pm109.2$ & 0.28 \\
$[$S II$]$ & narrow & 6718.3 & $19.1\pm1.5$ & 0.03 \\
$[$S II$]$ & narrow & 6732.7 & $19.1\pm1.5$ & 0.03 \\
He I & broad & 7067.1 & $129.9\pm2.5$ & 0.17 \\
O I & narrow & 7256.4 & $17.9\pm0.2$ & 0.02 \\
$[$O II$]$ & narrow & 7321.9 & $19.3\pm0.2$ & 0.03 \\
$[$O II$]$ & narrow & 7332.2 & $16.0\pm0.2$ & 0.02 \\
O I & narrow & 8446.7 & $43.4\pm4.9$ & 0.08 \\
O I & broad & 8446.7 & $57.0\pm10.8$ & 0.11 \\
Ca II & broad & 8500.4 & $82.1\pm3.8$ & 0.11 \\
Ca II & broad & 8544.4 & $82.1\pm3.8$ & 0.11 \\
Ca II & broad & 8664.5 & $82.1\pm3.8$ & 0.11 \\
Pa10 & narrow & 9017.4 & $8.3\pm2.6$ & 0.02 \\
$[$S III$]$ & narrow & 9071.1 & $2.3\pm1.6$ & 0.00 \\
Pa9 & narrow & 9231.6 & $0.8\pm1.9$ & 0.00 \\
$[$S III$]$ & narrow & 9533.2 & $5.6\pm3.8$ & 0.01 \\
Pa$\epsilon$ & broad & 9548.6 & $76.0\pm6.4$ & 0.15 \\
Pa$\epsilon$ & narrow & 9548.6 & $14.1\pm4.1$ & 0.03 \\
%\enddata
\bottomrule
\end{tabular}
\tablecomments{Note that at rest-frame UV wavelengths, broad and narrow components cannot be reliably distinguished at NIRSpec PRISM resolution. The Fluxes are corrected for magnification $\mu=1.7\pm0.2$. Relative intensities are to the broad component of \hbeta.}
\end{table}
\begin{table}[h!]
\centering
\caption{Model configuration.}
\label{tab:results}
\begin{tabular}{lllll}
\toprule
\multicolumn{5}{c}{III. STARS + AGN} \\
\midrule
{Component} & {Parameter} & {Type} & {Value} & {Prior} \\
\midrule
all & dust\_index & Free & (-1.8, -0.4) & Linear \\
all & z & Fixed & 4.4655 & -- \\
all & lsf\_scale & Fixed & 1.3 & -- \\
agn\_blue & broad\_fwhm & Fixed & 4850 & -- \\
agn\_blue & narrow\_fwhm & Fixed & 650 & -- \\
agn\_blue & Av & Free & (0.0, 1.0) & Linear \\
agn\_blue & EW\_Balmer\_10kK\_b & Free & (-1.0, 4.0) & Log$_{10}$ \\
agn\_blue & EW\_Balmer\_10kK\_n & Free & (-1.0, 4.0) & Log$_{10}$ \\
agn\_blue & beta\_1 & Free & (-3.0, -1.0) & $N(-1.7, 0.3)$ \\
agn\_blue & L\_beta\_1 & Free & (6.0, 15.0) & Log$_{10}$ \\
agn\_red & Av & Free & (1.0, 5.0) & Linear \\
agn\_red & L\_beta\_1 & Free & (6.0, 15.0) & Log$_{10}$ \\
sps & lmass & Free & (7.0, 12.0) & Log$_{10}$ \\
sps & ltau & Free & (6.5, 9.0) & Log$_{10}$ \\
sps & lage & Free & (7.5, 9.2) & Log$_{10}$ \\
sps & logzsol & Free & (-1.0, 0.3) & Log$_{10}$ \\
sps & Av & Free & (0.0, 4.0) & Linear \\
sps & dust\_type & Fixed & 0 & -- \\
sps & sigma\_smooth & Tied &  & 0.5 * (Log$_{10}$(M) - Log$_{10}$(r$_e$) - 5.9) \\
sps & r\_e & Fixed & 0.07 & -- \\
\bottomrule
\end{tabular}
\tablecomments{}
\end{table}

\section{Optical-NIR PRISM line fits}\label{appendix:lines}
\begin{figure}
\vspace{-10mm}
$$
\begin{array}{cc}
\includegraphics[width=0.4\textwidth]{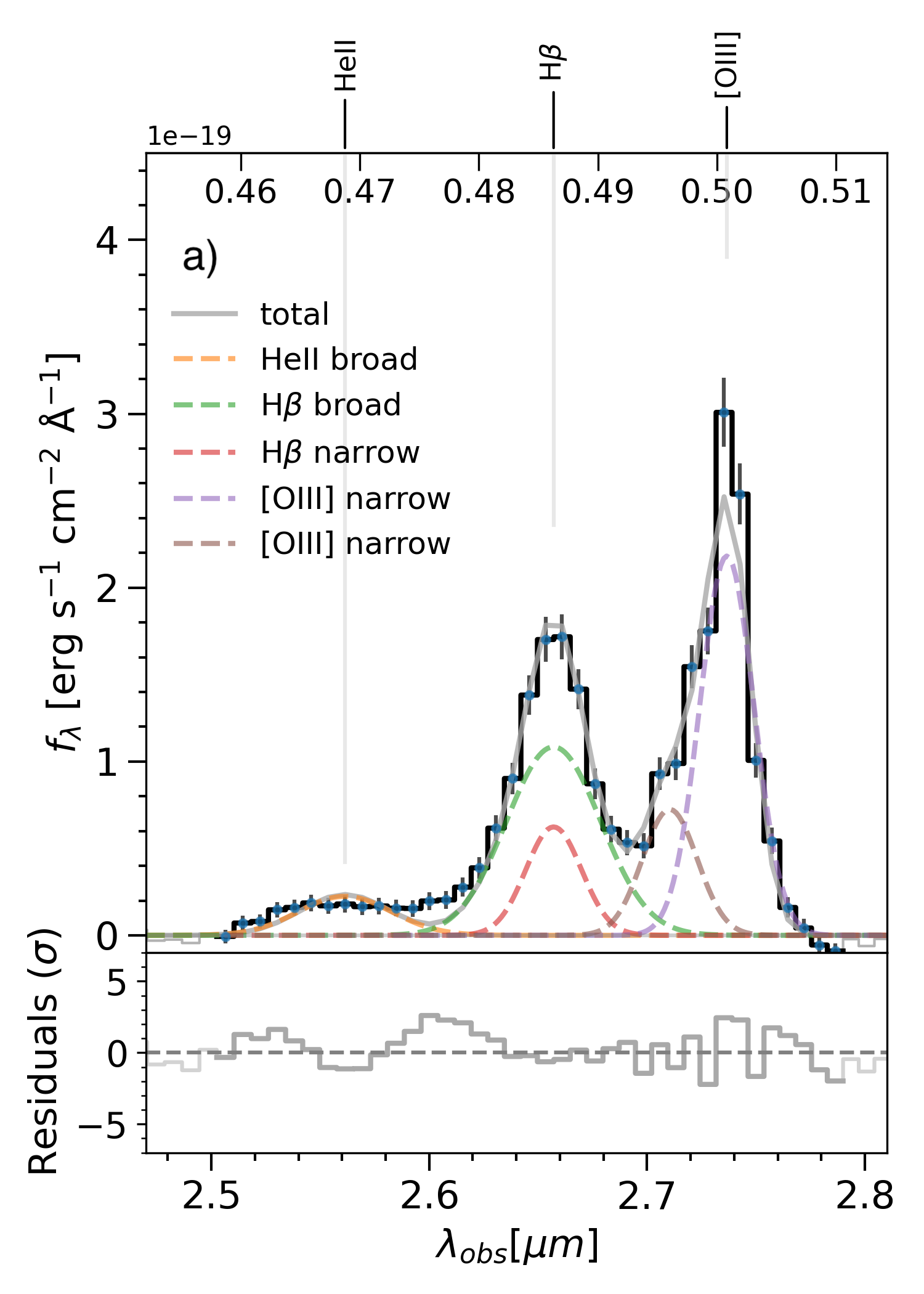} &
\includegraphics[width=0.4\textwidth]{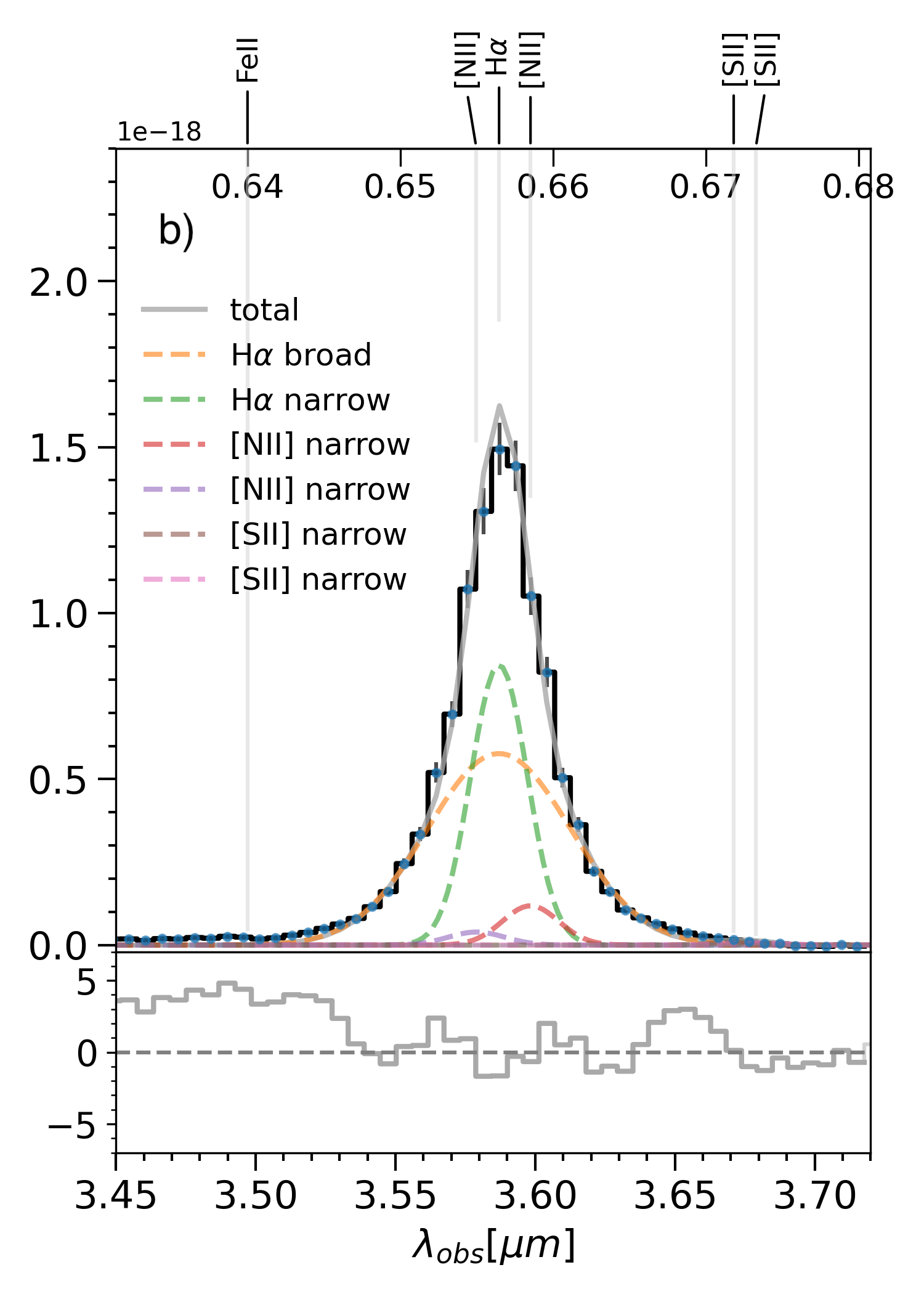} \\
\includegraphics[width=0.4\textwidth]{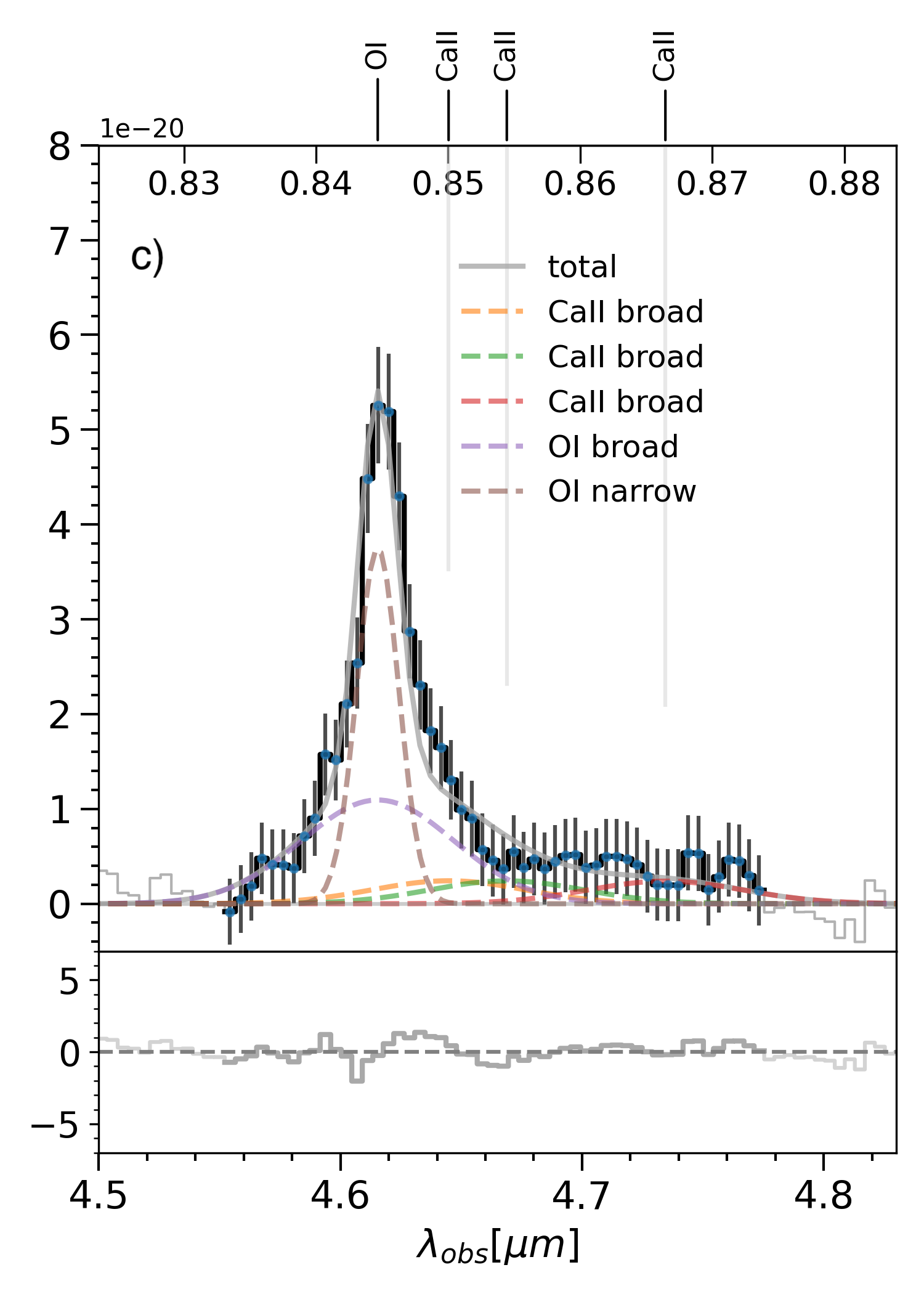} &
\includegraphics[width=0.4\textwidth]{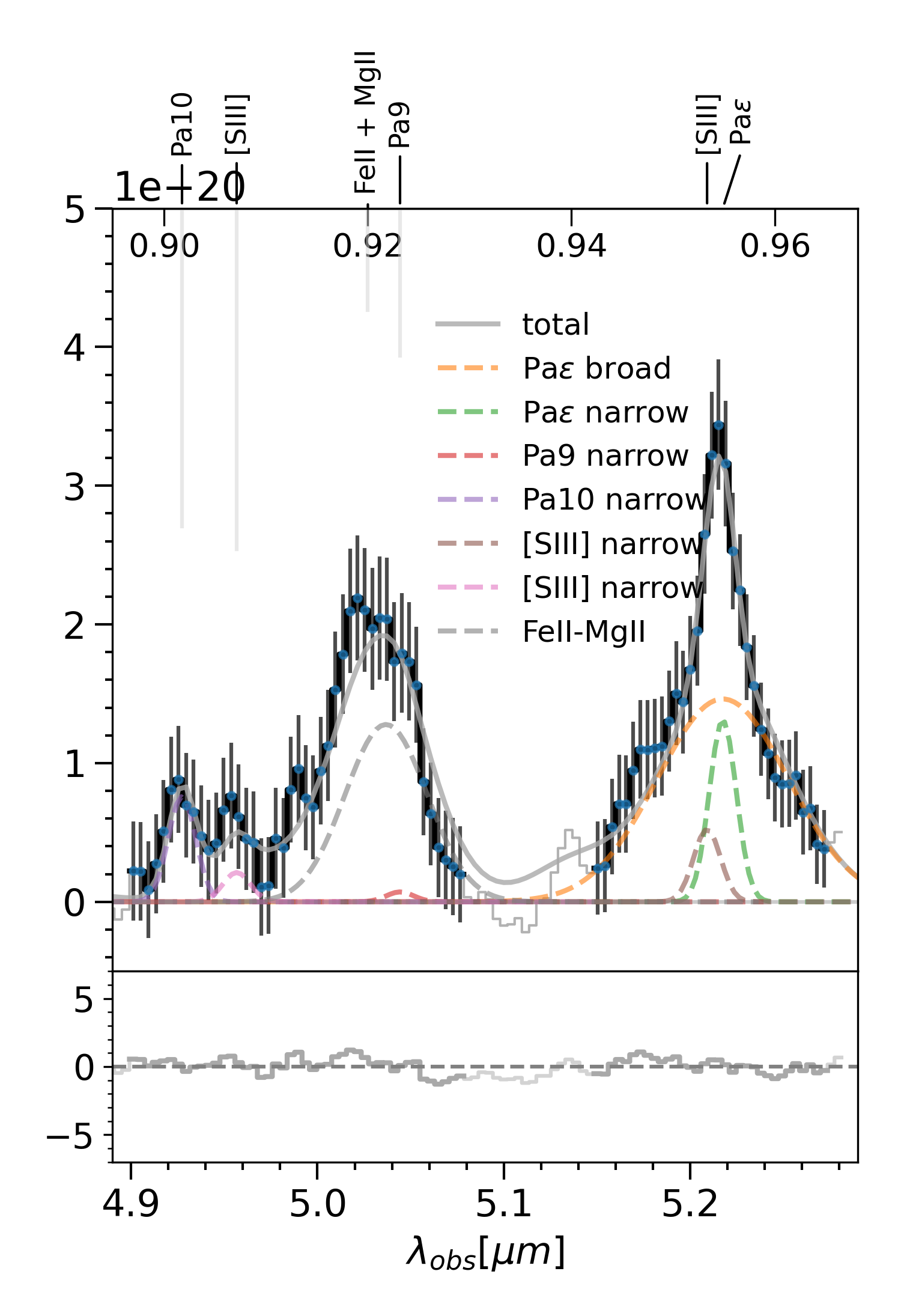} \\
\end{array}
$$
\caption{Model fits to the optical and near-IR lines. {\em Top panels:} optical emission lines: \hbeta, \oiii doublet, and \ion{He}{2}$\lambda4686$ ({\em a}) and \halpha, \ion{N}{2} doublet, \ion{S}{2} doublet ({\em b}). Narrow and broad lines are indicated. A background was removed by subtracting a powerlaw fit to the continuum near the lines, and these areas are were then masked (thin lines). Emission lines are modeled with Gaussians. Even at limited PRISM resolution a broad line component is required to model \halpha and \hbeta, while the \ion{O}{3} doublet is well reproduced by an unresolved line alone, as is typically seen in broad line AGN. There is a clear evidence for excess flux blueward of \hbeta, likely a blend of broad \ion{He}{2}$\lambda4686$ and optical \ion{Fe}{2} emission (e.g., \ion{Fe}{2}$\lambda4570$\angstrom bump). There is no evidence \ion{N}{2}, \ion{S}{2} emission, consistent with the higher-resolution NIRCam GRISM data from ALT. There is a broad flux excess at $\lambda\lambda6200-6450$\angstrom, possibly indicating broad iron line emission \ion{Fe}{2}$\lambda4570$\angstrom  \citep{veroncetty:2004}. {\em Bottom panels:} NIR emission lines, showing \ion{O}{1}, \ion{Ca}{2} (Ca triplet, CaT) ({\em c}) Paschen-series, and \ion{S}{3} ({\em d}). Pa$\epsilon$, \ion{O}{1}, and likely CaT have a broad component. Blueward of Pa$9$ there is a strong excess at $9200\AA$, which we identify as the \ion{Fe}{2}$\lambda9200$\angstrom bump. This emission is highly correlated with \ion{Fe}{2}$\lambda4570$\angstrom in AGN, and its ratio can be used to constrain excitation mechanisms \citep{marinello2006}. Taken together, the optical-NIR line spectrum shows evidence for prominent broad line region emission lines.}
\label{fig:appendix:lines}
\end{figure}

We follow the same procedure as in \S \ref{sec:nearuv} to model rest-frame optical and near-infrared emission lines. The results are shown in Figure \ref{fig:appendix:lines}. In addition to clear evidence for broad Balmer and Paschen lines, the most notable features are evidence for broad \ion{He}{2}$\lambda4686$, \ion{O}{1}$\lambda8446$, and \ion{Fe}{2} emission, all indicating emission originating from the Broad line region. 

\section{SED fits}\label{appendix:sed}
\begin{figure}
\vspace{-10mm}
$$
\begin{array}{cc}
\includegraphics[width=0.55\textwidth]{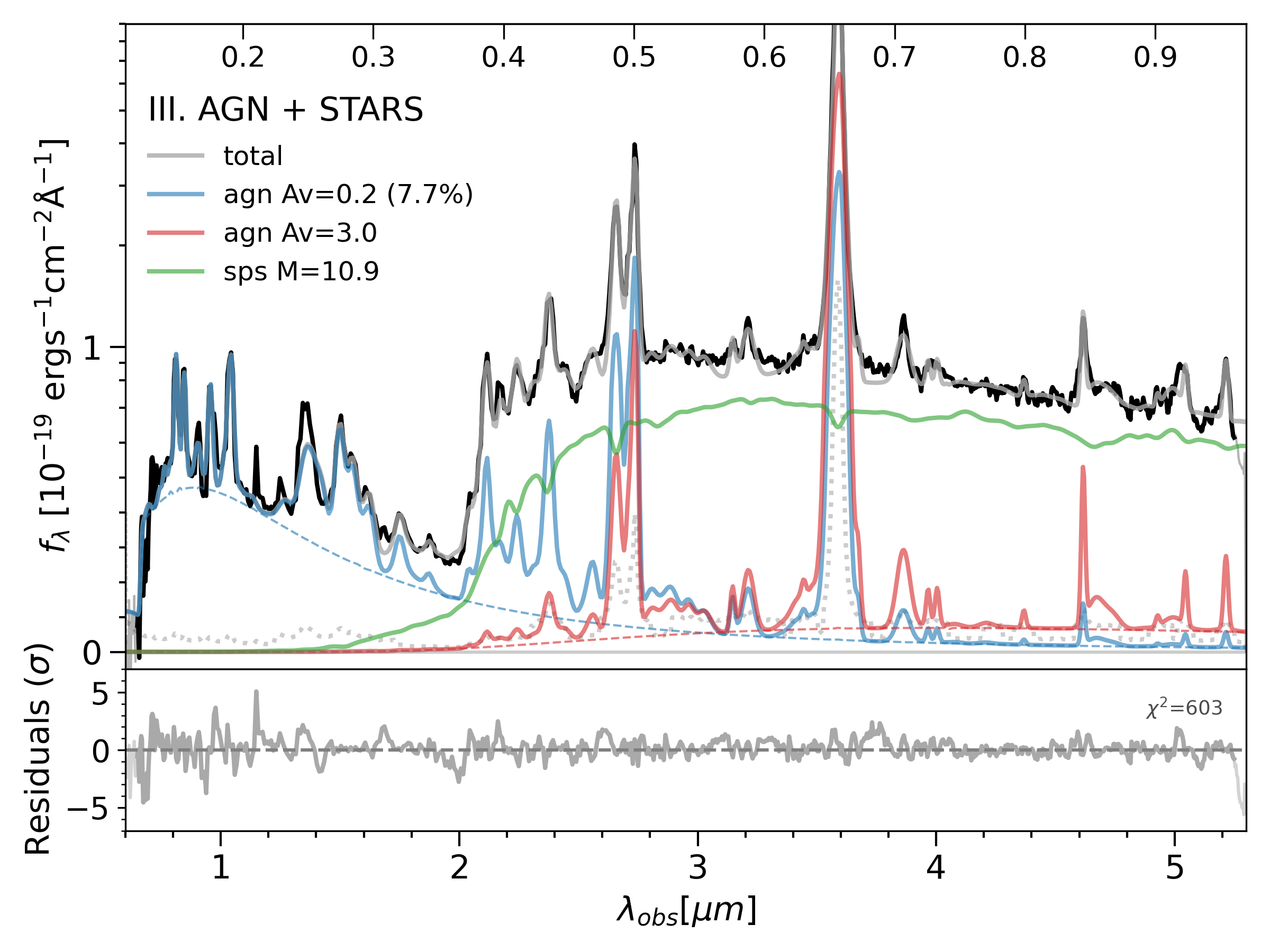} &
\includegraphics[width=0.41\textwidth]{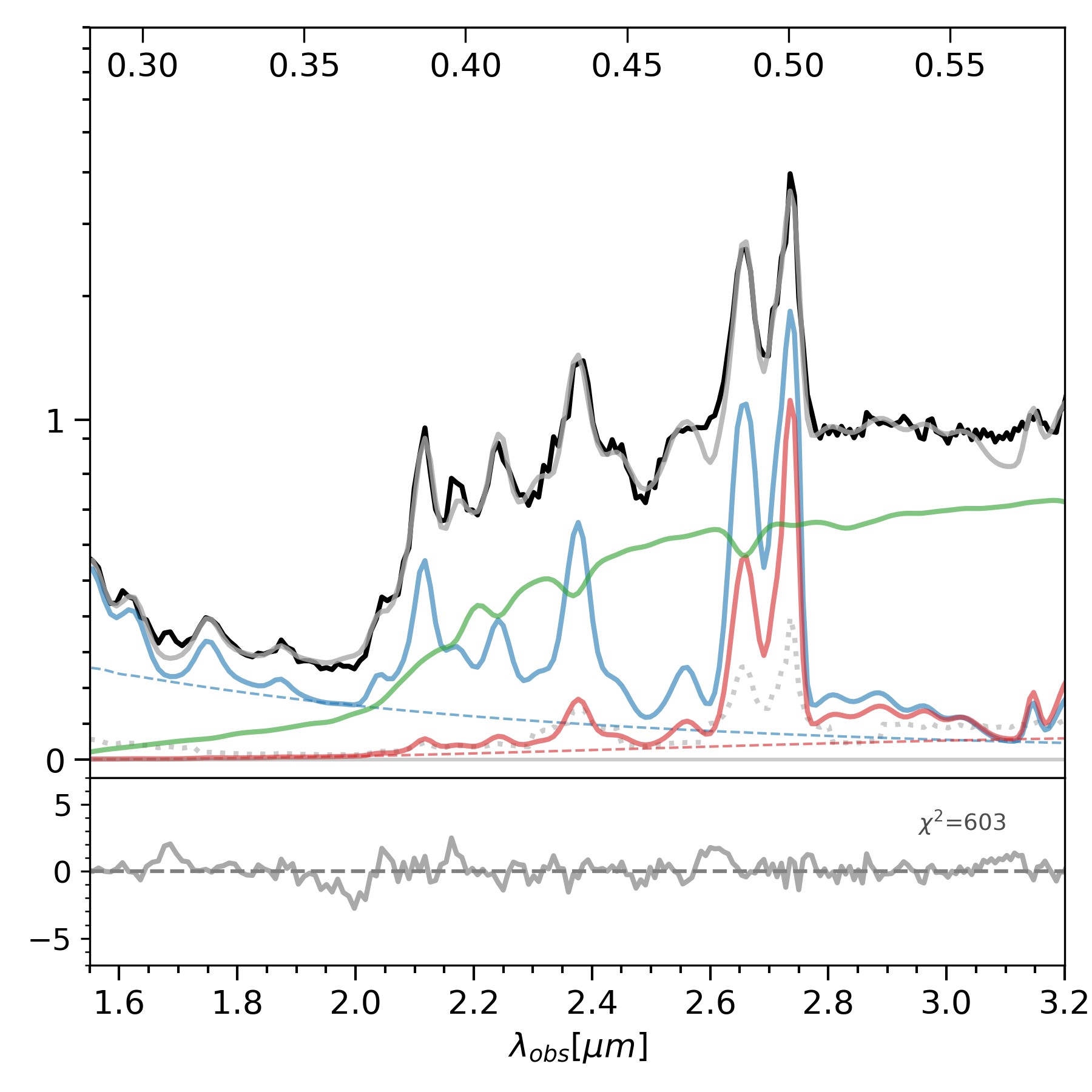} \\
\includegraphics[width=0.55\textwidth]{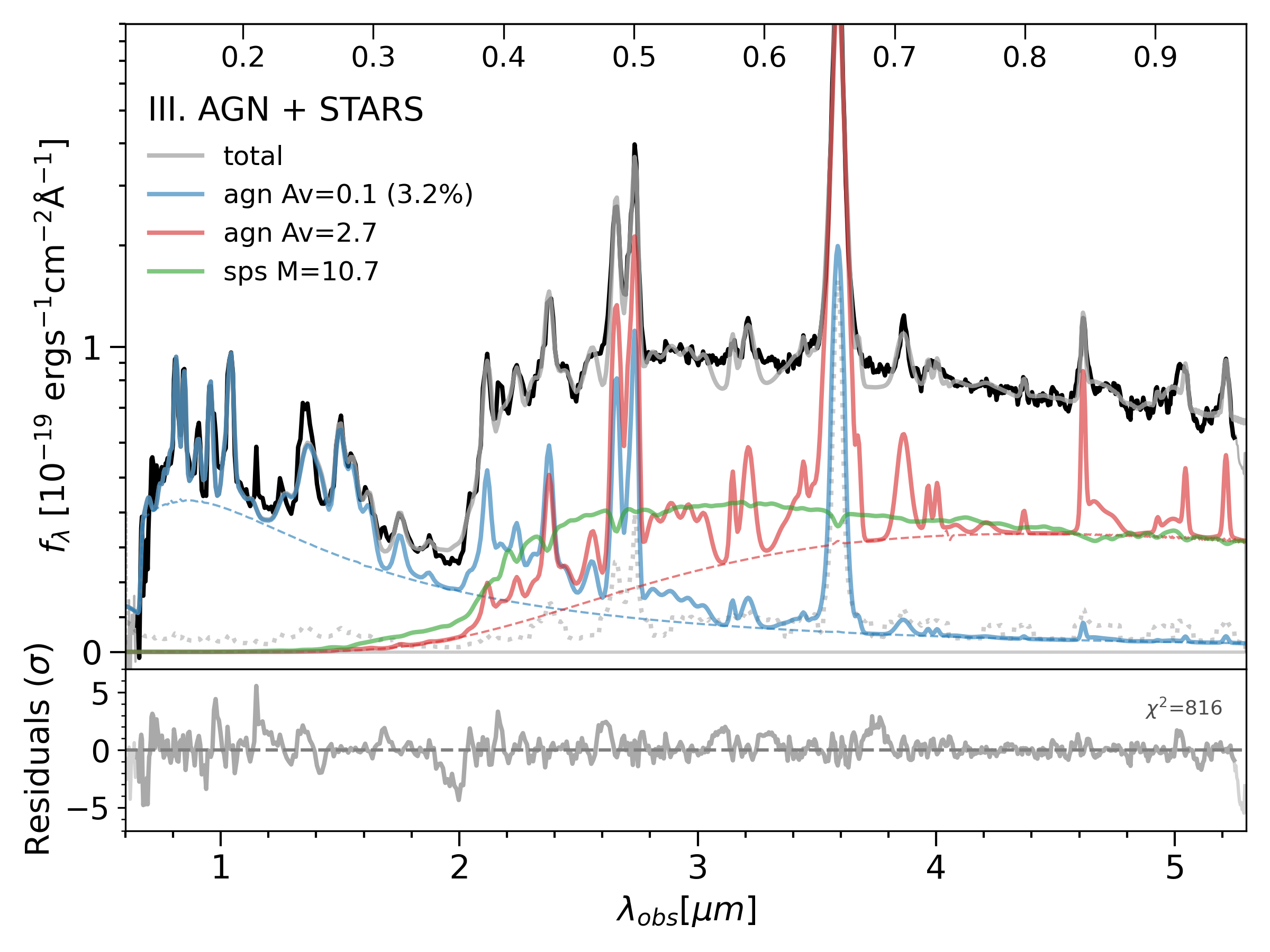} &
\includegraphics[width=0.41\textwidth]{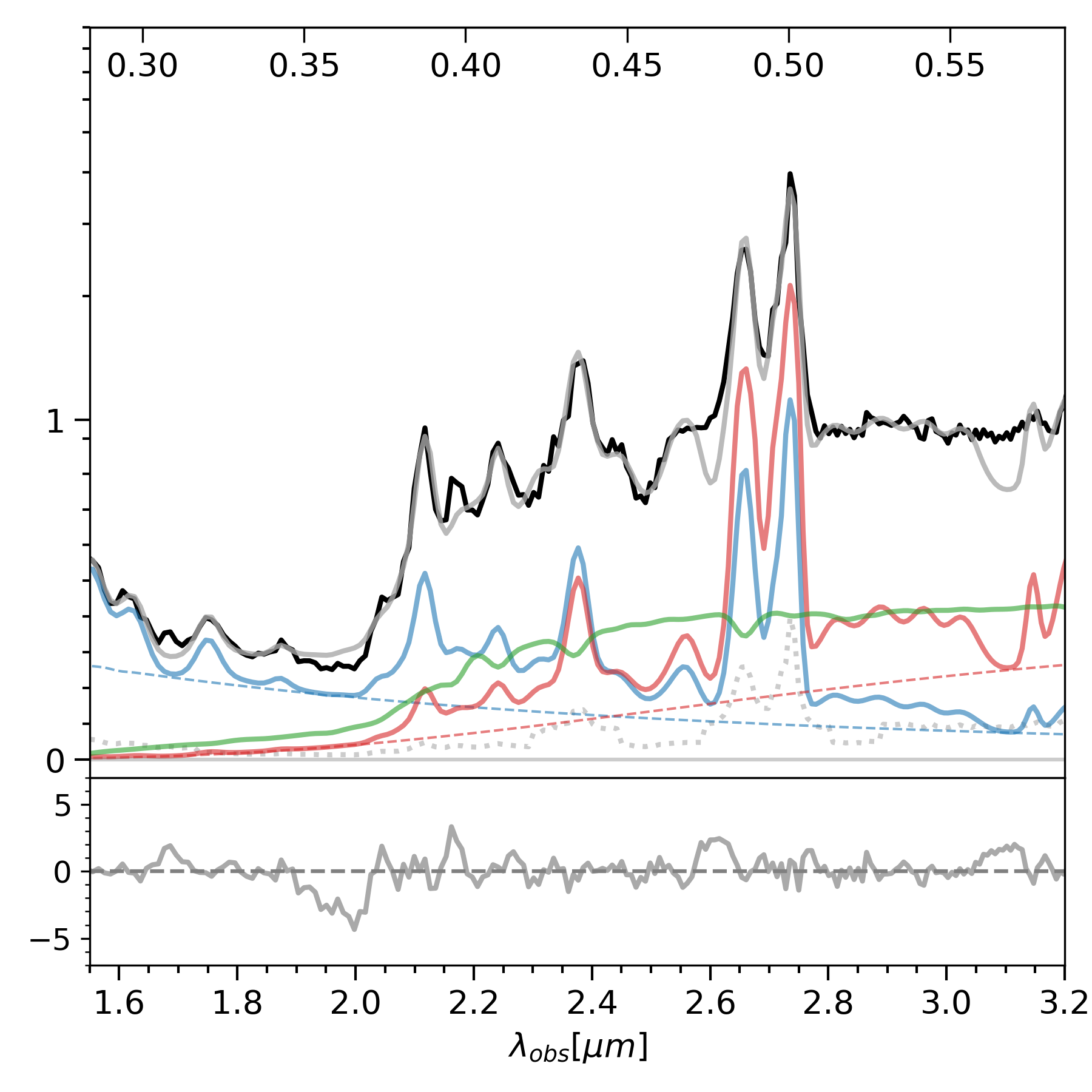} \\
\includegraphics[width=0.55\textwidth]{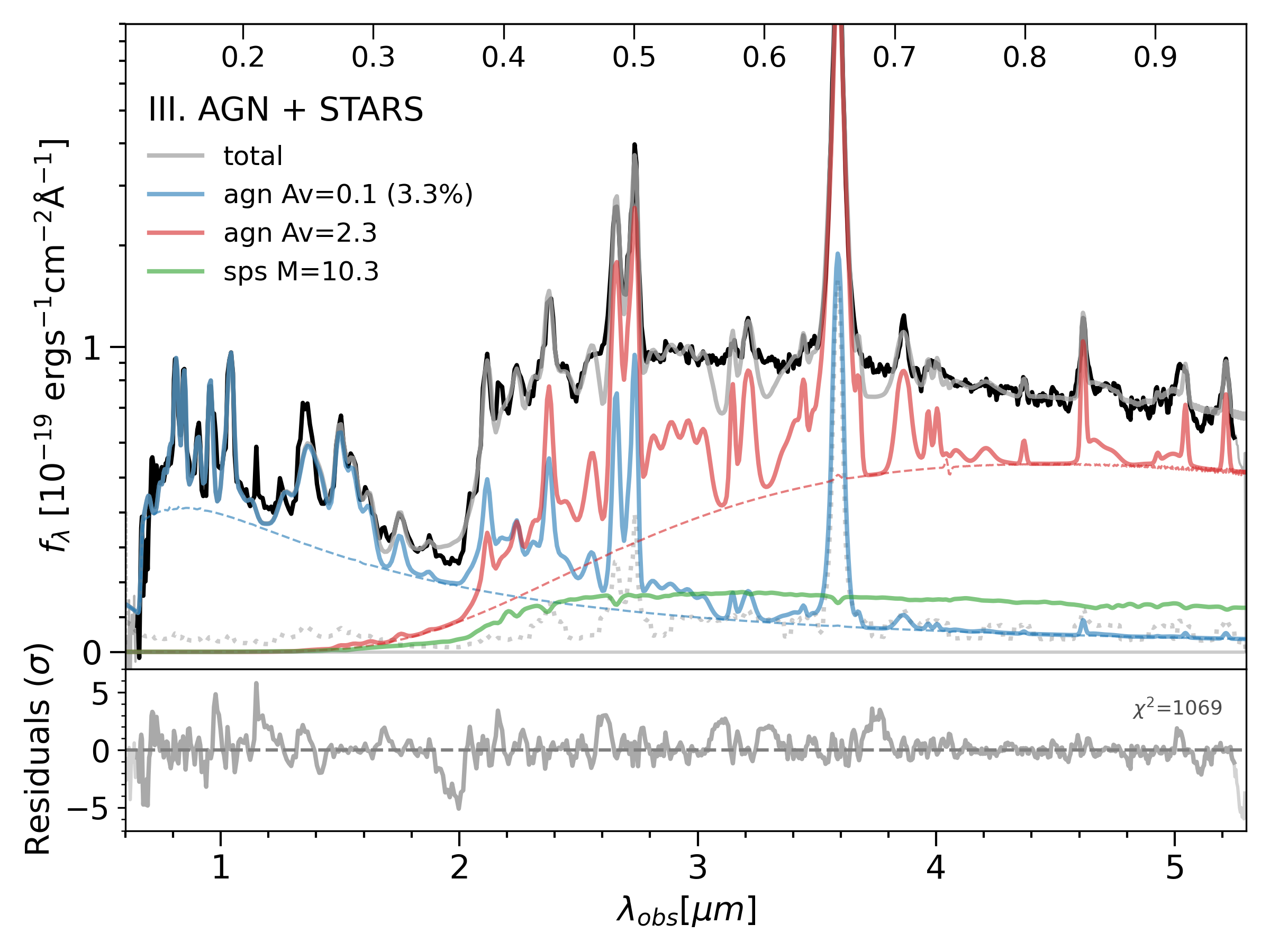} &
\includegraphics[width=0.41\textwidth]{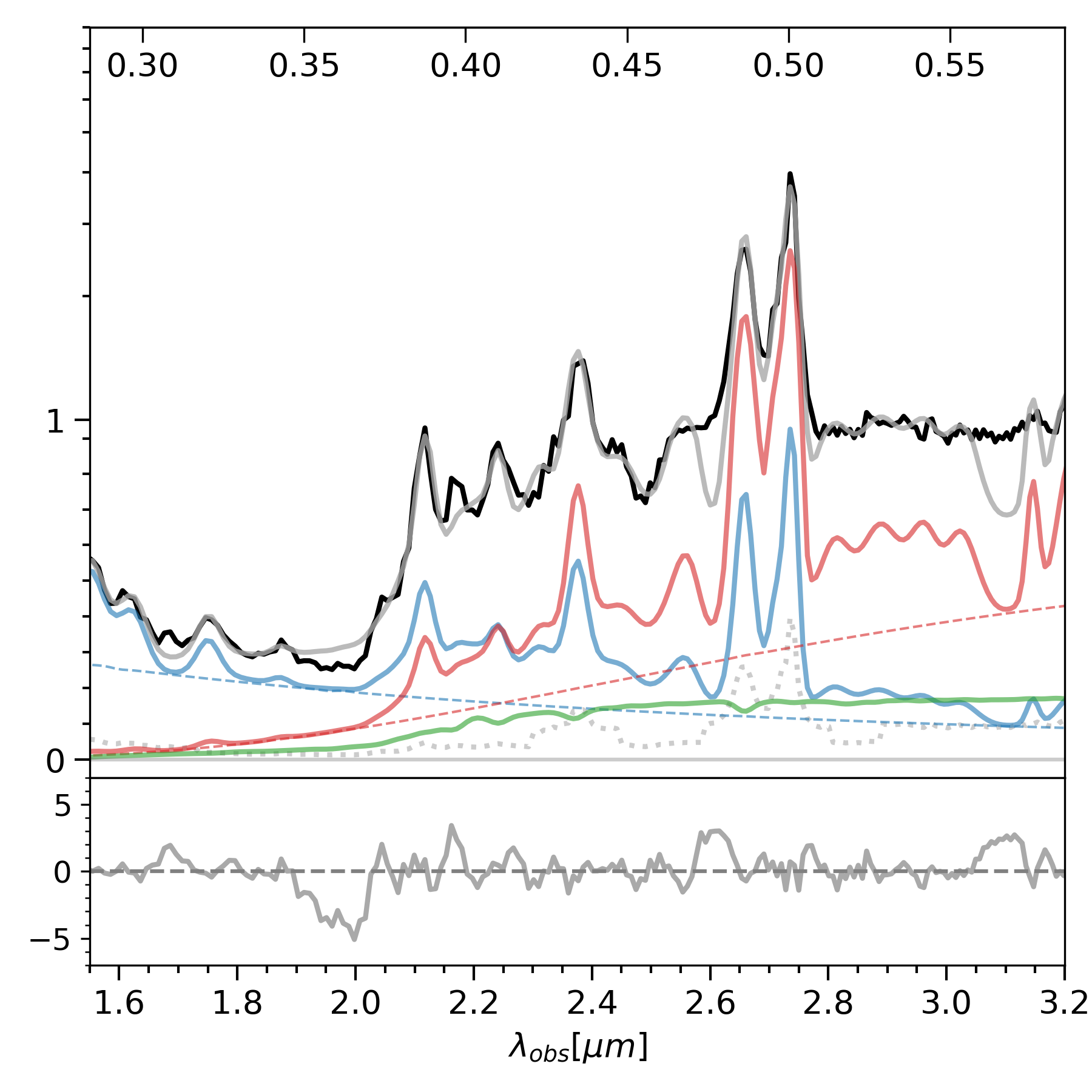} \\
\end{array}
$$
\caption{Model fits to the optical and near-IR lines, using the joint model III (AGN+Stars, see \S 5.2.3), but reducing the contribution of the evolved stellar population ({\em green}) in steps of 0.3 dex ($log(M/M_\sun)={10.9,10.6,10.3}$) and refitting. The fit residuals become significantly larger as soon as the contribution of the Balmer break component is reduced, especially in regions where \ion{Fe}{2} is expected to be weak, which produces spurious absorption-like features. This highlights that a strong break at $3650$\angstrom is difficult to produce with typical AGN models based on a power-law continuum and emission lines, and that the Balmer break component dominates the emission at rest-frame optical wavelengths.}
\label{fig:appendix:sed}
\end{figure}

Here we explore further how fit quality changes with varying contributions of the old stellar population. The main purpose is to inspect how sensitive fit quality the relative contribution of a Balmer break component. As a starting point we take the joint model III (AGN+Stars, see \S 5.2.3) best-fit solution for the evolved stellar component $log(M/M_\sun)=10.9$. Then we keep all parameters except stellar mass the same (e.g., age, $\tau$, and A$_V$), and fix stellar mass to ($log(M/M_\sun)={10.9,10.7,10.3}$) while refitting for the AGN component. The results are shown in Figure \ref{fig:appendix:sed}. As the contribution of the Balmer break component is reduced, the fit quality becomes rapidly worse. This highlights that a strong break at $3650$\angstrom is difficult to produce with typical AGN models (based on a power-law continuum and emission lines). It is likely that the Balmer break component dominates the emission at rest-frame optical wavelengths.

\end{document}